\documentclass[sigconf,nonacm]{acmart}

\usepackage[utf8]{inputenc}
\usepackage{amsmath}
\usepackage{amsthm}
\usepackage{bm}
\usepackage{bbm}
\usepackage{mathrsfs}
\usepackage{verbatim}
\usepackage {enumerate}
\usepackage{setspace}
\usepackage{graphicx}
\graphicspath{{fig/}}
\usepackage{float}
\usepackage{subfigure}
\usepackage{tikz}
\usepackage{soul}   
\usetikzlibrary{positioning} 
\usepackage{xcolor}
\usepackage{multirow}
\newtheorem{Def}{Definition}
\newtheorem{theorem}{Theorem} 
\newtheorem{proposition}{Proposition} 
\newtheorem{lemma}{Lemma}

\newcommand{\Sthree}{} 
\newcommand{\Sfour}{} 
\newcommand{\eyeone}{{
i}} 
\newcommand{\eyetwo}{{
{i+1}}} 
\newcommand{\Ltm}{L_{t^-}}

\newcommand{\dg}[1]{{
#1}}
\newcommand{\sj}[1]{{\color{black}#1}}
\newcommand{\sjc}[1]{{\color{orange}#1}}

\usetikzlibrary {shapes.symbols}
\usetikzlibrary{positioning,tikzmark}
\usetikzlibrary{overlay-beamer-styles}
\usetikzlibrary{shapes.geometric} 
\usetikzlibrary{shadings}

\makeatletter

\pgfdeclareshape{hornedsquare}{
    \savedanchor\centerpoint{
        \pgf@x=0pt
        \pgf@y=0pt
    }
    \inheritsavedanchors[from=rectangle]  
    \inheritanchorborder[from=rectangle]
    \inheritanchor[from=rectangle]{center}
    \inheritanchor[from=rectangle]{north}
    \inheritanchor[from=rectangle]{south}
    \inheritanchor[from=rectangle]{east}
    \inheritanchor[from=rectangle]{west}
    \inheritanchor[from=rectangle]{north east}
    \inheritanchor[from=rectangle]{north west}
    \inheritanchor[from=rectangle]{south east}
    \inheritanchor[from=rectangle]{south west}

    \backgroundpath{
        \pgf@xa=-0.2cm 
        \pgf@xb=0.4cm  
        \pgf@ya=-0.19cm 
        \pgf@yb=0.39cm  

        \pgfpathmoveto{\pgfpoint{\pgf@xa+0cm}{\pgf@yb+0cm}}
        \pgfpathlineto{\pgfpoint{\pgf@xb}{\pgf@yb+0cm}}
        \pgfpathlineto{\pgfpoint{\pgf@xb+0cm}{\pgf@ya+0cm}}
        \pgfpathlineto{\pgfpoint{\pgf@xa+0cm}{\pgf@ya+0cm}}
        \pgfpathclose

        \pgfpathmoveto{\pgfpoint{\pgf@xa+0.031cm}{\pgf@yb+0.125cm}}
        \pgfpathlineto{\pgfpoint{\pgf@xa}{\pgf@yb+0cm}}
        \pgfpathlineto{\pgfpoint{\pgf@xa+0.125cm}{\pgf@yb+0cm}}
        \pgfpathclose
        
        \pgfpathmoveto{\pgfpoint{\pgf@xb-0.031cm}{\pgf@yb+0.125cm}}
        \pgfpathlineto{\pgfpoint{\pgf@xb+0cm}{\pgf@yb+0cm}}
        \pgfpathlineto{\pgfpoint{\pgf@xb-0.125cm}{\pgf@yb+0cm}}
        \pgfpathclose
    }
}
\pgfdeclareshape{hornedsquareempty}{
    \savedanchor\centerpoint{
        \pgf@x=0pt
        \pgf@y=0pt
    }
    \inheritsavedanchors[from=rectangle]  
    \inheritanchorborder[from=rectangle]
    \inheritanchor[from=rectangle]{center}
    \inheritanchor[from=rectangle]{north}
    \inheritanchor[from=rectangle]{south}
    \inheritanchor[from=rectangle]{east}
    \inheritanchor[from=rectangle]{west}
    \inheritanchor[from=rectangle]{north east}
    \inheritanchor[from=rectangle]{north west}
    \inheritanchor[from=rectangle]{south east}
    \inheritanchor[from=rectangle]{south west}

    \backgroundpath{
        \pgf@xa=-0.25cm 
        \pgf@xb=0.25cm  
        \pgf@ya=-0.25cm 
        \pgf@yb=0.25cm  

        \pgfpathmoveto{\pgfpoint{\pgf@xa+0cm}{\pgf@yb+0cm}}
        \pgfpathlineto{\pgfpoint{\pgf@xb}{\pgf@yb+0cm}}
        \pgfpathlineto{\pgfpoint{\pgf@xb+0cm}{\pgf@ya+0cm}}
        \pgfpathlineto{\pgfpoint{\pgf@xa+0cm}{\pgf@ya+0cm}}
        \pgfpathclose

        \pgfpathmoveto{\pgfpoint{\pgf@xa+0.031cm}{\pgf@yb+0.125cm}}
        \pgfpathlineto{\pgfpoint{\pgf@xa}{\pgf@yb+0cm}}
        \pgfpathlineto{\pgfpoint{\pgf@xa+0.125cm}{\pgf@yb+0cm}}
        \pgfpathclose
        
        \pgfpathmoveto{\pgfpoint{\pgf@xb-0.031cm}{\pgf@yb+0.125cm}}
        \pgfpathlineto{\pgfpoint{\pgf@xb+0cm}{\pgf@yb+0cm}}
        \pgfpathlineto{\pgfpoint{\pgf@xb-0.125cm}{\pgf@yb+0cm}}
        \pgfpathclose
    }
}
\pgfdeclareshape{hornedsquare2nb}{
    \savedanchor\centerpoint{
        \pgf@x=0pt
        \pgf@y=0pt
    }
    \inheritsavedanchors[from=rectangle]  
    \inheritanchorborder[from=rectangle]
    \inheritanchor[from=rectangle]{center}
    \inheritanchor[from=rectangle]{north}
    \inheritanchor[from=rectangle]{south}
    \inheritanchor[from=rectangle]{east}
    \inheritanchor[from=rectangle]{west}
    \inheritanchor[from=rectangle]{north east}
    \inheritanchor[from=rectangle]{north west}
    \inheritanchor[from=rectangle]{south east}
    \inheritanchor[from=rectangle]{south west}

    \backgroundpath{
        \pgf@xa=-0.13cm 
        \pgf@xb=0.5cm  
        \pgf@ya=-0.19cm 
        \pgf@yb=0.39cm  

        \pgfpathmoveto{\pgfpoint{\pgf@xa+0cm}{\pgf@yb+0cm}}
        \pgfpathlineto{\pgfpoint{\pgf@xb}{\pgf@yb+0cm}}
        \pgfpathlineto{\pgfpoint{\pgf@xb+0cm}{\pgf@ya+0cm}}
        \pgfpathlineto{\pgfpoint{\pgf@xa+0cm}{\pgf@ya+0cm}}
        \pgfpathclose

        \pgfpathmoveto{\pgfpoint{\pgf@xa+0.031cm}{\pgf@yb+0.125cm}}
        \pgfpathlineto{\pgfpoint{\pgf@xa}{\pgf@yb+0cm}}
        \pgfpathlineto{\pgfpoint{\pgf@xa+0.125cm}{\pgf@yb+0cm}}
        \pgfpathclose
        
        \pgfpathmoveto{\pgfpoint{\pgf@xb-0.031cm}{\pgf@yb+0.125cm}}
        \pgfpathlineto{\pgfpoint{\pgf@xb+0cm}{\pgf@yb+0cm}}
        \pgfpathlineto{\pgfpoint{\pgf@xb-0.125cm}{\pgf@yb+0cm}}
        \pgfpathclose
    }
}
\pgfdeclareshape{smallhornedsquare}{
    \savedanchor\centerpoint{
        \pgf@x=0pt
        \pgf@y=0pt
    }
    \inheritsavedanchors[from=rectangle]  
    \inheritanchorborder[from=rectangle]
    \inheritanchor[from=rectangle]{center}
    \inheritanchor[from=rectangle]{north}
    \inheritanchor[from=rectangle]{south}
    \inheritanchor[from=rectangle]{east}
    \inheritanchor[from=rectangle]{west}
    \inheritanchor[from=rectangle]{north east}
    \inheritanchor[from=rectangle]{north west}
    \inheritanchor[from=rectangle]{south east}
    \inheritanchor[from=rectangle]{south west}

    \backgroundpath{
        \pgf@xa=-0.13cm 
        \pgf@xb=0.27cm  
        \pgf@ya=-0.15cm 
        \pgf@yb=0.25cm  

        \pgfpathmoveto{\pgfpoint{\pgf@xa+0cm}{\pgf@yb+0cm}}
        \pgfpathlineto{\pgfpoint{\pgf@xb}{\pgf@yb+0cm}}
        \pgfpathlineto{\pgfpoint{\pgf@xb+0cm}{\pgf@ya+0cm}}
        \pgfpathlineto{\pgfpoint{\pgf@xa+0cm}{\pgf@ya+0cm}}
        \pgfpathclose

        \pgfpathmoveto{\pgfpoint{\pgf@xa+0.0155cm}{\pgf@yb+0.0625cm}}
        \pgfpathlineto{\pgfpoint{\pgf@xa}{\pgf@yb+0cm}}
        \pgfpathlineto{\pgfpoint{\pgf@xa+0.0625cm}{\pgf@yb+0cm}}
        \pgfpathclose
        
        \pgfpathmoveto{\pgfpoint{\pgf@xb-0.0155cm}{\pgf@yb+0.0625cm}}
        \pgfpathlineto{\pgfpoint{\pgf@xb+0cm}{\pgf@yb+0cm}}
        \pgfpathlineto{\pgfpoint{\pgf@xb-0.0625cm}{\pgf@yb+0cm}}
        \pgfpathclose
    }
}
\pgfdeclareshape{small2nbhornedsquare}{
    \savedanchor\centerpoint{
        \pgf@x=0pt
        \pgf@y=0pt
    }
    \inheritsavedanchors[from=rectangle]  
    \inheritanchorborder[from=rectangle]
    \inheritanchor[from=rectangle]{center}
    \inheritanchor[from=rectangle]{north}
    \inheritanchor[from=rectangle]{south}
    \inheritanchor[from=rectangle]{east}
    \inheritanchor[from=rectangle]{west}
    \inheritanchor[from=rectangle]{north east}
    \inheritanchor[from=rectangle]{north west}
    \inheritanchor[from=rectangle]{south east}
    \inheritanchor[from=rectangle]{south west}

    \backgroundpath{
        \pgf@xa=-0.1cm 
        \pgf@xb=0.3cm  
        \pgf@ya=-0.15cm 
        \pgf@yb=0.25cm  

        \pgfpathmoveto{\pgfpoint{\pgf@xa+0cm}{\pgf@yb+0cm}}
        \pgfpathlineto{\pgfpoint{\pgf@xb}{\pgf@yb+0cm}}
        \pgfpathlineto{\pgfpoint{\pgf@xb+0cm}{\pgf@ya+0cm}}
        \pgfpathlineto{\pgfpoint{\pgf@xa+0cm}{\pgf@ya+0cm}}
        \pgfpathclose

        \pgfpathmoveto{\pgfpoint{\pgf@xa+0.0155cm}{\pgf@yb+0.0625cm}}
        \pgfpathlineto{\pgfpoint{\pgf@xa}{\pgf@yb+0cm}}
        \pgfpathlineto{\pgfpoint{\pgf@xa+0.0625cm}{\pgf@yb+0cm}}
        \pgfpathclose
        
        \pgfpathmoveto{\pgfpoint{\pgf@xb-0.0155cm}{\pgf@yb+0.0625cm}}
        \pgfpathlineto{\pgfpoint{\pgf@xb+0cm}{\pgf@yb+0cm}}
        \pgfpathlineto{\pgfpoint{\pgf@xb-0.0625cm}{\pgf@yb+0cm}}
        \pgfpathclose
    }
}

\makeatother

\tikzset{
  adversary/.style = {shape=house, minimum width=0.5cm, minimum height=0.5cm, fill=pink,
                     draw, align=center},
  honest/.style     = {shape=rectangle, minimum width=0.58cm, minimum height=0.58cm, fill=green, draw, align=center},
root/.style     = {honest},
	blkhonest/.style     = {shape=rectangle, minimum width=0.58cm, minimum height=0.58cm,fill=blue!0, draw, align=center},
 	blkad/.style     = {shape=hornedsquare, minimum width=0.58cm, minimum height=0.58cm, fill=red!0, draw, align=center, text centered},
   	blkad2nb/.style     = {shape=hornedsquare2nb, minimum width=0.58cm, minimum height=0.58cm, fill=red!0, draw, align=center, text centered},
    blksmall/.style = {shape=rectangle, minimum width=0.4cm, minimum height=0.4cm, fill=red!0, draw, align=center, text centered},
    blkadsmall/.style = {shape=smallhornedsquare, minimum width=0.4cm, minimum height=0.4cm, fill=red!0, draw, align=center, text centered},
    2nbblkadsmall/.style = {shape=small2nbhornedsquare, minimum width=0.4cm, minimum height=0.4cm, fill=red!0, draw, align=center, text centered},
     blkadempty/.style     = {shape=hornedsquareempty, minimum width=0.58cm, minimum height=0.58cm, fill=red!0, draw, align=center},
	rec/.style     = {shape=rectangle, minimum width=0.58cm, minimum height=0.58cm, fill=blue!0, draw, align=center},
 hidhonest/.style = {shape=rectangle, minimum width=0.58cm, minimum height=0.58cm, fill=black!15, 
  draw, align=center},
 smallhidhonest/.style = {shape=rectangle, minimum width=0.4cm, minimum height=0.4cm, fill=black!15, 
  draw, align=center},
  hidad/.style = {shape=hornedsquare, minimum width=0.58cm, minimum height=0.58cm, fill=black!15,
    draw, align=center},
    smallhidad/.style = {shape=smallhornedsquare, minimum width=0.4cm, minimum height=0.4cm, fill=black!15,
    draw, align=center}
}
\tikzstyle{block} = [draw,rectangle] 

\author{Shu-Jie Cao}
\orcid{0009-0004-0132-2586}
\affiliation{%
  \institution{Northwestern University}
  \city{Evanston}
  \state{Illinois}
  \country{USA}
}
\email{shujiecao2026@u.northwestern.edu}

\author{Dongning Guo}
\orcid{0000-0001-9000-3480}
\affiliation{%
  \institution{Northwestern University}
  \city{Evanston}
  \state{Illinois}
  \country{USA}
}
\email{dGuo@northwestern.edu}

\begin{document}
\title{How to Beat Nakamoto in the Race}

\begin{abstract}
This paper studies proof-of-work Nakamoto consensus \dg{protocols} under bounded network delays, settling two long-standing questions in blockchain security: \dg{What is the most effective attack on} block safety under a given block confirmation latency? And what is the resulting probability of safety violation? A Markov decision process (MDP) framework is introduced to precisely characterize the system state (including the \dg{block}tree and timings of all blocks mined), the adversary's potential actions, and the state transitions due to the adversarial action and the random block arrival processes. An optimal attack, called {\em bait-and-switch}, is proposed and proved to maximize the adversary's chance of violating block safety by ``beating Nakamoto in the race''. The exact probability of this violation is calculated for any \dg{given} confirmation depth using Markov chain analysis, offering fresh insights into the interplay of network delay, confirmation rules, and blockchain security.
\end{abstract}



\maketitle
\lhead{}
\rhead{}
\cfoot{\thepage} 
\thispagestyle{fancy}

\section{Introduction}
\label{s:intro}

The Nakamoto consensus, underpinned by the longest-chain fork-choice rule, serves as a foundational pillar in blockchain technology, facilitating trustless and decentralized consensus \dg{in} Bitcoin and numerous other systems. This pioneering mechanism has demonstrated consistency and liveness both theoretically and in practice, enabling consensus without a central authority. Provided that a sufficient majority of the network adheres to the consensus \dg{protocols}, the Nakamoto consensus ensures that the probability of a block's safety (or blockchain consistency) being violated vanishes as it gains more confirmations within a longest chain.

While the safety of Nakamoto consensus is well understood when all honest nodes are always fully synchronized with identical views~\cite{nakamoto2008bitcoin, garay2015bitcoin}, its precise safety under weaker synchrony assumptions and finite confirmation latency, crucial for real-world scenarios, remains an open question. Garay et al.~\cite{garay2015bitcoin} conducted a formal security analysis of Bitcoin's backbone protocol, followed by diverse approaches in later works~\cite{bagaria2019prism, pass2017analysis,kiffer2018better,gazi2020tight,garay2020full,li2021close,gazi2022practical,gazi2023practical,cao2025security,doger2024refined}, including renewal process races~\cite{li2021close, cao2025security}, Markov chains~\cite{kiffer2018better, doger2024refined}, and numerical evaluation using dynamic programming~\cite{gazi2022practical, gazi2023practical}. However, previous works only offer upper and lower bounds on block safety, where the gap between those bounds can be very large when the adversarial power is near the ultimate fault tolerance. Notably, no attack has been identified as optimal under network delays, leaving a critical aspect of the problem unresolved.

This paper adopts the widely \dg{used} $\Delta$-synchrony model, which assumes that any block or message generated or received by an honest node is delivered to all other honest nodes within a fixed delay of $\Delta$ seconds. In the proof-of-work (PoW)~\cite{nakamoto2008bitcoin} setting, we propose an attack, termed ``bait-and-switch'', and prove its optimality under the $k$-confirmation commitment rule. In this attack, the adversary privately maintains a fork and delays the propagation of honest blocks for the maximum allowed duration ($\Delta$). At a strategic moment---specifically, when the private chain is behind but matches the length of \dg{a highest} public chain visible to all honest nodes---the adversary reveals the private fork as ``bait''. This tactic causes honest nodes to perceive the adversary's chain as a longest and redirect their efforts to extend it. In effect, the adversary opportunistically harnesses honest mining power to increase their chances of winning Nakamoto's longest-chain race. Although the adversary has countless possible strategies, we prove that no other attack offers a better chance of ``beating Nakamoto in the race''!

By defining an explicit attack that is optimal, we directly address the long-standing question: ``How long must I wait for a block to settle?'' We provide an exact safety guarantee for any given confirmation depth $k$. That is, for a target safety level, one will know the exact sufficient and necessary number of block confirmations before committing a block.

It is commonly understood that the adversary's problem is to determine how to allocate resources and influence honest participants at every moment in time based on all information available until then. While prior work has modeled selfish mining using Markov decision processes (MDPs)~\cite{sapirshtein2017optimal,bai2023blockchain,feng2019selfish,zur2020efficient,doger2025when}, the rapid expansion of the state space has been a major roadblock. Suboptimal learning algorithms have been proposed to yield a solution with constrained complexity~\cite{bar2022werlman}. In this paper, we fully develop a general infinite-horizon MDP framework for the Nakamoto consensus and prove for the first time that a Markov definite policy (or attack) maximizes the probability of eventual safety violation. We also reduce the state representation---from a full blocktree and per-node views---to a compact, variable-length tuple that retains all the necessary information for assessing block safety. Using this streamlined framework, the bait-and-switch attack is proved to be an optimal policy.


The remainder of this paper begins with a rigorous model for Nakamoto consensus in Sec.~\ref{s:model}, followed by theorems for height-1 safety in Sec.~\ref{s:main}. The MDP framework is developed in Sec.~\ref{sec:MDP model}, and a simplified zero-delay model is analyzed in Sec.~\ref{s:zero} to build intuition. Sec.~\ref{s:delta} generalizes to the case with delays. A further generalization to safety at arbitrary heights is presented in Sec.~\ref{s:target}. Appendix~\ref{a:pr_of_lemma} provides proofs of some basic lemmas from the main text. Proofs related to the zero-delay case and the arbitrary delay case appear in Appendices~\ref{a:proof_proposition12} and~\ref{a:pow_delta}, respectively. Appendix~\ref{a:trade-off} presents calculations for height-1 safety, while Appendix~\ref{a:general} contains the analysis of safety at arbitrary heights.

\section{Model}
\label{s:model}

We introduce a succinct notation for referring to blocks and chains.

\begin{Def}[block] 
\label{def:block}
    The genesis block, also referred to as block~0, arrives first. Subsequent blocks are numbered sequentially in the order they arrive, starting from block 1, block 2, and so on. We let $t_b$ denote the time when block $b$ arrives. From block 1 onward, each block has a unique parent block, which arrives strictly before it. We use $f_b \in \{0,\dots,b-1\}$ to denote block~$b$'s parent block number. The parent of block 1 is the genesis block, i.e., $f_1=0$. If multiple blocks arrive simultaneously, they are assigned sequential numbers based on the order of their parents' block numbers, where those sharing the same parent are ordered arbitrarily.
\end{Def}  

\begin{Def}[blockchain and height]
\label{def:chain}
   A sequence of $n + 1$ block numbers, $(b_0, b_1, \dots, b_n)$, defines a blockchain (or simply a chain) if $b_0 = 0$ and block $b_{i-1}$ is block $b_i$'s parent for every $i=1,\dots,n$. Since every chain can be uniquely identified by its last block, we also refer to the preceding chain as chain $b_n$. As a convention, we say the height of this chain is $n$. Also, we say block $b_{i+1}$ extends chain $b_i$ and that chain $b_i$ is a prefix of chain $b_n$ for every $i=0,\dots,\dg{n}$. In general, we use $h_b$ to denote the height of chain $b$, which is also referred to as the height of block $b$. Evidently, the genesis block is the only block \dg{at} height $0$.
\end{Def}


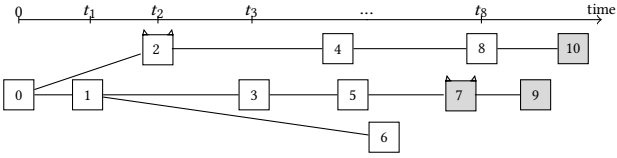
\begin{figure}
    \centering
    {\scriptsize
    \begin{tikzpicture}
    \draw[->] (0,1.0)--(7.75,1.0)node[above] {
    time};
    \draw[-] (0,1.05)--(0,0.95) node[above] {$0$}; 
    \draw[-] (0.95,1.05)--(0.95,0.95) node[above] {$t_1$};
    \draw[-] (1.85,1.05)--(1.85,0.95)node[above] {$t_2$};
    \draw[-] (3.1,1.05)--(3.1,0.95)node[above] {$t_3$};
    \draw[-] (4.625,1.0)--(4.625,1.0)node[above] {$...$};
    \draw[-] (6.15,1.05)--(6.15,0.95)node[above] {$t_8$};
  \node(0) [blksmall]{0};
    \node[ right=.5cm of 0] (1) [blksmall] {1};
    \draw (0)--(1);
    \node[above right=0.2cm and 0.5cm of 1] (2) [blkadsmall] {2};
  \draw (0)--(2);
  \node[ right=1.8cm of 1] (3) [blksmall] {3};
  \draw (1)--(3);
      \node[right=2cm of 2] (4) [blksmall] {4};
\draw (2)--(4);
  \node[ right=0.9cm of 3] (5) [blksmall] {5};
  \draw (3)--(5);
   \node[below right=0.15cm and 0cm of 5] (6) [blksmall] {6};
   \draw (1)--(6);
 \node[right=1.0cm of 5] (7) [smallhidad] {7};
 \draw (5)--(7);
  \node[right=1.5cm of 4] (8) [blksmall] {8};
 \draw (4)--(8);
   \node[ right=0.6cm of 7] (9) [smallhidhonest] {9};
 \draw (7)--(9);
\node[right=0.8cm  of 8] (10) [smallhidhonest] {10};
 \draw (8)--(10);
 \end{tikzpicture}}
     \caption{\dg{An example blocktree.}}
     \label{fig:blocks_feature}
\end{figure}

Fig.~\ref{fig:blocks_feature} depicts a snapshot of the blocktree upon block~10's arrival, where chain~9 is defined by $(0,1,3,5,7,9)$ and its height is 5.

\subsection{Decentralized Blockchain System}
\label{s:system}

A Nakamoto consensus protocol enables pseudonymous participants to achieve probabilistic state replication with high reliability~\cite{garay2015bitcoin, nakamoto2008bitcoin}. We assume a well decentralized system with a large number of participants (nodes or miners). Without loss of generality, we consider a single adversarial participant, since multiple adversaries can be treated as a single coordinated entity---yielding a more powerful adversary. The protocol can be characterized by the following key components and principles.

\subsubsection{Proof of Work}

Nodes compete to solve cryptographic puzzles, and each successful solution grants the right to propose a new block extending an existing chain. 
The puzzle---essentially a brute-force search for a solution to a hash inequality based on SHA-256---is memoryless nature.
Since nodes typically perform an \dg{astronomical} number of independent attempts per second, the block arrival process is well modeled as a continuous-time Poisson point process (see, e.g.,~\cite{guo2022bitcoin, dembo2020everything}).

\subsubsection{Longest-Chain Fork Choice}

When an honest node gets the opportunity to propose a block, the node always extends a highest valid chain in its view. In the event that multiple highest chains exist in the view, we adopt a worst-case assumption, allowing the adversary to resolve the tie in any manner it prefers for that node. \dg{Throughout this paper, the terms ``longest'' and ``highest'' are used interchangeably.}

\subsubsection{Block Propagation Delay}
\label{s:ss:delay}

When an honest node mines a new block or receives a new valid block for the first time, it relays the block to its peers through a peer-to-peer network. 
Due to network delays, different honest nodes may maintain different views of the block\dg{tree}. The widely used $\Delta$-synchrony model posits an upper bound of $\Delta$ seconds on end-to-end communication delays among nodes, such that any information known to an honest node at time $t$ becomes known to all honest nodes by time $t+\Delta$.

For this worst-case analysis, we assume that honest mining is sufficiently decentralized in the sense that the chance that a block mined by an honest node is followed by another block mined by the same node within $\Delta$ time is negligible.

\subsubsection{Commitment rule}

This paper adopts the $k$-confirmation commitment rule, namely, a node commits to a block when at least $k-1$ blocks have been mined on top of it as part of a longest chain \dg{in that node's view}, where $k$ is an integer selected by the node. If, after a block at height $\eta$ is committed by some honest node, it remains permanently included in all longest chains in all honest nodes' views, then the block---and thus height $\eta$---is (forever) safe; otherwise the safety of the block (and its height) is violated (at some point). 

\subsubsection{Adversarial Capabilities} 
\label{s:adversary}

The adversary is assumed to be omniscient in the sense that it knows everything up to the present moment. However, the adversary cannot foresee future events and is therefore limited to mounting {\em causal} attacks. Specifically, the adversary possesses the following capabilities:
\begin{itemize}
    \item It can allocate its own mining power to extend any existing block and may arbitrarily split this power across an arbitrary set of candidate parent blocks.

    \item It has full control over message propagation delays between any pair of nodes, subject to the $\Delta$-synchrony constraint (see Sec.~\ref{s:ss:delay}).
    
    \item It may withhold its own chains for an arbitrary duration and selectively reveal any chain to any honest node at any time\dg{.}
    
    \item Combined with its ability to break ties, the adversary can effectively dictate how each honest node allocates---or even splits---its mining power across multiple viable chains.
\end{itemize}
Although its capabilities are bounded by its mining power and the $\Delta$-synchrony model, the adversary nonetheless has access to a vast space of possible strategies over continuous time. This makes identifying its optimal attack strategy a highly nontrivial challenge.

\subsection{Enhanced Adversary}
\label{s:enhanced}

In this paper, we develop a technique to reduce the space of adversarial strategies, making the search for an optimal attack tractable. The approach consists of the following steps:
\begin{enumerate}
    \item[i)] Enhancing the adversary's capabilities in a specific manner, so that it suffices for the adversary to act only at discrete block arrival times instead of over continuous time.
    \item[ii)] Formulating the attack problem as a Markov decision process to compute an optimal policy for the enhanced adversary.
    \item[iii)] Proving that the optimal policy makes no use of the enhanced capabilities, and is therefore also optimal for the original adversary.
\end{enumerate}
In addition to the original adversarial powers described in Sec.~\ref{s:adversary}, the enhanced adversary is granted the following extra capabilities:
\begin{itemize}
    \item At the arrival time of a block mined by the adversary, it may choose any existing block as the parent;

    \item At the arrival time of a block $b$ not mined by the adversary, if there exists a chain $c$ that is no lower than a highest chain in any honest node's view, then the adversary may designate the honest node as the miner of block $b$ and assign block $c$ as its parent. (As a result, chain $b$ enters the honest miner's view.)
\end{itemize}

The enhancement appears subtle: For a block that arrives at $t$, the original adversary must decide which node(s) and chain(s) are the potential miner(s) and parent(s) by allocating (adversarial and honest) mining power based on available information \dg{strictly before $t$}. In contrast, the enhanced adversary may choose both the miner and the parent of the block upon its arrival, based on all available information in $[0,t]$ (which is more than  $[0,t)$). In particular, if the original adversary splits its mining power between two chains, it cannot predict which one will be extended, whereas the enhanced adversary can decide which one is extended upon the block's arrival.

Another distinction is that the enhanced adversary can delay revealing a chain to honest nodes, yet still letting them unknowingly mine to extend it. The chain is only revealed if an honest node successfully extends it. In contrast, the original adversary has to reveal the chain to at least one honest node to attract its mining power, which causes the chain to become publicly known within $\Delta$ time---even if no honest node succeeds to extend it.


A key insight is that if the adversary is allowed to select the parent of each new block at the moment it arrivals, then it no longer needs to pre-allocate mining power toward specific chains. Consequently, the enhanced adversary need not take any action until a block arrives. Moreover, to preserve strategic flexibility, it suffices for the enhanced adversary to maximally delay the propagation of every honest block. This ensures that if an honest node mines a block to extend a chain according to the original adversary's orchestration, the enhanced adversary can designate the same honest node and chain as the miner and parent of that block.

\subsection{A Tree-based Mathematical Model}
\label{s:mathmodel}

Following Definitions~\ref{def:block} and~\ref{def:chain}, we \dg{introduce} a succinct \dg{{\em tree-based}} mathematical model for \dg{the view-based} blockchain system with an enhanced adversary \dg{described in Secs.~\ref{s:system} and~\ref{s:enhanced}.}

\begin{Def}[public]
\label{def:public}
    A block or chain may become public at a specific point in time strictly after its creation and will remain public at all times afterward. \dg{Block $b$ is public if and only if chain $b$ is public.} A chain being public at $t$ is equivalent to all of its blocks being public at $t$. As a convention, the genesis block becomes public at time 0 and block 1 is mined afterward, i.e., $t_1 > 0$.
    At time $t$, the height of a highest public chain is referred to as the public height. 
\end{Def}

At any given time, every existing chain comprises a public prefix, possibly extended by nonpublic blocks. \dg{As an example, in Fig.~\ref{fig:blocks_feature}, at $t_{10}$, all} nonpublic blocks (7, 9, and 10) are shaded. \dg{It is easy to see that the} public height is 3.

\begin{Def}[credible]
\label{def:credible}
    A chain~$c$ is said to be credible at time $t$, or $t$-credible, if it is no lower than the public height at time $t$.
\end{Def}

In Fig.~\ref{fig:blocks_feature}, chains~5, 7, 8, 9, and 10 are $t_{10}$-credible. No other chain in Fig.~\ref{fig:blocks_feature} is $t_{10}$-credible.

\begin{Def}[A-block and H-block]
\label{def:AH}
    By convention, the genesis block is neither an A-block or an H-block. Each subsequent block is either an A-block or an H-block. An H-block that arrives at $t$ must be strictly higher than the public height at time $t$.
\end{Def}

\dg{In Fig.~\ref{fig:blocks_feature}, each A-block is distinguished by a pair of horns on top. Blocks~2 and~7 are A-blocks; blocks 1, 3--6, and 8--10 are H-blocks.} 

\begin{Def}[propagation delay bound $\Delta$]
\label{def:delta}
    If block $b$ is an H-block, 
    then chain $b$ must become public no later than $t_b+\Delta$.
\end{Def}

\begin{Def}[$k$-confirmation]
\label{def:confirmation}
    If there exists a $t$-credible chain $c$ for some $t>0$ \dg{that} includes a block $b$, where \dg{the heights satisfy} $h_c \ge h_b+k-1$, then we say chain $c$ confirms block $b$ \dg{at} depth $k$, or block $b$ is $k$-confirmed by chain $c$.
\end{Def}

\dg{In Fig.~\ref{fig:blocks_feature}, chain~7 confirms block~1 at depth~4, even though block~7 is nonpublic. At the same time, chain~8 confirms block~2 at depth 3, even though chain~8 is not the longest.}

\begin{Def}[safety violation]
\label{def:v}
    If there ever exist times $t>0$ and \dg{$t'\ge t$}, when one $t$-credible chain confirms block~$c$ and another $t'$-credible chain confirms a different block at the same height, $h_c$, we say the safety of height $h_c$ is violated, or a safety violation occurs at height $h_c$ at time $t'$.
\end{Def}

\dg{In Fig.~\ref{fig:blocks_feature}, only block~1 is 5-confirmed by $t_{10}$, thus there is no violation yet under a 5-confirmation rule. However, if a 4-confirmation rule is adopted instead, then block~7 confirms block~1 at $t_7$ and block~10 confirms block~2 at $t_{10}$, causing a violation at height~1.}

\begin{Def}[arrival processes]
\label{def:mining process}
    Let $H_t$ (resp.\ $A_t$) denote the number of honest (resp.\ adversarial) arrivals during $(0,t]$. 
    \dg{Let} $(H_t)_{t>0}$ and $(A_t)_{t>0}$ \dg{be} independent Poisson point processes with rates $h$ and $a$, respectively.
    An honest arrival at $t$ corresponds to the mining of an H-block at $t$. An adversarial arrival corresponds to the mining of an A-block.
\end{Def}

Since time is modeled as a continuum, the probability of two or more arrivals ever occurring at exactly the same instant is zero; hence we assume that no two blocks arrive simultaneously at any point in time. Indeed, the continuous-time model avoids an artifact inherent in discrete-time models, where multiple arrivals may occur within the same time slot, regardless of its brevity.

\subsection{Tree-Based Model for View-Based System}
\label{sec:sufficiency-model}

The decentralized system described in \dg{Secs.~\ref{s:system} and~\ref{s:enhanced}} is {\em view-based} in the sense that its evolution can be fully described using the trajectory of all nodes' views (including the adversary's). In the literature, nodes are often treated as interactive Turing machines which follows an algorithm (see, e.g.,~\cite{garay2015bitcoin, garay2020full}). \dg{We next draw connections between the view-based system and its tree-based model.}

A chain known to all honest nodes \dg{in the view-based system} is termed public \dg{in the tree-based model}, per Definition~\ref{def:public}. A chain that is at least as high as any public chain is deemed credible, per Definition~\ref{def:credible}.
A-blocks and H-blocks \dg{in the tree-based model} represent blocks mined by honest and adversarial nodes, respectively, \dg{in a view-based system}. The adherence of honest nodes to the longest-chain fork-choice rule is encapsulated by the requirement \dg{(in the tree-based model)} that each new H-block must be higher than all existing public chains, as specified in Definition~\ref{def:AH}. In contrast, an A-block can have any existing block as its parent.

Evidently, a longest chain in a node's view \dg{(in a view-based system) is credible in the corresponding tree-based model} (Definition~\ref{def:confirmation}). For example, in Fig.~\ref{fig:blocks_feature}, with $k = 3$, block~2 is committed via a $t_8$-credible chain, chain~8, as two blocks are mined on top of it within chain~8, which is a longest chain visible to an honest node. However, block~2 is absent from the subsequent $t_9$-credible chain (chain~9), where $t_9>t_8$, resulting in a safety violation at height~1.

The public view represents the collective knowledge shared by all honest nodes. Rather than modeling each node’s individual view---which can vary due to network delays or message propagation issues---the public view serves as a common reference point for consensus. It also represents the ``weakest'' or most disadvantaged perspective an honest node can have, corresponding to the least up-to-date state of the blockchain due to factors including network latency and manipulation by the adversary. If a safety violation (e.g., a fork or rollback of a committed block) occurs in any honest node's view---potentially a more up-to-date or informed perspective---then it must also occur in the weakest view. Thus, the public view provides a worst-case perspective from the standpoint of honest nodes, ensuring that any violation observed in a stronger view is necessarily reflected in the weakest one.

We next show that the tree-based mathematical model 
\dg{is} sufficient for studying safety violations in the view-based \dg{system}.



\begin{proposition}
    For every realization (or sample path) of the A- and H-block arrival processes, if the enhanced adversary in the view-based \dg{system} can produce a certain blocktree, yielding a safety violation, so can the adversary in the tree-based mode, and vice versa.
\end{proposition}

\begin{proof}
    We prove by induction on the sequence of block arrivals. At time $t_0=0$, the two models share the same blocktree, which includes only the genesis block. In the view-based \dg{system}, all honest nodes' views include only this single-node blocktree. In the tree-based model, the genesis is public.
    
    Suppose the two models share the same blocktree immediately after the arrival time of block $(b-1)$, and that the public prefix of the blocktree in the tree-based model is included in all honest views in the view-based \dg{system}. As a consequence, every honest view includes at least one chain that is no lower than the public height.
    
    First, if an honest node mines block $b$ to extend chain $c$ in the view-based \dg{system}, then chain $c$ is no lower than the public height, so that block $b$ can be made a child of block $c$ in the tree-based model.

    Second, by delaying the propagation of all honest blocks maximally in the view-based system, all honest nodes hold the same (public) view except for those who have mined blocks within the most recent $\Delta$ period, which are yet to become public.
    As a consequence, if an H-block $b$ arrives to extend chain $c$ in the tree-based model, chain $c$ must be credible (with respect to the public view), which implies that chain $c$ is as high as a highest chain in some honest views in the view-based \dg{system}, so that the enhanced adversary can pick such an honest node as the miner of block $b$ to extend chain $c$.

    The preceding arguments guarantee that the view-\dg{based system and the tree-based model} share the same blocktree immediately after the arrival of block $b$. Moreover, the public prefix of the blocktree in the tree-based model is included in all honest views in the view-based \dg{system}.

    Consider a safety violation in the tree-based model, where one $t$-credible chain $c$ confirms block $b$ \dg{at} height $s$ and another $t'$-credible chain $c'$ confirms a different block $b'$ \dg{at} height $s$. Evidently, chain $c$ can be made a highest in some honest view at time $t$ in the view-based \dg{system}, and $c'$ can be made a highest in another honest view at time $t'$, creating a safety violation in the view model with the same blocktree. It is easy to see that the converse is also true. Hence the two models are equivalent in the sense that if the enhanced adversary can create a safety violation in the view-based \dg{system}, then so can the adversary in the tree-based model, and vice versa.
\end{proof}


\subsection{Discussion}

Traditional approaches often employ a view-based \dg{system} incorporating elements of the {\em universally composable} security framework, executed in discrete time steps~\cite{pass2017analysis,gazi2020tight,dembo2020everything,canetti2020universally}. However, modeling the views of individual nodes necessitates accounting for intricate interactions and dependencies, thereby incurring substantial mathematical and computational overhead. The tree-based mathematical model in Sec.~\ref{s:mathmodel} is considerably simpler yet self-contained. Rather than tracking each honest node's view over time, we focus on the ``weakest possible'' view across all honest nodes. This weakest view, defined precisely in Definitions~\ref{def:public} and~\ref{def:credible}, includes only public chains. This tree-based model is sufficient to capture and evaluate all adversarial strategies that could lead to safety violations. In a nutshell, our insight on the weakest possible view eliminates the need to define individual views and their complex interactions. We maintain that our mathematical model is succinct, sufficient, and sound.

As mentioned previously, we shall show that granting the enhanced adversary some specific additional privileges does not increase the adversary's chance of causing a safety violation. In fact, we shall establish an optimal ``bait-and-switch'' attack, in which the adversary strategically makes certain A-blocks public to attract honest mining power when it benefits the adversary to do so, without relying on the extra privileges of choosing the block' parent after its mining. This bait-and-switch attack is entirely causal and hence realizable.

We remark that all existing latency-security analyses of the Nakamoto consensus allow a wide range of non-causal attacks by giving the adversary foresight. For example, previous works~\cite{garay2015bitcoin, pass2017rethinking, dembo2020everything, cao2025security, guo2022bitcoin} all focus on the event where a greater number of certain types of blocks are mined compared to others during the confirmation period of interest. This is tantamount to allowing the adversary to optimally assemble blocks into chains with foresight of the timings of all future block arrivals. Those analyses generally do not achieve the ultimate fault tolerance for finite confirmation depths. An exception is~\cite{cao2025security}, where the gap is closed by introducing a crucial causality requirement: Once the adversary falls behind the public height, honest mining power is concentrated on the public front, and the adversary can only catch up using freshly mined A-blocks, where no other blocks can be used to catch up with the public block.

Our mathematical model does not incorporate cryptographic mechanisms. For instance, arrival processes are exogenous rather than resulting from cryptographic hash computations. Consequently, we do not introduce a cryptographic security parameter, such as the probability of hash collisions, as used in related works to quantify block safety (see e.g.,~\cite{pass2017analysis,garay2020full}). This omission does not compromise practicality, as the probability of a block safety violation of interest is orders of magnitude higher than that of a cryptographic failure.

\subsection{Adversarial Policy (Attack)}
\label{sec:attack}

In all but Sec.~\ref{s:target}, 
we restrict our attention to the safety of height 1 for clarity and simplicity.
A well-known attack is the following:

\begin{Def}[private mining attack on height 1]
\label{def:private mining}
    The adversary delays the propagation of every H-block maximally (for $\Delta$ seconds). Starting from time 0, the adversary always mines in private to extend the highest chain which does not include the first H-block.
\end{Def}

Under this private mining attack, all A-blocks form a chain, referred to as the A-chain. Height 1's safety is violated as soon as the A-chain is both credible and consists of at least $k$ blocks. At this point, the A-chain confirms the first A-block \dg{at} a depth of $k$ or more, while an H-block at height 1 is either already $k$-confirmed or will be once an H-block reaches height $k$.

\begin{Def}[agree and disagree]
    Blocks $b$ and $d$ are said to agree on a given height if chains $b$ and $d$ both contain the same block $e$ on that height. In this case, we also say that blocks $b$ and $d$ agree on block~$e$. Otherwise, we say blocks or chains $b$ and $d$ disagree at the height, or disagree on block $e$.
\end{Def}

\begin{Def}[jumper]
\label{def:jumper}
    The first H-block to arrive at its height is called a jumper. 
\end{Def}

\begin{Def}[branch]
\label{def:branch}
    Agreement on height-1 blocks is an equivalence relation, which divides all blocks into equivalence classes according to all the height-1 blocks. Each equivalence class is called a branch. If block $b$ is the height-1 block in a branch, the branch is referred to as branch $b$. At time $t$, we use branch $b$ to refer to all of its blocks that have arrived by $t$, where the height of the branch is equal to the height of a highest block in it, and the branch is said to be $t$-credible if it is no lower than the public height.
\end{Def}

For example, the blocktree in Fig.~\ref{fig:blocks_feature} contains two branches, branches 1 and 2. Branch 1 includes only block 1 at $t_1$. By $t_6$, branch 1 includes also blocks 3, 5, and \dg{6}.

In order to ease our analysis, we use the following lemma to equate the event of interest from safety violation 
to the existence of two credible branches higher than confirmation depth $k$, which by Definition~\ref{def:v} constitutes a safety violation.

\begin{lemma}[The equivalence of safety violation]
\label{lm:SV->absorb}
    Under the $ k $-confirmation rule, if a safety violation occurs at height 1 at time $ t $, then there must exist a time $ s \leq t $ at which two distinct $ s $-credible branches are at least as high as $ k $.
\end{lemma}

The proof is relegated to Appendix~\ref{a:proof_lm1}.

By Lemma~\ref{lm:SV->absorb}, each safety violation event is encompassed by the occurrence of two distinct credible branches with heights of at least $k$. Hence, the analysis can be simplified by focusing on the event where the two highest credible branches have heights at least $k$.

Recall that the adversary can break ties arbitrarily for longest-chain fork choice. 
Thus it can make a credible chain the fork choice of all honest nodes. Formally we introduce the following notion:

\begin{Def}[public fork choice]
    We say that the adversary makes a chain or branch the public fork choice at time $t$ if the first H-block that arrives after $t$ extends the chain or branch.
\end{Def}

\begin{Def}[bait-and-switch attack on height 1]
\label{def:bait-and-switch}
    Upon every adversarial arrival,
    if there is a unique highest branch and a second highest branch is strictly lower than the highest jumper, the adversary creates one A-block extending the second highest branch; otherwise, the adversary creates one A-block extending a highest branch.
    All A-blocks are kept private except when there is a unique highest branch and the height of a second highest branch is equal to the highest public height, then the second branch is made the public fork choice (this is referred to as baiting). The propagation of every H-block is delayed maximally (for $\Delta$ seconds) unless it becomes public in a public fork choice.
\end{Def}

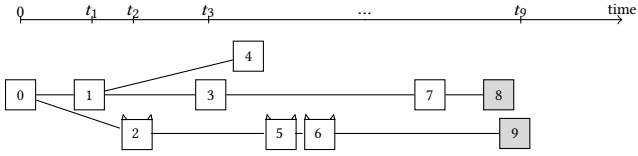
\begin{figure}
    \centering
    \scriptsize
    \begin{tikzpicture}
    \draw[->] (0,1.0)--(8,1.0)node[above] {
    time};
    \draw[-] (0,1.05)--(0,0.95) node[above] {$0$}; 
    \draw[-] (0.95,1.05)--(0.95,0.95) node[above] {$t_1$};
    \draw[-] (1.5,1.05)--(1.5,0.95)node[above] {$t_2$};
    \draw[-] (2.5,1.05)--(2.5,0.95)node[above] {$t_3$};
    \draw[-] (4.575,1.0)--(4.575,1.0)node[above] {$...$};
    \draw[-] (6.65,1.05)--(6.65,0.95)node[above] {$t_9$};
    \node(0) [blksmall]{0};
    \node[ right=.5cm of 0] (1) [blksmall] {1};
    \draw (0)--(1);
    \node[below right=0.1cm and 0.2cm of 1] (2) [blkadsmall] {2};
    \draw (0)--(2);
     \node[ right=1.2cm of 1] (3) [blksmall] {3};
     \draw (1)--(3);
     \node[above right=0.1cm and 1.7cm of 1] (4) [blksmall] {4};
     \draw (1)--(4);
     \node[ right=1.5cm of 2] (5) [blkadsmall] {5};
     \draw (2)--(5);
     \node[ right=0.1cm of 5] (6) [blkadsmall] {6};
     \draw (5)--(6);
     \node[ right=2.5cm of 3] (7) [blksmall] {7};
     \draw (3)--(7);
     \node[ right=0.5cm of 7] (8) [smallhidhonest] {8};
     \draw (7)--(8);
     \node[ right=2.2cm of 6] (9) [smallhidhonest] {9};
     \draw (6)--(9);
    \end{tikzpicture}
    \caption{Example of a \dg{successful bait-and-switch attack}.}
    \label{fig:PoW balancing+}
\end{figure}

A snapshot of a successful bait-and-switch attack is \dg{illustrated} in Fig.~\ref{fig:PoW balancing+}, assuming confirmation depth $k=4$. At $t_8$, chain~8 is on \dg{a} unique highest branch, and it $k$-confirms block 1. Chain~7 is public. Chain~6 is a second highest branch and equal to public height~3, thus it is made the public fork choice at $t_8$.
H-block 9 arrives by $t_8+\Delta$ and is used to extend chain~6, confirming block 2 
to cause a safety violation.

Unlike selfish mining, where the adversary maintains a lead 
and reveals 
its blocks 
when honest miners approach 
parity, bait-and-switch reveals 
the adversarial chain 
when the newest nonpublic honest block is one height ahead, baiting 
honest nodes to extend the adversarial chain. Although both strategies hinge on revealing a hidden 
fork, their 
triggers and impacts on mining incentives differ markedly. 

\section{Main Results on Safety of Height 1}
\label{s:main}

We recall that $a$ \dg{represents} the adversarial arrival rate, $h$ \dg{represents} the honest arrival rate, and $\Delta$ \dg{represents the upper bound on network delays. In the remainder of this paper, we denote the} total arrival rate \dg{as} $\lambda=a+h$, and the fraction of adversarial arrivals \dg{as} $\beta=a/\lambda$.

\begin{theorem}
    \label{th:opt_pow}
    Under the proof-of-work Nakamoto consensus protocol, the bait-and-switch attack maximizes the probability of violating the safety of height 1.
\end{theorem}

We shall develop a proof of Theorem~\ref{th:opt_pow} in Sec.~\ref{s:delta}. We next evaluate the probability of violation achieved by the bait-and-switch attack.

Define the moment generating function of a certain random variable as
    \begin{align}\label{eq:E(r)}
        \mathcal{E}(r)
        = \frac{(1-r)(h - a -h a \Delta)}{ h - e^{(1-r) a \Delta}( h+ a- a r)r} ,
    \end{align}    
so the corresponding probability mass function (pmf) is
    \begin{align} \label{eq:e(i)}
        e(i) &= \frac{1}{i!} \mathcal{E}^{(i)}(0) , \qquad i=0,1,\dots .
    \end{align}

Let $f_1(\cdot;\lambda)$ denote the pmf of Poisson distribution with expectation $\lambda$. Let $f_2(\cdot;l,a)$ denote the probability dense function (pdf) of Erlang distribution with shape parameter $l$ and rate $a$ if $l>0$. We define a conditional pmf for integer-valued random variables $W$ and $L$:
  \begin{align}
  \label{eq:PW}
    \begin{split}
    &    P_{W|L}(w|l)\\
    &    =
        \begin{cases}
            f_1(w,a\Delta), &\text{if } l < 0 \text{ or } w< l; \\
           e^{-\lambda\Delta}, &\text{if }  l= 0, w=l;
            \\
            \int_{0}^{\Delta}f_2(t;l,a)e^{-\lambda(\Delta-t)}dt, &\text{if }  l> 0, w=l;
            \\
            \int_{0}^{\Delta} \lambda e^{-\lambda s} f_1(w-l-1;a(\Delta-s))ds, &\text{if }  l= 0, w>l;\\
            \int_{0}^{\Delta}f_2(t;l,a)\int_{0}^{\Delta-t} \lambda e^{-\lambda s} \\
            \qquad \cdot
            f_1(w-l-1;a(\Delta-t-s))dsdt, &\text{if }  l > 0, w>l.
        \end{cases}
    \end{split}
    \end{align}

We now define $k$ transition probability matrices of dimension $k+1$. For $y,y'\in\{0,\dots,k\}$, let $P^{(j)}_{y,y'}$ denote the probability of transitioning to state $y'$ from state $y$ in the $j$-th matrix. We further specify that $P^{(j)}_{y,y'}=0$ if $y'<y$. For all $y'\in\{y,\dots,k-1\}$, let
\begin{align}
   P^{(j)}_{y,y'}
   &= \sum_{i=0}^{y'-y}
   \frac{h}{\lambda} \left(\frac{a}{\lambda} \right)^i
   P_{W|L}( y'-y-i | j-1-y-i) \label{eq:P^{j}_1}
\end{align}
where $P_{W|L}$ is given by~\eqref{eq:PW}. Moreover, for every $y$,
\begin{align}
\label{eq:P^{j}_2}
    P^{(j)}_{y,k}
    = 1 - \sum_{i=y}^{k-1} P^{(j)}_{y,i} .
\end{align}
We have thus completed the description of transition matrices $P^{(j)}$, $j=1,\dots,k$, whose entry on the $(y+1)$-st row and $(y'+1)$-st column is equal to $P^{(j)}_{y,y'}$.

\begin{theorem}
\label{th:pow}
    If 
    \begin{align}
    \label{eq:a>}
      \frac{1}{a}>\frac{1}{h}+\Delta  ,
    \end{align}
    then the probability of safety violation at height 1 is equal to
    \begin{align}
    \label{eq:exact_bitcoin}
         1 - 
       \sum_{i=0}^{k-1} \, \sum_{j=1}^{k-i} e(i) \left[ P^{(1)} \times P^{(2)} \times \dots \times P^{(k)} \right]_{1,j} ,
    \end{align}
    where $e(\cdot)$ is given by~\eqref{eq:e(i)}
    and 
    $P^{(1)},\dots,P^{(k)}$ denote some $(k+1)\times(k+1)$
    transition probability matrices defined in~\eqref{eq:P^{j}_1} and~\eqref{eq:P^{j}_2}, and $[\cdot]_{1,j}$ takes the entry of a matrix on the first row and the $j$-th column.
\end{theorem}

Theorem~\ref{th:pow} is proved in Appendix~\ref{a:trade-off}.

\begin{figure}[t]
    \centering
    \includegraphics[width = \columnwidth ]{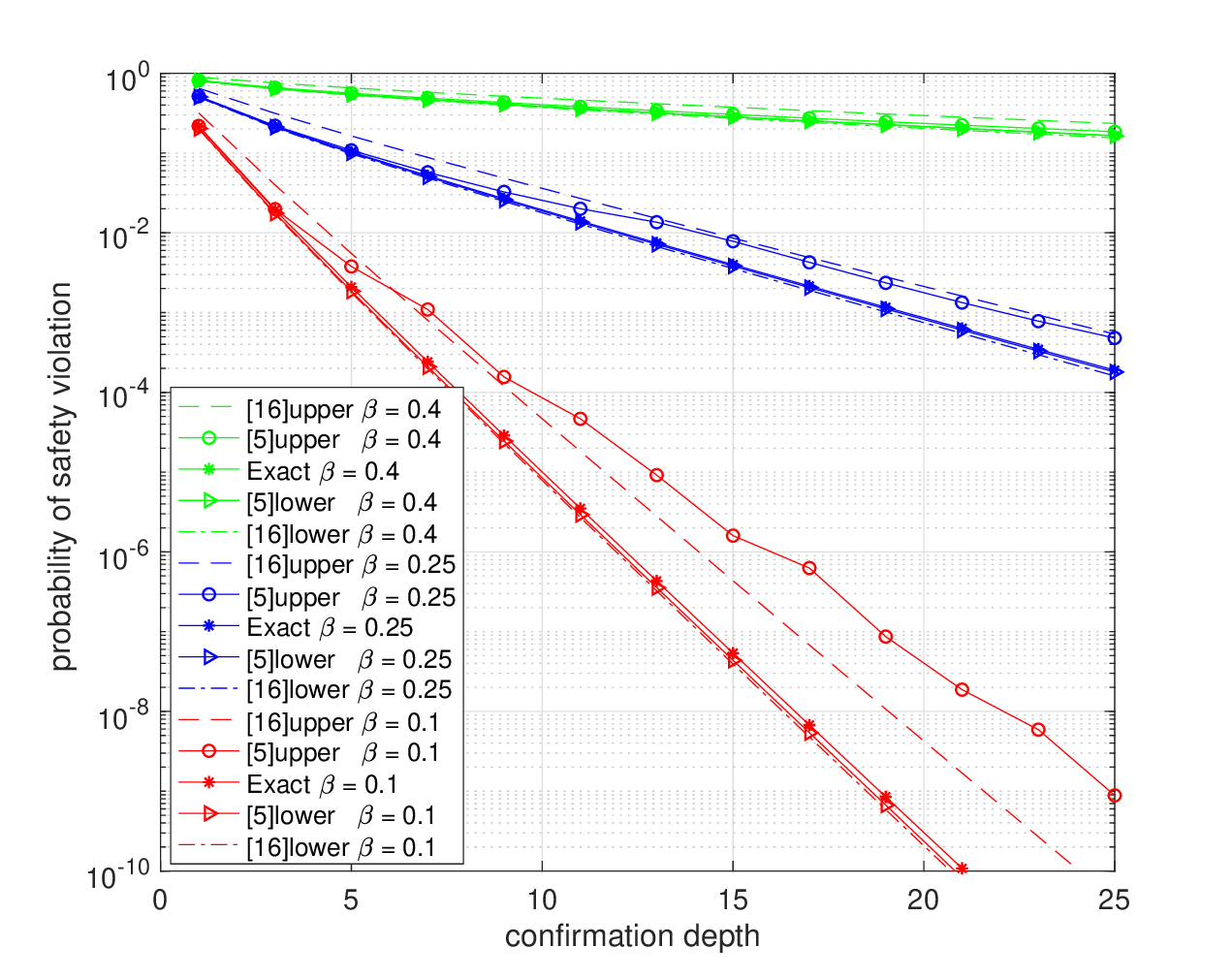}
    \caption{The total mining rate is $1/600$ blocks per second. The propagation delay upper bound is $10$ seconds.}
    \label{fig:BTC}
\end{figure}

\begin{figure}[t]
    \centering
    \includegraphics[width = \columnwidth ]{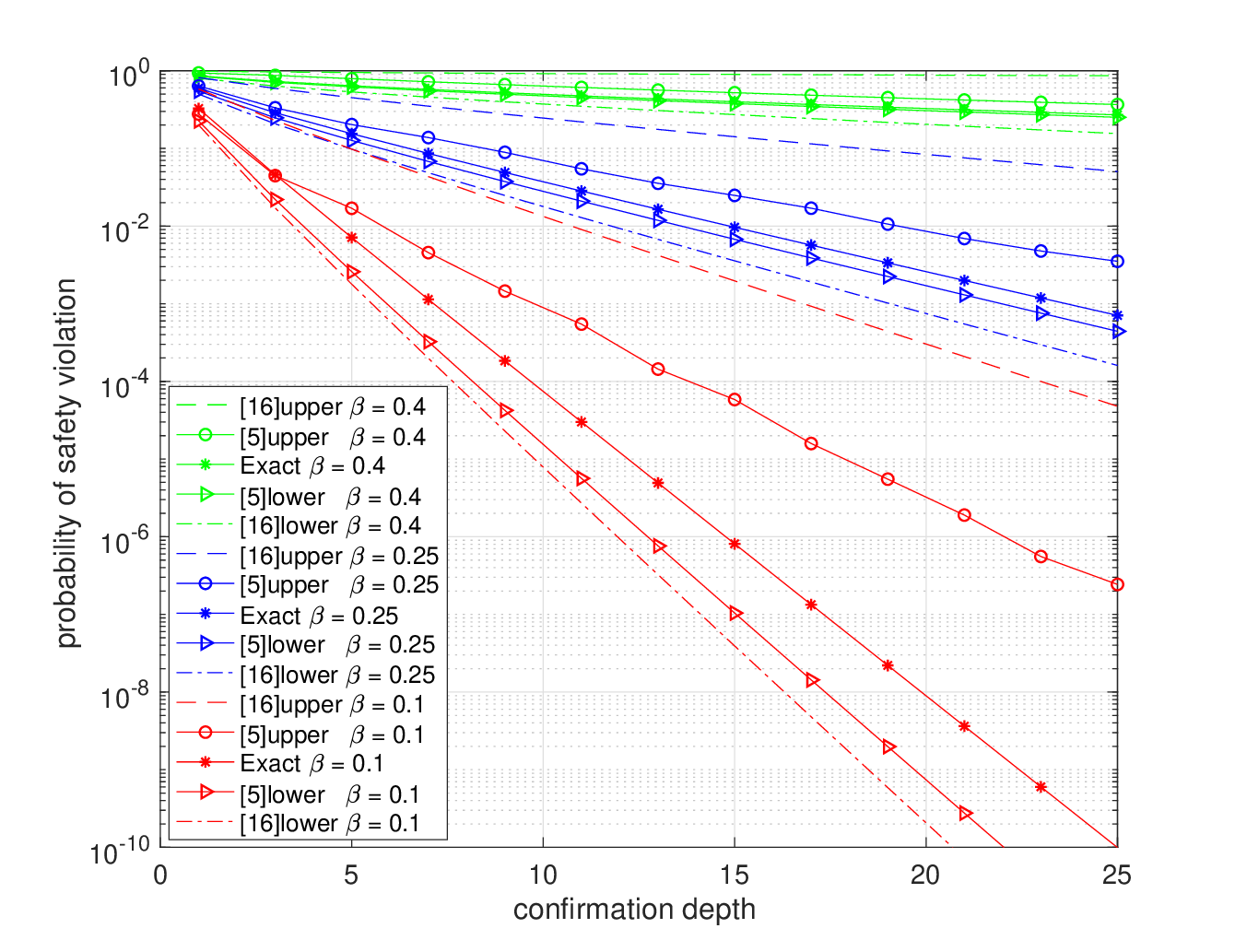}
    \caption{The total mining rate is $1/13$ blocks per second. The propagation delay upper bound is $2$ seconds.}
    \label{fig:ETC}
\end{figure}

Figs.~\ref{fig:BTC} and~\ref{fig:ETC} plot the security-latency trade-off for Bitcoin ($\lambda=1/600$ blocks/second, $\Delta=10$ seconds) and Ethereum Classic (PoW, $\lambda=1/13$ blocks/second, $\Delta=2$ seconds), respectively. Computing the exact probabilities entails multiplying \(k\) matrices of size \((k+1)\times(k+1)\), yielding a time complexity of \(\mathcal{O}(k^4)\).
In Fig.~\ref{fig:BTC}, we plot the trade-off for Bitcoin given by~\eqref{eq:exact_bitcoin} under three different adversarial fractions, with $\beta=0.4$, $0.25$, and $0.1$, respectively. Similarly, in Fig.~\ref{fig:ETC}, we plot the trade-off of Ethereum Classic (PoW) given by~\eqref{eq:exact_bitcoin} under three different adversarial fractions. We also include upper and lower bounds based on results in~\cite{guo2022bitcoin} and~\cite{cao2025security} under the same parameters for reference. 

\section{An MDP Framework}
\label{sec:MDP model}

\subsection{State and Transition}
\label{sec:MDP-I}

Within the constraints of mining power and propagation delay, the adversary is allowed to {\em continuously} reallocate mining effort and arbitrarily delay block propagation. However, once we grant the enhanced adversary the privilege of placing each block at the moment of its arrival (cf.~Sec.~\ref{s:enhanced}), the action simplifies: it suffices for the adversary to act only at discrete times---specifically, upon block arrivals. This naturally formulates the adversary's task as a discrete-time control problem: Given the full history of the system up to the current arrival, where should the adversary place \dg{each} new block to optimize its objective? Each adversarial action taken upon a block’s arrival transitions the system to a new state, which will be further updated at the next arrival. To formalize this setting, we introduce an MDP that captures the adversary’s decision-making problem.

The central challenge lies in crafting a state representation that is expressive enough to capture all relevant information, yet compact enough to render the MDP tractable. The most complete description of system state would include the full local views of all nodes. However, as discussed in Sec.~\ref{sec:sufficiency-model}, when analyzing safety violations, it suffices to consider only the public view---the weakest view shared by all honest nodes. Accordingly, we \dg{first consider defining} the system state to consist of the following components: 1) the full blocktree, including the A/H classification of each block, 2) the tree's public prefix, and 3) the arrival times of all non-public blocks.

We introduce a timer associated with each block to track, at any moment in time, the remaining duration before the block must become public.

\begin{Def}[block timer]
    \label{def:block timer}
    For every $b\in\mathbb{N}$, let $\tau_b$ denote block $b$'s timer. At time 0, all block timers, $\tau_1,\tau_2,\dots$, are initialized to $+\infty$. When an H-block $b$ arrives, its timer is immediately set to $\Delta$. In addition, for every block contained in chain $b$, if its timer exceeds $\Delta$ (it must be an A-block), it is also set to $\Delta$ immediately. Each timer decreases continuously over time. When block $b$'s timer expires (i.e., reaches zero), chain $b$ becomes public.
\end{Def}

By convention, a timer remains at $+\infty$ unless it is explicitly set to a finite value at some later point. Once a timer expires, its timer remains zero permanently. Each timer represents a time-to-expiration value in $[0,\infty]$. The adversary's ability to control block propagation is functionally equivalent to reducing block timers at will. However, the propagation delay constraint ensures that the adversary cannot delay the expiration of any timer beyond its current value.

In lieu of recording all individual arrival times, it now suffices to maintain a blocktree, where each block is equipped with a timer. Blocks whose timers have expired collectively form the public prefix tree. To fully define the state, we include an additional element indicating any new block arrival. Specifically, a state $s$ is represented as $(\mathcal{T},I)$, where $\mathcal{T}$ is a blocktree with an associated timer for each block, and $I$ indicates whether a new arrival is present---along with its A/H attribute, if applicable.

Upon a new arrival, the state is $(\mathcal{T},I)$, where $I\in\{A,H\}$. The adversary's action consists of choosing a parent for the new block arrival. This action transitions the MDP to a new state $(\mathcal{T}',\emptyset)$, where the blocktree $\mathcal{T}'$ incorporates the newly added blocks and updates certain timers (per \dg{Definition}~\ref{def:block timer} in the case of an H-arrival). The $\emptyset$  \dg{in $(\mathcal{T}',\emptyset)$} indicates  that this state immediately follows an action and \dg{awaits} the next arrival. Upon the subsequent arrival, the system transitions to a new state $(\mathcal{T}'',I'')$, and so on.

\begin{table}
    \caption{An example of the trajectory of an MDP. We assume the inter-arrival time $t_2-t_1<\Delta$. The action $Y_i$ points to the parent block number for the new arrival.}
    \begin{tabular}{|l|l|l|}
    \hline
    time & \qquad\qquad\qquad state & action \\
    \hline
           0 &
           $X_1=$ \Big(
            \begin{tikzpicture}
            \node[] (0) [blksmall] {0};
            \node[above right=-2mm of 0] {0};
        \end{tikzpicture},
        $\emptyset$ \Big)& $Y_1=\emptyset$ \\ 
         \hline
         $t_1$ &
         $X_2=$ \Big(
        \begin{tikzpicture}
            \node[] (0) [blksmall] {0};
            \node[above right=-1.5mm of 0] {0};
        \end{tikzpicture},
        $H$ \Big)& $Y_2=0$ \\ 
         \hline
         $t_1$ &
         $X_3=$ \Big(
        \begin{tikzpicture}
            \node[] (0) [blksmall] {0};
            \node[right = 5mm of 0](1)[blksmall] {1};
            \draw (0)--(1);
            \node[above right=-1.5mm of 0] {0};
            \node[above right=-1.5mm of 1] {$\Delta$};
        \end{tikzpicture},
        $\emptyset$ \Big)& $Y_3=\emptyset$ \\ 
         \hline
         $t_2$ &
         $X_4=$ \Big(
        \begin{tikzpicture}
            \node[] (0) [blksmall] {0};
            \node[right = 5mm of 0](1)[blksmall] {1};
            \draw (0)--(1);
            \node[above right=-1.5mm of 0] {0};
            \node[above right=-1.5mm of 1] {$\Delta-(t_2-t_1)$};
        \end{tikzpicture},
        $A$ \Big)& $Y_4=0$ \\ 
         \hline
         $t_2$ &
         $X_5=$ \Big(
        \begin{tikzpicture}
            \node[] (0) [blksmall] {0};
            \node[right = 5mm of 0](1)[blksmall] {1};
            \node[below right = 0.5mm of 1](2)[blksmall] {2};
            \draw (0)--(1);
            \draw (0)--(2);
            \node[above right=-1.5mm of 0] {0};
            \node[above right=-1.5mm of 1] {$\Delta-(t_2-t_1)$};
            \node[above right=-1mm of 2] {$\infty$};
        \end{tikzpicture},
        $\emptyset$ \Big)& $Y_5=\emptyset$ \\ 
         \hline
         $t_3$ &
         $X_6=$ \Big(
        \begin{tikzpicture}
            \node[] (0) [blksmall] {0};
            \node[right = 5mm of 0](1)[blksmall] {1};
            \node[below right = 0.5mm of 1](2)[blksmall] {2};
            \draw (0)--(1);
            \draw (0)--(2);
            \node[above right=-1.5mm of 0] {0};
            \node[above right=-1.5mm of 1] {$0$};
            \node[above right=-1mm of 2] {$\infty$};
        \end{tikzpicture},
        $H$ \Big)& $Y_6=1$ \\ 
         \hline
         $t_3$ &
                  $X_7=$ \Big(
        \begin{tikzpicture}
            \node[] (0) [blksmall] {0};
            \node[right = 5mm of 0](1)[blksmall] {1};
            \node[below right = 0.5mm of 1](2)[blksmall] {2};
            \node[right = 12mm of 1][blksmall] {3};
            \draw (0)--(1);
            \draw (0)--(2);
            \draw (1)--(3);
            \node[above right=-1.5mm of 0] {0};
            \node[above right=-1.5mm of 1] {$0$};
            \node[above right=-1.3mm of 3] {$\Delta$};
            \node[above right=-1mm of 2] {$\infty$};
        \end{tikzpicture},
        $\emptyset$ \Big)& $Y_7=\emptyset$ \\ 
         \hline
    \end{tabular}
    \label{tb:MDP_demo}
\end{table}

We illustrate this MDP formulation with a simple example depicted in Table~\ref{tb:MDP_demo}. Let the initial state $X_1$ consist solely of the genesis block with an expired timer. Upon the arrival of block 1 at $t_1$ with $I=H$ (i.e., an H-block), the system transitions to state $X_2$. The adversary takes action $Y_2=0$ to attach block 1 as a child of the genesis block, resulting in state $X_3$, where block 1's timer is set to $\Delta$. Next, an A-block (2) arrives at $t_2<t_1+\Delta$, leading to state $X_4$, in which the timer for block 1 decreases to $\Delta-(t_2-t_1)$. The adversary then takes action $Y_4=0$ to place block 2 as a child of the genesis block, initializing its timer to $\infty$, yielding state $X_5$. Subsequently, an H-block (3) arrives at $t_3>t_1+\Delta$, after block 1's timer expires, to cause a transition to state $X_6$. Block 3 must be attached to block 1 by action $Y_6=1$, yielding state $X_7$. In this example, even-numbered states represent new arrivals awaiting placement, while odd-numbered states reflect the state immediately after the corresponding placement, awaiting the next arrival.

\subsection{MDP, Reward, and Markov Policy}
\label{s:reward}

This subsection defines state transition probabilities and a reward function for safety violations, and outlines the mathematical principles showing that a Markov \dg{policy} maximizes the probability of such violations. It also develops several lemmas which serve as the foundation for subsequent analysis.

We consider an infinite-horizon MDP as follows. Let $\mathcal{S}$ denote the set of all possible states. At each decision epoch, the system is in a state $s \in \mathcal{S}$, and the adversary selects an action $\alpha$ from the admissible set $\mathcal{A}_s$. The action yields a real-valued reward $r(s,\alpha)$ and induces a transition to the next state according to a conditional probability distribution $p(\cdot|s,\alpha)$. 

A decision rule $d_i$ prescribes how actions are chosen at decision epoch $i$. In general, a randomized, history-dependent rule defines a distribution over actions $Y_i$ based on the full history $H_i = (X_1,Y_1,\dots,X_{i-1},Y_{i-1},X_i)$, where $X_i$ denotes the state at epoch $i$. A policy $\pi = (d_1,d_2,\dots)$ is a sequence of decision rules that induces a reward process $(X_i,r(X_i,Y_i))$, for $i= 1,2,\dots$. A Markov definite decision rule is a special case where the action is deterministic and depends only on the {\em current} state, i.e., \dg{$Y_i = d(X_i)$}. Such policies are particularly important, as they can be shown to achieve optimality in our problem.

Let $\mathscr{A}$ denote the set of states whose blocktrees contain more than one credible branch that is no lower than $k$. By Definition~\ref{def:v}, each such state violates the safety at height 1. Moreover, Lemma~\ref{lm:SV->absorb} implies that any safety violation must pass through some state in $\mathscr{A}$. Therefore, the probability of reaching a state in $\mathscr{A}$ is exactly the probability of violating safety at height 1.

Since we are only interested in whether a safety violation occurs---not what happens afterward---we may, without loss of generality, treat every state in $\mathscr{A}$ as absorbing. That is, once the system reaches a state $s\in\mathscr{A}$, it remains there indefinitely:
\begin{align}\label{eq:absorb}
   p(s|s,\alpha) = 1, \forall  s \in \mathscr{A}, \, \alpha\in \mathcal{A}_s.
\end{align}
Finding an optimal attack is henceforth equivalent to finding a policy that maximizes the probability of reaching $\mathscr{A}$.

Our MDP exhibits a particular structure: in a state $X_i=(\mathcal{T},I)$ with a new arrival (i.e., $I\in\{A,H\}$), the action $Y_i$---which specifies the placement of the block---{\em deterministically} determines the next state: $p(X_{i+1}|X_i,Y_i) = 1$. In  this resulting state $X_{i+1}=(\mathcal{T}',\emptyset)$, no action is taken (so $Y_{i+1}=\emptyset$), and the transition to the next state depends randomly on the time and type of the next arrival (cf.~Table.~\ref{tb:MDP_demo}).

A safety violation occurs \dg{only when some blocks is placed onto} the blocktree, not upon a new arrival that is yet to be placed. With this in mind, we assign a unit reward the first time the system transitions into a state in $\mathscr{A}$:
\begin{align} 
\label{eq:reward}
     r(x,y) =\begin{cases}
         1, \quad &\text{if } 
         y \neq \emptyset,\, x \not\in \mathscr{A},\,
         \text{ and } \\
         &  \quad      p(x'|x,y)=1 \text{ for some } x'\in\mathscr{A} , \\ 
         0, \quad &\text{otherwise}.
     \end{cases}  
\end{align}
Once the system reaches a state in $\mathscr{A}$, it remains there (by~\eqref{eq:absorb}) and collects no further reward under~\eqref{eq:reward}. Thus, for any trajectory, the total reward equals 1 if the trajectory ever enters $\mathscr{A}$, and 0 otherwise.

Let $v^{\pi}(s)$ denote the expected total reward under policy $\pi$ starting from state $X_1=s$:
\begin{align}
   v^{\pi}(s)
   = E_s^{\pi}\left[\sum_{i=1}^{\infty}r(X_i,Y_i)\right].
\end{align}
We define the probability of reaching $\mathscr{A}$---i.e., the probability of a safety violation---as:
\begin{align} \label{eq:def_vbar}
    \bar{v}^{\pi}(s) = v^{\pi}(s)+1_{s \in \mathscr{A}}
\end{align}
where the indicator $1_{s \in \mathscr{A}}$ ensures that $\bar{v}^{\pi}(s) = 1$ if the initial state $s$ already lies in $\mathscr{A}$. Due to the reward structure in \eqref{eq:reward} and the absorbing property of $\mathscr{A}$, we have $\bar{v}^{\pi}(s) = v^{\pi}(s)$ if $s \notin \mathscr{A}$ and $\bar{v}^{\pi}(s)=1$ if $s \in \mathscr{A}$. 

In the remainder of this subsection, we apply classical results from MDP theory to establish that a Markov policy maximizes the probability of safety violation. To support this conclusion, we verify a set of technical conditions, all of which are naturally satisfied by our blockchain problem.

We define the maximum total reward over all policies as:
\begin{align} \label{eq:v*}
    v^*(s) = \sup_{\pi}v^{\pi}(s).
\end{align}
According to~\cite[Theorem 7.1.3]{puterman1994markov}, the function $v^*$ must satisfy the Bellman optimality equation:
\begin{align} \label{eq:Bellman}
    v(s) = \max_{\alpha \in \mathcal{A}_s}
    \left\{
     r(s,\alpha)+\int_{\mathcal{S}}v(u)p(du|s,\alpha)
    \right\},
\end{align}
where $v(\cdot)$ is Lebesgue-Stieljes integrable with respect to $p(du|s,\alpha)$ for every $s$ and $\alpha\in\mathcal{A}_s$.

\begin{Def}[Eq.~7.1.9 in \cite{puterman1994markov}]
    A Markov definite decision rule $d$ is conserving if for all $s \in \mathcal{S}$,
    \begin{align} \label{eq:conserving}
        r(s,d(s))+\int_{\mathcal{S}}v^*(u)p(du|s,d(s))
        =
        v^*(s).
    \end{align}
\end{Def}

By definition, a conserving Markov definite decision rule selects an action in each state that satisfies the Bellman equation and achieves the optimal value $v^*$.

Within the ultimate fault tolerance~\eqref{eq:a>}, the probability of violating height-1 safety under the $k$-confirmation commitment rule vanishes as $k\to\infty$. Since H-block heights necessarily increase without bound over time, it follows that the probability of a first safety violation occurring after $n$ arrivals also vanishes as $n\to\infty$ under condition~\eqref{eq:a>}. Consequently, a stationary policy $\pi = (d,d,\dots)$ formed by repeating a conserving decision rule $d$ is an {\em equalizing policy} in the sense of~\cite[Eq.~7.0.950]{puterman1994markov}, i.e.,
\begin{align}
    \limsup_{n \to \infty} E_s^{\pi} \left[v^*(X_{n+1})\right]\leq 0 .
\end{align}

With the preceding technical conditions verified, we invoke~\cite[Theorem 7.1.7 b]{puterman1994markov} to conclude the following: 

\begin{lemma}
With $d$ being a conserving Markov definite decision rule, the equalizing stationary attack $\pi^* = (d,d,\dots)$ is optimal, i.e., $v^{\pi^*}=v^*$.
\end{lemma}

It is straightforward to verify that this Markov policy $\pi^*$ also maximizes the probability of violation, $\bar{v}^\pi(s)$: 
\begin{itemize}
    \item If $s \in \mathscr{A}$, then by definition $\bar{v}^\pi(s) = 1$ for all policies $\pi$.
    \item If $s \notin \mathscr{A}$, then $\bar{v}^{\pi^{*}}(s) = v^{\pi^{*}}(s)$ by~\eqref{eq:def_vbar}.
\end{itemize}
In both cases, $\pi^*$ maximizes $\bar{v}^\pi(s)$.
For convenience, we define the maximum probability of violation as a value function $V(s)$:
\begin{align} \label{eq:Vs}
    V(s) = \bar{v}^{\pi^{*}}(s) . 
\end{align}
That is, $V(s)$ represents the probability of eventually reaching $\mathscr{A}$ from state $s$ under the optimal policy $\pi^*$.

\subsection{Lemmas for Comparing States}

The remaining challenge is to determine an optimal action to take upon each arrival. A Markov definite decision rule uniquely determines the next state, and an optimal action is one that transitions the system to a state with the highest value. For $s \notin \mathscr{A}$, the right-hand side of \eqref{eq:Bellman} is equivalent to $\max_{\alpha \in \mathcal{A}_s}\{ \dg{r(s,\alpha)+} V(u)\}$, where $u$ denotes the resulting next state from action $\alpha$. Therefore, \dg{if no action transitions the state $s$ to violation in one step (i.e., $r(s,\alpha)=0$ for all $\alpha\in\mathcal{A}_s$),} finding an optimal \dg{action} boils down to comparing the values of all feasible next states. 

We say state $s$ is {\em strictly better} than state $s'$ if the probability of violation starting from state $s$ is higher, henceforth denoted as
\begin{align}
    s\succ s' 
    \quad \Leftrightarrow \quad
    V(s) > V(s').
\end{align}
The related comparisons $\succeq$, $\prec$, $\preceq$ and $\asymp$ are defined analogously. If states $S$ and $S'$ are random, by $S \asymp S'$, we mean $E[V(S)] = E[V(S')]$, with the other relations extended in the natural way to expectations. To describe the evolution of the system over time, we use $V(X_n)$ to denote the value of the state at epoch $n$. 
In Appendix~\ref{a:lmlimV}, we establish the following:

\begin{lemma}
\label{lm:lim V(X_n)}
    $\lim_{n \to \infty} V(X_n)$ exists with probability 1.
\end{lemma}

In certain situations, comparing the values of two states becomes easier by examining the outcomes of a sequence of actions up to a stopping time (a random time that is determined based on information available up to that point). This approach may be reminiscent of how a chess player evaluates a line of play. Let $T$ denote a stopping time taking values in the extended nonnegative integers (which includes $+\infty$). We define
\begin{align}
    Z_T = \begin{cases}
        V(X_T) & \text{if } T<\infty , \\
        \lim_{n \to \infty}V(X_n) \quad &\text{if } T=\infty .
    \end{cases}
\end{align}

Starting from state $s$ at epoch~1, suppose an optimal policy 
$\pi^*$ is followed, resulting in a state $X_T$ at stopping time $T$, then the expectation of $Z_T$ is equal to $V(s)$. Conversely, if another policy $\pi$ is followed (not necessarily Markovian), the expected value of the corresponding $Z_T$ cannot exceed $V(s)$. Formally, we have the following result
(proved in Appendix~\ref{a:lemma7}).

\begin{lemma}\label{lm:Vbar=EVbar}
    For the stationary conserving policy $\pi^* = (d,d,\dots)$, any stopping time $T$, and any policy $\pi$,
    \begin{align}
    V(s)
    &=  E_s^{\pi^*} \left[Z_T \right] \label{eq:V=EVXT} \\
    &\geq E_s^{\pi} \left[Z_T \right]. \label{eq:Vbar*>EVbar}
    \end{align}
\end{lemma}

Let us denote the realization of the adversarial and honest arrival processes as $q$. 
Starting from state $s$ at epoch $t$, if the adversary takes actions according to a deterministic policy $\pi$ until time $T$ in response to the arrival processes $q$, then the state at $T$ is represented as $x_T^\pi(s,q)$. 
Evidently, this state depends on the arrival processes only up until $T$.

\begin{lemma} \label{lm:compare}
    Consider two systems $C$ and $C'$ which start at time 0 from states $s$ and $s'$, respectively, and both systems have the same arrival processes over time. Let $\pi^*$ denote an optimal attack for $C'$. Let $\overline{T}$ denote the first time $C'$ arrives in $\mathscr{A}$. If there exists an attack $\pi$ for $C$ and a stopping time $T \leq \overline{T}$, such that 
    $
        x_T^\pi \left(s,q\right)
        \succeq
        x_T^{\pi^*} \left(s',q\right)
    $
    for every realization $q$ of the arrival processes as long as $T<\infty$, then $s \succeq s'$. As a special case, we have $s\succeq s'$ if, depending on the arrival processes, either $C$ arrives in $\mathscr{A}$ no later than $C'$ or $C$ arrives at a state with no lower value than the state of $C'$ at some stopping time $T$.
\end{lemma}

Proved in Appendix~ \ref{a:lemma7} and \ref{a:lemmacompare}, 
Lemmas~\ref{lm:Vbar=EVbar} and~\ref{lm:compare} are applied repeatedly in this paper for state comparisons in order to find an optimal attack $\pi^*$.

\subsection{Simplification of State Representation}
\label{s:pru}

As far as a $\pi^*$ for maximizing the probability of violation is concerned, the state representation can be dramatically reduced as shown next using lemmas from the previous subsection. First, we provide an intuitive state comparison rule:

\begin{lemma} 
    \label{lm:h/l rule'}
    If states $s$ and $s'$ have identical blocktrees except that, for every block, \dg{its} timer in $s$ is greater than or equal to \dg{its} timer in $s'$, then $s \succeq s'$.
\end{lemma}
\begin{figure}
    \centering
 \begin{tikzpicture}
            \node[] (0) [blksmall] {0};
            \node[right = 5mm of 0](1)[blksmall] {1};
            \node[below right = 3mm of 1](2)[blksmall] {2};
            \node[right = 10mm of 1][blksmall](3) {3};
            \node[below right = 1mm of 3](4)[blksmall] {4};
            \draw (0)--(1);
            \draw (0)--(2);
            \draw (1)--(3);
            \draw (1)--(4);
            \node[above right=-1.5mm of 0] {0};
            \node[above right=-1.5mm of 1] {$0$};
            \node[above right=-1.3mm of 3] {$\Delta/4$};
            \node[above right=-2mm of 2] {$\infty$};
            \node[above right=-2mm of 4] {$\Delta$};
            \node[above = 2mm of 0]{state $s$};
        \end{tikzpicture}
        \begin{tikzpicture}
            \node[] (0) [blksmall] {0};
            \node[right = 5mm of 0](1)[blksmall] {1};
            \node[below right = 3mm of 1](2)[blksmall] {2};
            \node[right = 10mm of 1][blksmall] {3};
            \node[below right = 1mm of 3](4)[blksmall] {4};
            \draw (0)--(1);
            \draw (0)--(2);
            \draw (1)--(3);
            \draw (1)--(4);
            \node[above right=-1.5mm of 0] {0};
            \node[above right=-1.5mm of 1] {$0$};
            \node[above right=-1.3mm of 3] {$\Delta/4$};
            \node[above right=-2mm of 2] {$0$};
            \node[above right=-2mm of 4] {$\Delta/2$};
            \node[above = 2mm of 0]{state $s'$};
        \end{tikzpicture}   
    \caption{An example for Lemma~\ref{lm:h/l rule'}, where $s\succeq s'$.}
    \label{f:h/l rule'}
\end{figure}
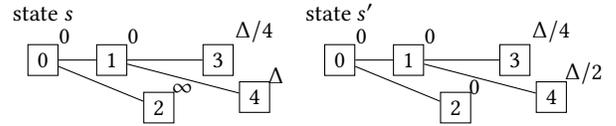

\dg{Lemma~\ref{lm:h/l rule'} is proved} in Appendix~\ref{a:lm_h/l}. \dg{An example illustrating the lemma is provided in Fig.~\ref{f:h/l rule'}.}

To simplify the state $s = (\mathcal{T}, I)$, we partition the entire blocktree $\mathcal{T}$ into distinct branches according to Definition~\ref{def:branch}. The number of branches is equal to the number of height-1 blocks in the blocktree. 
\dg{It can be shown that only} the two highest of these branches need to be recorded in the state representation and that it suffices to store only the minimum block timer on each height. We demonstrate this in the following lemma, which is proved in Appendix~\ref{a:two_branches}.

\begin{lemma}
\label{lm:two_branches}
    We have $s = (\mathcal{T}, I) \asymp s'=(\mathcal{T}',I)$ if \;$\mathcal{T}'$ retains only two highest branches of \;$\mathcal{T}$, and all timers of \;$\mathcal{T}'$ at every height are set to the minimum timer of all blocks on that height in \,$\mathcal{T}$.
\end{lemma}

With the observation above, the state can be concisely represented as $[m,n,(l_0,l_1,l_2,\dots,l_d),I]$, where $n$ denotes the height of a highest branch, $m$ denotes the height of a second highest branch, $l_i$ represents the minimum of the timers of the blocks in $s$ with height $i$, referred to as the timer of height $i$, $d$ denotes the highest height whose timer is no greater than $\Delta$, and $I$ is the same as in $s=(\mathcal{T},I)$. We omit all timers higher than height $d$ as they are all $\infty$. By Definition~\ref{def:block timer}, the timer sequence is non-decreasing: $l_0 \leq l_1 \leq \dots \leq l_d$.

From this point forward, we refer to the two branches as {\em the higher branch} and {\em the lower branch}, corresponding to heights $n$ and $m$, respectively. In the special case where the two branches have equal heights ($m=n$), the state representation does not distinguish between them. In such cases, the higher branch refers to an arbitrary one of those branches, while the lower branch refers to another one of them.

For convenience, let $m\land d$ denote the smaller number between $m$ and $d$. To continue simplifying state representation, we reason that timers of height lower than $m\land d$ can be regarded as $0$ without loss of generality:

\begin{lemma}
\label{lm:l_{mland d}-1 = 0}
    $[m,n,(l_1,\dots,l_d),I] \asymp [m,n,(0,\dots,0,l_{m\land d},\dots,l_d),I]$.
\end{lemma}

The proof is relegate to Appendix~\ref{a:lemma l_{mland d}-1 = 0}.

\begin{table*} 
\centering
\caption{The state transition rule starting from state $s=[m,d,n,(l_{m\land d},\dots,l_d),I]\notin\mathscr{A}$. (If $s\in\mathscr{A}$, the state remains the same regardless of arrivals and actions.)}
\begin{tabular}{|c|l|l|l|}
    \hline
    \textbf{I} & \textbf{Action} & \textbf{Condition} & \textbf{New state} \\
        \hline
    \multirow{6}{*}{$\emptyset$}&
    \multirow{6}{*}{$\emptyset$}& Not applicable. &$[m,d,n,(l_{m\land d}',\dots,l_d'),H]$  
    if an H-arrival (at $t$),
    \\
    && &$[m,d,n,(l_{m\land d}',\dots,l_d'),A]$ if an A-arrival (at $t$),
    \\
     &&& where $l_j' = \max\{0,l_j - t\}$ for $j = m\land d, \dots, d$.\\
    &&& The time of arrival $t$ is exponential with rate $\lambda$. \\
    &&& The probability of an A-arrival is $a/\lambda$;\\
    &&& the probability of an H-arrival is $h/\lambda$.\\
    \hline
    \multirow{5}{*}{A }&&  $i<n+1$  &  $[m,d,n,(l_{m\land d},\dots,l_d)]$ \\
    &  $\overline{i}$ &  $i=n+1$  &  $[m,d,n+1,(l_{m\land d},\dots,l_d)]$ \\
    \cline{2-4}
     && $i<m+1$
    & $[m,d,n,(l_{m\land d},\dots,l_d)]$\\
    &$\underline{i}$ & $i=m+1$, $m<n$ 
    &$[m+1,d,n,(l_{(m+1)\land d},\dots,l_d)]$\\
    && $i=m+1$, $m=n$
    &$[m,d,m+1,(l_{m\land d},\dots,l_d)]$\\
    \hline
     & &  $P+1\leq i \leq d$  &  $[m,d,n,(l_{m\land d},\dots,l_d)]$ \\
    && $m\leq d< i \leq n$ & $[m,i,n,(l_m,\dots,l_d,\Delta\dots ,l_i=\Delta)]$\\
    \multirow{6}{*}{H} 
    &$\overline{i}$ & $d< i \leq n$, $d<m$ & $[m,i,n,(l_{m \land i}=\Delta,\dots,l_i=\Delta)]$\\
    && $ i=n+1$, $m \leq d$& $[m,n+1,n+1,(l_{m},\dots,l_d,\Delta\dots,l_{n+1}=\Delta)]$ \\
   & & $ i=n+1$, $ d<m$& $[m,n+1,n+1,(l_{m}=\Delta,\dots,l_{n+1}=\Delta)]$ \\
    \cline{2-4}
   & & $P+1\leq i \leq \min\{d,m\}$& $[m,d,n,(l_{m\land d},\dots,l_d)]$\\
    && $d+1\leq i \leq m$ & $[m,i,n,(l_{i}=\Delta)]$ \\
    &$\underline{i}$& $P+1\leq i= m+1$, $m+1\leq d \leq n$& $[m+1, d,n,(l_{m+1}\dots l_{d})]$\\
    && $P+1\leq i= m+1$, $ d \leq m < n $& $[m+1, m+1,n,(l_{m+1}=\Delta) ]$\\
    && $P+1\leq i= m+1$, $d\leq m=n$& $[m,m+1,m+1,(l_{m}=\Delta, l_{m+1}=\Delta)]$\\
    \hline
\end{tabular}
\label{tb:I=H}
\end{table*}

We can thus simplify the state representation to:
\begin{align} \label{eq:mdnli}
    [m,d,n,(l_{m\land d},\dots,l_d),I]
\end{align}
where we omit timers from height $0$ to $(m\land d) -1$ by regarding them as $0$. Moreover, in the special case of $I=\emptyset$, we may omit $I$ to express the state as $[m,d,n,(l_{m\land d},\dots,l_d)]$.
Throughout the paper, by writing~\eqref{eq:mdnli}, we implicitly assume that it is a valid state, i.e., $I\in\{A,H,\emptyset\}$, $m\le n$, $d\le n$, $l_{d\land m} \le \dots \le l_d\leq \Delta$, and if $n>d$, also $l_{d+1}=\dots=l_n=\infty$.

\subsection{Public Height, Violation, and Transitions}

In state $s=[m,d,n,(l_{m\land d},\dots,l_d),I]$, the public height
\begin{align}
    P = \max\{ \eta\in \{(m\land d)-1
    , \dots, n\} \,|\, l_\eta=0 \}
\end{align}
where we use the fact that the timer \dg{at} height $(m\land d)-1$ is set to zero. By definition of $\mathscr{A}$ in Sec.~\ref{s:reward}, \dg{the state is in violation, i.e.,} $s\in\mathscr{A}$\dg{,} if and only if
\begin{align} \label{eq:in_A}
    m \geq \max\{k,P\}.
\end{align}

We introduce the following shorthand notations for actions:

\begin{Def}
    Let $\overline{i}$ denote the action of adding a newest block to a higher branch at height $i$;
    let $\underline{i}$ denote the action of adding the newest block to a lower branch at height $i$. 
\end{Def}

In any given state $[m, d, n, (l_{m \land d}, \dots, l_d), I]$, we explicitly describe the state transition rule in Table~\ref{tb:I=H} for all allowable actions, depending on the conditions of the state under which the actions are applied. For $I = \emptyset$, there is no block to place and the ``action'' is to await an arrival, so we let $\emptyset$ denote the action and $A_s=\{\emptyset\}$ in this case. The subsequent arrival transitions the state probabilistically as described in Table~\ref{tb:I=H}. For $I \in \{A,H\}$, the action deterministically transitions the system to the next state. For $I=H$, the H-block can be placed on either branch strictly higher than the public height, with the action set $  A_s=\{\overline{P+1}, \dots, \overline{n+1} \} \cup \{\underline{P+1},\dots,\underline{m+1}\} .$
For $I = A$, the A-block can be placed \dg{at} any height, so the action set $  A_s=\{\overline{1},\dots, \overline{n+1}\} \cup \{\underline{1}, \dots, \underline{m + 1}\} .$

It is not difficult to verify the transition rules in Table~\ref{tb:I=H}. We note that in the special case of $m=n$, the action $\underline{m+1}$ of putting a block on the lower branch swaps the lower and higher branches, as is seen in the last lines corresponding to $\underline{i}$ for both an A-arrival and an H-arrival.

\begin{figure}
    \centering

 \begin{tikzpicture}
            \node[] (0) [blksmall] {0};
            \node[right = 5mm of 0](1)[blksmall] {1};
            \node[below right = 1mm and 7mm of 0](2)[blksmall] {2};
            \node[right = 8mm of 1](3)[blksmall] {3};
            \node[right = 5mm of 3](4)[blksmall] {4};
            \draw (0)--(1);
            \draw (0)--(2);
            \draw (1)--(3);
            \draw (3)--(4);
            \node[above right=-1.5mm of 0] {0};
            \node[above right=-1.5mm of 1] {$0$};
            \node[above right=-1.3mm of 3] {$\Delta$};
            \node[above right=-2mm of 2] {$\infty$};
            \node[above right=-2mm of 4] {$\infty$};
              \draw[->] (1.5,-0.6) -- (1.5,-0.8);
        \end{tikzpicture}        

 \begin{tikzpicture}
            \node[] (0) [blksmall] {0};
            \node[right = 5mm of 0](1)[blksmall] {1};
            \node[below right = 1mm and 7mm of 0](2)[blksmall] {2};
            \node[right = 8mm of 1](3)[blksmall] {3};
            \node[right = 5mm of 3](4)[blksmall] {4};
            \node[right = 17mm of 2](5)[blksmall] {5};
            \draw (0)--(1);
            \draw (0)--(2);
            \draw (1)--(3);
            \draw (3)--(4);
            \draw (2)--(5);
            \node[above right=-1.5mm of 0] {0};
            \node[above right=-1.5mm of 1] {$0$};
            \node[above right=-1.3mm of 3] {$\frac23\Delta$};
            \node[above right=-2mm of 2] {$\infty$};
            \node[above right=-2mm of 4] {$\infty$};
            \node[above right=-2mm of 5] {$\infty$};
             \draw[->] (1.7,-0.6) -- (1.7,-0.8);  
        \end{tikzpicture}    

 \begin{tikzpicture}
            \node[] (0) [blksmall] {0};
            \node[right = 5mm of 0](1)[blksmall] {1};
            \node[below right = 1mm and 7mm of 0](2)[blksmall] {2};
            \node[right = 8mm of 1](3)[blksmall] {3};
            \node[right = 5mm of 3](4)[blksmall] {4};
            \node[right = 17mm of 2](5)[blksmall] {5};
            \node[ right = 15mm of 5](7)[blksmall] {6};
            \draw (0)--(1);
            \draw (0)--(2);
            \draw (1)--(3);
            \draw (3)--(4);
            \draw (2)--(5);
            \draw (5)--(7); 
            \node[above right=-1.5mm of 0] {0};
            \node[above right=-1.5mm of 1] {$0$};
            \node[above right=-1.3mm of 3] {$0$};
            \node[above right=-2mm of 2] {$\infty$};
            \node[above right=-2mm of 4] {$\infty$};
            \node[above right=-2mm of 5] {$\Delta$};
            \node[above right=-2mm of 7] {$\Delta$};
        \end{tikzpicture}
    \caption{Example of state transition.}
    \label{fig:Sec4.5}
\end{figure}
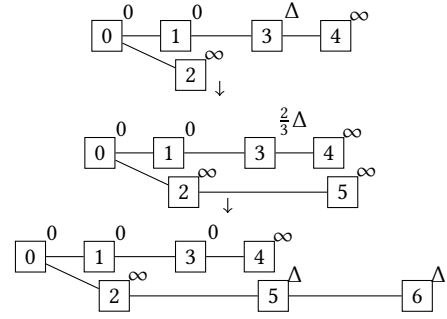

\dg{We illustrate the evolution of the blocktree with associated timers using an example in Fig.~\ref{fig:Sec4.5}. The corresponding arrivals, actions, and state transitions are}
as follows: Starting from state $[1,2,3,(0,\Delta)]$, an A-arrival occurs after time $\frac{1}{3}\Delta$, transitioning the state to $[1,2,3,(0,\frac{2}{3}\Delta),A]$. With $\underline{2}$ taken, the system transitions to state $[2,2,3,(\frac{2}{3}\Delta)]$. An H-arrival occurs $\dg{2} \Delta$ later, transitioning the state to $[2,2,3,(0),H]$. The adversary elects to take action $\underline{3}$, causing a transition to $[3,3,3,(\Delta)]$. We have $3\geq \max\{3,2\}$, thus state $[3,3,3,(\Delta)] \in \mathscr{A}$ is a violation if the depth $k\le 3$. 

\subsection{State Comparison Rules}

Here we collect some simple rules for comparing the values of different states. Intuitively, if two states, $s$ and $s'$, have identical branch heights, where $s$ has larger timers at  every height than $s'$, then the adversary would favor $s$ as $s\succeq s'$. 
Alternatively, if $s$ and $s'$ have identical timers at  all heights, and $s$ has higher branch heights than $s'$, then $s\succeq s'$. These observations are made precise in the following lemma:

\begin{lemma}\label{lm:h/l rule}
    If $d\ge d'$ and $l_i\le l'_i$ 
    for $i\in\{d\land m ,\dots,d'\}$,
    then
    \begin{align} \label{eq:l'>l}
        [m,d,n,(l_{d\land m},\dots,l_d),I]
        \preceq
        [m,d',n,(l'_{d'\land m},\dots,l'_{d'}),I ] .
    \end{align}
    Also, if $m\le m'$ and $n\le n'$, then
    \begin{align} \label{eq:m'n'>mn}
        [m,d,n,(l_{d\land m},\dg{\dots},l_d),I]
        \preceq
        [m',d,n',(l_{d\land m'},\dots,l_d),I] .
    \end{align}
\end{lemma}

\begin{proof}
    Equation~\eqref{eq:l'>l} is true by Lemma~\ref{lm:h/l rule'}. \dg{Moreover, having longer} lower and higher branches improve the probability of satisfying \eqref{eq:in_A}, which leads to \eqref{eq:m'n'>mn}.
\end{proof}

It is also intuitive that, for two states with the same representation except for the type of arrival, the state with $I = A$ is superior to the one with $I = \emptyset$ or $I = H$. This is because an A-arrival does not affect the timer and can increase the branch height. Formally, we have the followings:

\begin{lemma} \label{lm:A>=H}
    \begin{align}
        [m,d,n,(l_{d\land m},\dots,l_d),A]
        &\succeq
        [m,d,n,(l_{d\land m},\dots,l_d)]  \label{eq:A>=} \\
        [m,d,n,(l_{d\land m},\dots,l_d),A]
        &\succeq
        [m,d,n,(l_{d\land m},\dots,l_d),H] . \label{eq:>=H}
    \end{align}
\end{lemma}

\begin{proof}
    For \eqref{eq:A>=}, the left-hand side can take the action $\overline{d}$, transitioning into the right-hand side.
    For \eqref{eq:>=H}, the left-hand side can take the same action as the right-hand side, transitioning into the same state as the one the right-hand side becomes.
    Both equations hold due to~\eqref{eq:Vbar*>EVbar}.
\end{proof}

The following is a simple corollary of Lemma~\ref{lm:A>=H}.


\begin{lemma} \label{lm:A>=Hss'}
    In state $s=[m,d,n,(l_{d\land m},\dots,l_d)]$, 
    if, at the subsequent arrival time, there exists a placement of an H-block that transitions the system into state $s'$, 
    then $s\succeq s'$.
\end{lemma}

We can classify states based on the timer of the height just above the lower branch, $l_{m+1}$. Reaching $\mathscr{A}$ requires the lower branch to be no lower than the public height. The value of $l_{m+1}$ provides information about the race between the \dg{jumper's height (or public height) and the lower branch height.}

\begin{Def}[ahead, on-time, and behind states] \label{def:State}
Depending on $l_{m+1}$,
\begin{itemize}
    \item if $l_{m+1} = \infty$, it indicates that no H-block 
    is higher than the lower branch. In this case, the state is classified as \textbf{``ahead''}, signifying that the lower branch is at least as high as the highest jumper;
    \item if $l_{m+1} \in (0, \infty)$, it means there are nonpublic H-blocks above the lower branch, so that new H-blocks can still be used to extend the lower branch. States like this are classified as \textbf{``on-time''};
    \item if $l_{m+1} = 0$, the lower branch has fallen behind both the highest jumper and the public height. In this case, only A-blocks can increase the height of the lower branch, and the state is classified as \textbf{``behind''}.
\end{itemize}
\end{Def}

Our categorization of states bears resemblance to the ``cold'', ``critical'' and ``hot'' regions defined in~\cite{gazi2020tight} in the sense that both 
reflect the adversary's current advantage.
The state categorization of \cite{gazi2020tight} 
is based on the entire blocktree and all honest views, specifically the length difference between the adversary's best chain and the longest honest chain visible to all honest nodes. 
In contrast, our state categorization is based on a simplified, sufficient state representation, determined by the timer just above the second-highest branch, which allows a precise analysis of the MDP.

\begin{lemma} \label{lm:must take the form of}
    An ahead state must take the form of
    \begin{align}
        [m,d,n,(l_d),I]
    \end{align}
    with $d\le m$ 
    and $l_{m+1}=\infty$.
    An on-time state must take the form of
    \begin{align}
        [m,d,n,(l_m,\dots,l_d),I]
    \end{align}
    with $m<d$ and $l_{m+1}\in(0,\infty)$.
    A behind state must take the form of 
    \begin{align} \label{eq:00l}
        [m,d,n,(0,0,l_{m+2},\dots,l_d),I ]
    \end{align}
    with $m<d$ and $l_{m+1}=0$. 
\end{lemma}

\begin{proof}
    Under the $\Delta$-synchrony model, every finite timer is no larger than $\Delta$.
    By the definition of height $d$, $l_d \in [0,\Delta]$ and $l_{d+1} = \infty$.
    For a valid ahead state with $l_{m+1}=\infty$, we must have $d \leq m$, and the timer sequence $(l_{d\land m},\dots,l_d)$ reduces to a single element $(l_d)$.
    
    For a valid on-time state with $0<l_{m+1}\leq \infty$, we must have $m<d$ and the timer sequence begins from $l_m$ and ends with $l_d$.
    
    For a valid behind state with $l_{m+1}=0$, we must have $m+1<d$. Since $l_m\le l_{m+1}$, we also have $l_m=0$. It is straightforward to see that the state must take the form of~\eqref{eq:00l}. 
\end{proof}

\section{The Zero-Delay Special Case}
\label{s:zero}

This section is devoted to the special case of $\Delta = 0$, where all honest nodes are fully synchronized at all times. An important consequence is that all H-blocks are on different heights. 
Also, all timers are either zero or infinity, so the public height $P=d$ \dg{in a general state representation~\eqref{eq:mdnli}}.  
The sequence of timers \dg{in}~\eqref{eq:mdnli} always takes the form of a sequence of $\max\{1,d-m+1\}$ zeros. In this section, we omit this sequence of timers to use $[m,d,n,I]$ as a sufficient representation of the state, which is further reduced to $[m,d,n]$ when $I=\emptyset$.

\subsection{Preliminaries}

The following is a direct corollary of Lemma~\ref{lm:h/l rule}.

\begin{lemma} \label{lm:h/l0}
    If $d\ge d'$, then $ [m,d,n,I]
        \preceq
        [m,d',n,I ] .$
    Also, if $m\le m'$ and $n\le n'$, then $  [m,d,n,I]
        \preceq
        [m',d,n',I] .$
\end{lemma}


We use Lemma~\ref{lm:h/l0} to narrow down the search for an optimal action upon an H-arrival:

\begin{lemma}\label{lm:posHaction_0}
    In state $[m,d,n,H]$,
    either $\overline{d+1}$, or $\underline{m+1}$ (applicable when $d\le m$), or $\overline{n+1}$ is optimal.
\end{lemma}

\begin{proof} 
The state $[m,d,n,H]$ includes a new H-block arrival, which can only be added above the public height, $d$.  Specifically, it must be added to either the higher branch on a height between $d+1$ and $n+1$ or the lower branch on a height between $d+1$ and $m+1$ (this is applicable only if $d\le m$).
If the block is added to the higher branch \dg{at} some height $i\in\{d+2,\dots,n\}$, then the system transitions to state $[m,i,n]$.
If the block is added to the lower branch \dg{at} some height $j\in\{d+2,\dots,m\}$,
then the
system transitions to state $[m,j,n]$.
By Lemma~\ref{lm:h/l0}, we have 
\begin{align}
[m,i,n] &\preceq [m,d+1,n] \\
[m,j,n] &\preceq [m,d+1,n] .
\end{align}
This implies that neither action $\overline{i}$ nor $\underline{j}$ is better than $\overline{d+1}$ (with a transition to $[m,d+1,n]$) in those respective cases.
Hence the proof of the lemma.
\end{proof}


We next use Lemma~\ref{lm:h/l0} again to narrow down the search for an optimal action upon an A-arrival.

\begin{lemma}\label{lm:Aaction-0}
    In state $[m,d,n,A]$,
    either $\underline{m+1}$ (applicable when $m<n$) or $\overline{n+1}$ is 
    optimal. 
\end{lemma}

\begin{proof} 
The state $[m,d,n,A]$ includes a new A-block arrival, which must be added to either the higher branch \dg{at} any height up to $n+1$ or, in the case of $m<n$, the lower branch \dg{at} any height up to $m+1$.
Unless the block is added to the higher branch \dg{at} height $n+1$ (by taking action $\overline{n+1}$ and transitioning to $[m,d,n+1]$) or the lower branch \dg{at} height $m+1$ (by taking action $\underline{m+1}$ and transitioning to $[m+1,d,n]$),
the system transitions to state $[m,d,n]$.
In any case,
all timers remain unchanged.
By Lemma~\ref{lm:h/l0}, we have
\begin{align}
    [m,d,n] &\preceq [m,d,n+1] \\
    [m,d,n] &\preceq [m+1,d,n] .
\end{align}
This implies that either $\overline{n+1}$ or $\underline{m+1}$ is optimal.
\end{proof}

\subsection{Optimal Actions in the \dg{Zero-Delay} Case}
\label{ss:pow0}

When $\Delta=0$, the private mining attack has been shown to be optimal by block-counting arguments (e.g.,~\cite{guo2022bitcoin}).
Here we adopt the MDP framework to directly derive an optimal placement upon each block arrival that maximizes the probability of eventually violating the safety of height~1. While this alternative \dg{means of} proof is not easier, the new techniques introduced here will be further developed subsequently to address the general case of $\Delta\ge0$.

\begin{proposition}[optimal action upon an H-arrival] \label{prop:posdelay0H}
    In state $[m,d,n,H]$, we have:
    \begin{enumerate}
        \item If $d=m<n$, then $\underline{m+1}$ is optimal, where the system transitions to state $[m+1,m+1,\max\{m+1, n\}]$.
        \item If $d\ne m$ or $d=m=n$, then $\overline{d+1}$ is optimal, where the system transitions to state $[m,d+1, \max\{d+1, n\}]$.
    \end{enumerate}
\end{proposition}

\dg{We prove Proposition~\ref{prop:posdelay0H} in detail in Appendix~\ref{a:posdelay0H}. The proposition} is consistent with the bait-and-switch attack 
(Definition~\ref{def:bait-and-switch}).
Since $\Delta=0$, every H-blocks becomes public upon arrival. The only relevant part of Definition~\ref{def:bait-and-switch} is:
``All A-blocks are kept private except when there is a unique highest
branch and the height of a second highest branch is equal to the
highest public height, then the second branch is made the public fork choice.'' This means that if $n>m=d$, then the lower branch is made public as a ``bait'' to ensure that $\underline{m+1}$ is the preferred action, which is exactly rule~(1) in Proposition~\ref{prop:posdelay0H}. If a bait successfully attracts an H-block, subsequent H-blocks switch to this branch. Under all other conditions, bait-and-switch does not specify any particular action, except that A-blocks above the public height $d$ are nonpublic, so new H-blocks are always placed \dg{at} height $d+1$, which is consistent with rule~(2) in the proposition. In fact, $\underline{d+1}$ and $\overline{d+1}$ yield the same state $[m,d+1,n]$, so both are optimal under rule (2). Indeed, Definition~\ref{def:bait-and-switch} leaves this choice open. Thus, Proposition~\ref{prop:posdelay0H} establishes the optimality of bait-and-switch attack upon H-arrivals when $\Delta=0$.



\begin{proposition}[optimal action upon an A-arrival] \label{prop:delay0A}    
    In state $[m,d,n,A]$, we have:
    \begin{enumerate}
        \item if $d\le m$, 
        then $\overline{n+1}$ is optimal, where the system transitions to state $[m,d,n+1]$;
        \item if $d>m$, 
        then $\underline{m+1}$ is optimal, where the system transitions to state $[m+1,d,n]$;
    \end{enumerate}
\end{proposition}

The proof of Proposition~\ref{prop:delay0A} is relegated to Appendix~\ref{a:delay0A}.
Here we further demonstrate that this proposition for A-blocks is also consistent with the Definition~\ref{def:bait-and-switch}'s description: ``If there is a unique highest branch and the second-highest branch is strictly lower than the highest jumper, the adversary creates one A-block extending the second-highest branch; otherwise, the adversary creates one A-block extending a highest branch.'' Evidently, this means that if $d > m$, then take action $\underline{m+1}$; otherwise, take action $\overline{n+1}$, which is consistent with Proposition~\ref{prop:delay0A}.

We further note
that, in this case of $\Delta=0$, 
private mining (Definition~\ref{def:private mining}) places all A-blocks on a single chain and all H-blocks on another single chain from height~1 onward. If a shorter chain is as high as the public height, the chain must be formed by the H-blocks, so it is already the public fork choice, rendering baiting inconsequential. It is easy to see that private mining is consistent with bait-and-switch in this case. Hence if the latter is optimal, so is the former.

\section{The Case of Arbitrary Delays}
\label{s:delta}

In this section, we generalize the treatment of the zero-delay case in Sec.~\ref{s:zero} to arbitrary delays upper bounded by $\Delta\ge0$, and henceforth develop a proof of Theorem~\ref{th:opt_pow}.

\subsection{Preliminaries}
\label{s:prelim}

We first establish some preliminary results to narrow down potentially optimal adversarial actions, laying the groundwork for proving the optimality of bait-and-switch.

If the state is not ahead, i.e., $l_{m+1}$ is finite, then whether the lower branch is entirely public (aka $l_m=0$) or not yields the same state value.

\begin{lemma}\label{lm:lm=0}
    If 
    $l_{m+1}<\infty$, then
    \begin{align}
    \label{eq:lm=0}
        [m,d,n,(l_m,l_{m+1},\dots,l_d),I ]
        \asymp
        [m,d,n,(0,l_{m+1},\dots,l_d),I] .
    \end{align}
\end{lemma}

\begin{proof}
Let $s$ and $s'$ denote the left and right hand sides of~\eqref{eq:lm=0}, respectively. We have $s \succeq s'$ by Lemma~\ref{lm:h/l rule}, so it suffices to show $s \preceq s'$.

Consider two systems, referred to as $\mathcal{S}$ and $\mathcal{S}'$, which begin at states $s$ and $s'$, respectively.
Suppose the arrival processes in the two systems are identical. 
Suppose an {\em optimal} $\pi^*$ is adopted in system $\mathcal{S}$. Let $T$ denote a stopping time, which is equal to the first time $\mathcal{S}$ arrives in $\mathscr{A}$
or when $\mathcal{S}$'s height-$m$ timer expires, whichever occurs first. We define the following attack $\pi'$ for system $\mathcal{S}'$:
\begin{enumerate}
    \item Upon an H-arrival, if $\pi^*$ takes action $\overline{m}$ or $\underline{m}$, then $\pi'$ takes action $\overline{m+1}$.
    \item Under all other circumstances, $\pi'$ takes the same action as of $\pi$.
\end{enumerate}
We first justify that attack $\pi'$ is well-defined. The only time the two systems take different actions is when an H-block is placed \dg{at} height $m$ in $\mathcal{S}$, where a corresponding H-block needs to be placed \dg{at} height $m+1$ in $\mathcal{S}'$, in which case all timers in both systems remain unchanged. Hence the timer \dg{at} height $m$ in $\mathcal{S}$ is always dominated by the timer \dg{at} height $m+1$ in $\mathcal{S}'$. Moreover, $\mathcal{S}$ and $\mathcal{S}'$ always have identical timers from height $m+1$ onward. Therefore, $\pi'$ is well-defined as long as $\pi$ is. Additionally, both branches of $\mathcal{S}'$ are always no lower than the corresponding branches of $\mathcal{S}$ up to $T$.

If \dg{the stopping time $T$ is} when $\mathcal{S}$'s height-$m$ timer expires, then $\mathcal{S}$ and $\mathcal{S}'$ arrive at the same state at $T$.

If $\mathcal{S}$ arrives in $\mathscr{A}$ at $T$, it implies that, by Definition~\ref{def:v}, both the higher and lower branches of $\mathcal{S}$ are no lower than $k$, and so are both branches of system $\mathcal{S}'$. Furthermore, the public height in $\mathcal{S}'$ is $m$, which is no higher than the lower branch of $\mathcal{S}'$. Hence, $\mathcal{S}'$ must also have arrived in $\mathscr{A}$ by $T$.

By Lemma~\ref{lm:compare}, we have $s \preceq s'$, which completes the proof.
\end{proof}

We next narrow down the search for an optimal action upon each new \dg{A-arrival and H-arrival}.

\begin{lemma}[Potentially optimal actions for an A-arrival]\label{lm:potential_action_A>0}
in state $[m,d,n,(l_{d\land m},\dots,l_d),A]$,
at least one of the following actions (along with the corresponding transition) is optimal:
\begin{enumerate}[i)]
    \item 
    $\overline{n+1}$ (causing 
    a transition to $[m,d,n+1,(l_{d\land m},\dots,l_{d})]$),
    \item 
    $\underline{m+1}$ (applicable if $m<n$, causing 
    a transition to\\ $[m+1,d,n,(l_{d\land (m+1)},\dots,l_d)]$) .
\end{enumerate}
\end{lemma}

\begin{proof} 
The new A-block arrival must be added to either the longer branch \dg{at} any height up to $n+1$ or the lower branch \dg{at} any height up to $m+1$. If $m=n$, both $\underline{m+1}$ and $\overline{n+1}$ cause a transition to the same state. Hence making $\underline{m+1}$ applicable only if $m<n$ is without loss of generality. It is easy to verify that, if either $\underline{m+1}$ (in the case of $m<n$) or $\overline{n+1}$ is taken, the corresponding new state is as described in the lemma; otherwise, the system transitions to state $S=[m,d,n,(l_{d\land m},\dots,l_d)]$, i.e., both branches keep their heights and all timers remain unchanged. By Lemma~\ref{lm:h/l rule}, we have
\begin{align}
    S &\preceq [m+1,d,n,(l_{d\land (m+1)},\dots,l_d)] \\
    S &\preceq [m,d,n+1,(l_{d\land m},\dots,l_{d})] .
\end{align}
This implies that either $\underline{m+1}$ or $\overline{n+1}$ is optimal.
\end{proof}

\begin{lemma}[Potentially optimal actions for an H-arrival]\label{lm:potential_action_H>0}
    in state $s=[m,d,n,(l_{d\land m},\dots,l_d),H]$,
    at least one of the following actions (along with the corresponding transitions) is optimal:

    \begin{enumerate}[i)]
        \item $\overline{n+1}$ (the system transitions to state \\
        $[m,n+1,n+1,(l_{d\land m},\dots,l_d,\Delta,\dots,\Delta)]$),

        \item $\underline{m+1}$ (applicable only if $m<n$;
        if $m+1<d$, the system transitions to state $[m+1,d,n,(l_{m+1},\dots,l_d)]$;
        if $m+1\ge d$, the system transitions to state $[m+1,m+1,n,(\Delta)]$),
        
        \item $\overline{d}$ (applicable only if $l_d>0$; 
        the system transitions to state $[m,d,n,(l_{d\land m},\dots,l_d)]$),
        
        \item $\overline{d+1}$ (applicable only if $l_d=0$ and $d<n$ 
        ; the system transitions to state 
        $[m,d+1,n,(l_{(d+1) \land m},\dots,l_{d},\Delta)]$).
    \end{enumerate}
\end{lemma}

\begin{proof}
State $s$ includes a new H-block arrival, which can only be added above the public height. 
If $m=n$, then $\underline{m+1}$ is equivalent to $\overline{n+1}$, hence there is no loss of generality by letting $\underline{m+1}$ be applicable only if $m<n$.
Likewise, if $d=n$, then $\overline{d+1}$ is equivalent to $\overline{n+1}$, hence there is no loss of generality by letting $\overline{d+1}$ be applicable only if $d<n$.

It is not difficult to verify that the actions $\overline{n+1}$, $\underline{m+1}$, $\overline{d}$, and $\overline{d+1}$ cause the corresponding transitions given in (i)--(iv).

Any action that is different than $\overline{n+1}$, $\underline{m+1}$, $\overline{d}$, and $\overline{d+1}$ must be from one of the following categories:
\begin{enumerate}[\quad a)]
    \item either $\overline{j}$ or $\underline{j}$ for some $j<d$,
    \item $\overline{j}$ for some $j\in\{d+2,\dots,n\}$,
    \item 
    $\underline{j}$ for some $j\in\{d+1,\dots,m\}$.
\end{enumerate}

In Cases (a), since the H-block can be placed \dg{at} height $j<d$, the action $\overline{d}$ is also allowable. Moreover, both $\overline{j}$ and $\underline{j}$ cause a
transition to the same state as 
$\overline{d}$ would. 
Hence Case (a) is no better than $\overline{d}$.

By Lemma~\ref{lm:h/l rule}, if both $\overline{d}$ and $\overline{d+1}$ are applicable, $\overline{d}$ is strictly superior to $\overline{d+1}$. (Hence assuming $\overline{d+1}$ to be applicable only if $l_d>0$ is without loss of generality.) To establish the lemma, it suffices to show that the actions of Cases (b) and (c) are no better than $\overline{d+1}$, which is always an allowable action.

In Cases (b) and (c), the branches keep their heights, but the highest height with finite timer increases to $j$.

In Case (b),  which is applicable only if $n>d+1$, $\overline{j}$ causes a 
transition to $[m,j,n,(l_{j\land m},\dots,l_d,\Delta,\dots,\Delta)]$, whereas $\overline{d+1}$ causes a transition to $[m,d+1,n,(l_{(d+1)\land m},\dots,l_d,\Delta)]$.  
By Lemma~\ref{lm:h/l rule}, \dg{the latter state is more favorable}
for every $j>d+1$. Hence Case (b) is no better than $\overline{d+1}$.

In Case (c), which is applicable only if $d<m$, $\underline{j}$ causes a
transition to $[m,j,n,(\Delta)]$, 
whereas $\overline{d+1}$ 
causes a transition to $[m,d+1,n,(\Delta)]$. By Lemma~\ref{lm:h/l rule}, we have 
$    
    [m,j,n,(\Delta)]
    \preceq
    [m,d+1,n,(\Delta)]$. Hence Case (c) is no better than $\overline{d+1}$.

We have thus established that, unless $\overline{n+1}$ or $\underline{m+1}$ is taken, no action is better than either $\overline{d+1}$ or $\overline{d}$ if the latter is allowable. Hence the proof of the lemma.
\end{proof}


\subsection{Equivalent Propositions}

\dg{Building on the preliminary results in Sec.~\ref{s:prelim},} we establish the optimality of the bait-and-switch attack by proving the following three propositions.

\begin{proposition}\label{cl:ahead}
    In an ahead state $[m,d,n,(l),I]$, if $I=A$, then $\overline{n+1}$ is optimal, causing a transition to $[m,d,n+1,(l)]$; if $I=H$, the following are true:
    \begin{enumerate}[i)]
    \item if $l>0$, 
    then $\overline{d}$ is optimal, causing a transition to $[m,d,n,(l)]$;
    \item if $l=0$ and $d=m<n$, then 
    $\underline{m+1}$ is optimal, causing a transition to $[m+1,m+1,n,(\Delta)]$;
    \item if $l=0$ and $d<m$, then $\overline{d+1}$ is optimal, causing a transition to $[m,d+1,n,(\Delta)]$;
    \item if $l=0$ and 
    $d=m=n$, then $\overline{d+1}$ is optimal, causing a transition to     $[n,n+1,n+1,(0,\Delta)]$.
    \end{enumerate}
\end{proposition}

\begin{proposition}\label{cl:on-time}
    In an on-time state $[m,d,n,( 
    l_m, l_{m+1},\dots,l_d),I]$, with either $I=A$ or $H$, $\underline{m+1}$ is optimal, causing a transition to $[m+1,d,n,(l_{m+1},\dots,l_d)]$.
\end{proposition}

\begin{proposition}\label{cl:behind}
    In a behind state $[m,d,n,(0,0,l_{m+2},\dots,l_d),I ]$,
    if $I=A$, then $\underline{m+1}$ is optimal, causing the system to transition to state $[m+1,d,n,(0,l_{m+2},\dots,l_d)]$; if $I=H$, 
    the following are true:
    \begin{enumerate}[i)]
    \item if $l_d>0$, then action $\overline{d}$ is optimal, causing a transition to\\ $[m,d,n,(0,0,l_{m+2},\dots,l_d)]$;
    \item if $l_d = 0$, then $\overline{d+1}$ is optimal, causing a transition to $[m,d+1,\max\{n,d+1\},(0,0,l_{m+2},\dots,l_d,\Delta)]$.
    \end{enumerate}
\end{proposition}

We next demonstrate Propositions~\ref{cl:ahead}--\ref{cl:behind} are consistent with the bait-and-switch attack (Definition~\ref{def:bait-and-switch}). Specifically, we show that they describe a specific form of bait-and-switch.

By Definition~\ref{def:bait-and-switch}, a bait-and-switch attack is described under the following two situations:
\begin{enumerate}[i)]
    \item There is a unique highest branch and a second-highest branch at the public height, so this second-highest branch is chosen as the public fork choice. According to Lemma~\ref{lm:must take the form of}, this is either an ahead state where $d=m$ and $l_d=0$ or an on-time state. In both cases, upon an H-arrival, Propositions~\ref{cl:ahead} and~\ref{cl:on-time} prescribes action $\underline{m+1}$, which is identical to the action of bait-and-switch, where the H-block extends the public fork choice.

    \item In all other situations, no public fork choice is made and H-blocks are maximally delayed. 
    Bait-and-switch would place a new H-block right above the public height, which is \dg{at} height $d$ (if $l_{d} > 0$) or $d+1$ (if $l_{d} = 0$). In this case, Propositions~\ref{cl:ahead} and~\ref{cl:behind} dictate either $\overline{d}$ or $\overline{d+1}$, depending on whether $l_d>0$ or not, which is consistent with bait-and-switch. In fact, $\overline{d}$ and $\underline{d}$ yield the same state, and so do $\overline{d+1}$ and $\underline{d+1}$, hence Definition~\ref{def:bait-and-switch} leaves the choice of higher or lower branch open.
\end{enumerate}

Upon an A-arrival, noting that $d\le m$ in an ahead state and $d>m$ in an on-time or behind state, the optimal actions prescribed by Propositions~\ref{cl:ahead}--\ref{cl:behind} are simple:
\begin{itemize}
    \item $\overline{n+1}$ is optimal if $d\leq m$;
    \item $\underline{m+1}$ is optimal if $m<d$.
\end{itemize}
According to the bait-and-switch attack, when there is a unique highest branch ($m<n$) and a second highest branch is strictly lower than the highest jumper ($m<d$), A-block extends a second highest branch, making $\underline{m+1}$ optimal; otherwise ($m=n$ or $d\le m$), A-block extends a highest branch, making $\overline{n+1}$ optimal. As $m=n$ implies $d\le m$, the bait-and-switch attack follows the optimal actions given by Propositions~\ref{cl:ahead}--\ref{cl:behind}.

We relegate the rather technical proof of Propositions~\ref{cl:ahead}--\ref{cl:behind} to Appendices~\ref{a:A-ahead}--\ref{a:A-behind}.

\section{Safety at Arbitrary Target Height} 
\label{s:target}

We now generalize the \dg{analysis to allow the target block to appear at any height. However, if the adversary could} make unlimited attempts to mine a private fork that is $k$ blocks ahead of the public height before selecting a target block to attack, then a violation would be trivially guaranteed. To preclude such {\em adaptive attacks}, \dg{one may require the target block to be placed within} a finite execution horizon, \dg{enabling} a union bound \dg{(over all possible placements)} to obtain a safety guarantee (see, e.g.,~\cite{pass2017sleepy}).

\dg{In the literature, the timing of the target block is often tightly constrained. In this paper,} we follow Nakamoto's original approach~\cite{nakamoto2008bitcoin} and impose the following constraint \dg{on a high-value, high-priority transaction}: An honest node generates a new public key and provides it to the sender at time $s$, who must immediately generate a transaction using the new public key and make the transaction public. This ensures that the target block cannot arrive before time $s$, and will likely arrive soon after. Specifically, we assume \dg{the transaction's fee rate is high enough to ensure its inclusion} in the first H-block mined after $s$.

In contrast to analyzing safety at height~1, extending the analysis to arbitrary heights introduces additional complications and challenges. In particular, the adversarial may opportunistically gain extra advantage by mining a private fork that extends beyond the eventual height of the target block---even though the adversary has no control over the selection of time $s$.

\begin{Def}[pre-mining lead]
\label{def:pre}
    The pre-mining lead at time $t$, denoted $L_t$, is the difference between the height of a highest block (of any type) mined by time $t$ and the height of a highest H-block mined by time $t$.
\end{Def}

Evidently, $L_t\ge0$ for all $t$, with equality if and only if an H-block is among the highest at time $t$. Since $L_t$ is defined as the difference between two right-continuous counting processes, it is itself right-continuous as a function of $t$. We denote the pre-mining lead just before an arrival at time $t$ by $\Ltm$.

Given a target transaction created at time $s$, let the block that includes this transaction be denoted as block~$b$, where $h_b$ stands for its height, i.e., the target height. The target height is not fixed in advance---its height depends on when and where the target block is created. We are interested in the event of a safety violation occurring at the target height $h_b$, where block~$b$ is confirmed in one of the credible chains that participates in the violation.
Throughout the rest of this section, 
we use the term ``safety violation of the target block'' to refer to a safety violation at the target block’s height, with the target block confirmed in a credible chain.

To properly address safety at arbitrary heights, we modify several earlier definitions accordingly.

\begin{Def}[branch] 
\label{def:brancheta}
    Agreement 
    on a target block $b$ introduces an equivalence relation that partitions all blocks at height $h_b$ and above into two disjoint classes, depending on whether they agree on \dg{chain $b$ as a prefix}. 
    The set consisting of block $b$ and all its descendants is called the target branch. All other blocks at height $h_b$ and above form the challenger branch. The height of a branch is defined as the maximum height among the block(s) it contains. A branch is said to be $t$-credible if it is no lower than the public height at time $t$.
\end{Def}

The following result is a generalization of Lemma~\ref{lm:SV->absorb}.

\begin{lemma} 
\label{lm:eqtwobranches}
    Under the $ k $-confirmation commitment rule, 
    violating the safety of a target block $b$ is equivalent to the existence of a time $s$ at which two $ s $-credible branches that disagree on block $b$ both reach height \dg{at} least $h_b+k-1$.
\end{lemma}

\begin{Def}[bait-and-switch attack on target block $b$]
\label{def:bait-and-switch-target}
    Upon every A-block arrival, if either the target branch or the challenger branch is strictly lower than the highest jumper, the A-block extends the lower branch; otherwise, the A-block extends a highest branch. All A-blocks are kept private except when one branch is as high as the public height but strictly lower than the other branch, then the lower branch is made the public fork choice. The propagation of every H-block is delayed maximally (for $\Delta$) unless it becomes public in a public fork choice.
\end{Def}

\begin{theorem}
\label{th:general}
    Under the Nakamoto consensus protocol, for a target block that is the first H-block mined after a given time $s$,
    the bait-and-switch attack maximizes the probability of violating its safety. 
\end{theorem}

\begin{figure}
    \centering
    \includegraphics[width = 0.9\columnwidth ]{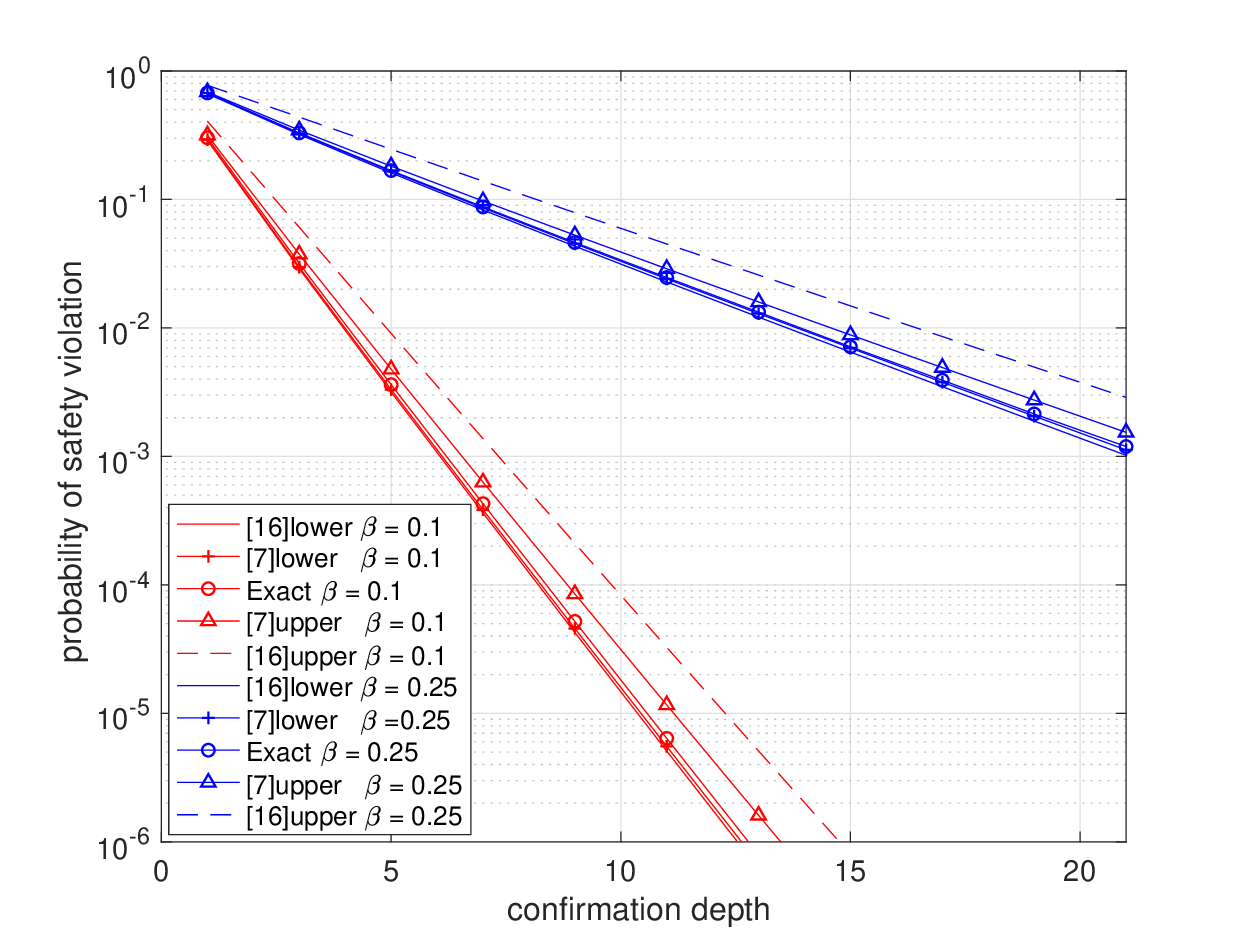}
    \caption{The total mining rate is $\lambda=1/600$ blocks per second. The propagation delay upper bound is $\Delta=10$ seconds.}
    \label{fig:BTC_l}
\end{figure}

\begin{theorem} 
\label{th:trade-offgeneral}
    Let $s\gg 0$ be a time at which the pre-mining lead process has reached steady state. 
    For a target transaction created at time $s$, the probability of violating the safety a target block including the target transaction is equal to
\begin{align}
\label{eq:exact_bitcoin_l}
1 - \sum_{l=0}^{k-1} \sum_{i=0}^{k-1} \sum_{y=l}^{k-1-i} e(i)
    & \Bigg(    f_3(l)\left[ P^{(1)} \times \cdots \times P^{(k)} \right]_{l+1, y+1} \notag \\
    &+ f_4(l)\left[ P'^{(2)} \times \cdots \times P'^{(k)} \right]_{l+2, y+1}
\Bigg)
\end{align}
where $f_3$ is given by \eqref{eq:L,J}, $f_4$ is given by \eqref{eq:L,Jc}, the transition probability matrix $P^{(i)}$ for $i = 1,\dots,k$ is given by~\eqref{eq:P1l}--
\eqref{eq:Pjyk}, and the transition probability matrix $P'^{(i)}$ for $i = 2,\dots,k$ is given by~\eqref{eq:P'2l}--
\eqref{eq:P'yk}.
\end{theorem}

Theorems~\ref{th:general} and~\ref{th:trade-offgeneral} are proved in Appendix~\ref{a:general}.

In Fig.~\ref{fig:BTC_l}, we plot the exact latency-security trade-off of Bitcoin at a general target height, 
given by~\eqref{eq:exact_bitcoin_l} under two different adversarial fractions, with $\beta=0.25$, and $0.1$, respectively. We also include upper and lower bounds based on results in \cite{guo2022bitcoin} and \cite{doger2024refined}.

\section{Conclusion}

This paper resolves the longstanding question of block safety in PoW Nakamoto consensus protocols by identifying an optimal attack strategy, termed ``bait-and-switch,'' and computing the precise probability of a safety violation after a given number of confirmations. Our analysis considers a baseline model where an omniscient adversary controls message propagation delays up to a maximum of $\Delta$. Central to our approach is the formulation of a tractable discrete-time Markov decision process that concisely captures the system state and the adversary's permissible actions.

The bait-and-switch attack naturally extends to scenarios with more constrained adversaries. Specifically, when the adversary has partial control over the views of honest nodes, it can delay the propagation of honest blocks within its capabilities and release the bait once it observes a second-best branch matching the public height. Determining the precise conditions under which this generalized attack remains optimal is an open question.

Reinforcement learning offers a promising framework for identifying attack strategies in settings where the adversary has incomplete state information or reduced capabilities. However, the exponential growth of the state space introduces significant computational challenges. Our proposed reward function and compact state representation may provide a foundation for efficient reinforcement learning-based methods in future research.

In a forthcoming work, we will demonstrate that in proof-of-stake variants of Nakamoto consensus, the bait-and-switch attack is also optimal under a broad set of conditions.



\section{Acknowledgment}

The authors sincerely appreciate the anonymous reviewers' constructive suggestions, which improved this paper significantly. This work is supported in part by the National Science Foundation under Grant no.~2132700.

\bibliographystyle{ACM-Reference-Format}
\balance
\bibliography{ref}


\begin{thebibliography}{26}


\ifx \showCODEN    \undefined \def \showCODEN     #1{\unskip}     \fi
\ifx \showISBNx    \undefined \def \showISBNx     #1{\unskip}     \fi
\ifx \showISBNxiii \undefined \def \showISBNxiii  #1{\unskip}     \fi
\ifx \showISSN     \undefined \def \showISSN      #1{\unskip}     \fi
\ifx \showLCCN     \undefined \def \showLCCN      #1{\unskip}     \fi
\ifx \shownote     \undefined \def \shownote      #1{#1}          \fi
\ifx \showarticletitle \undefined \def \showarticletitle #1{#1}   \fi
\ifx \showURL      \undefined \def \showURL       {\relax}        \fi
\providecommand\bibfield[2]{#2}
\providecommand\bibinfo[2]{#2}
\providecommand\natexlab[1]{#1}
\providecommand\showeprint[2][]{arXiv:#2}

\bibitem[Bagaria et~al\mbox{.}(2019)]%
        {bagaria2019prism}
\bibfield{author}{\bibinfo{person}{Vivek Bagaria}, \bibinfo{person}{Sreeram
  Kannan}, \bibinfo{person}{David Tse}, \bibinfo{person}{Giulia Fanti}, {and}
  \bibinfo{person}{Pramod Viswanath}.} \bibinfo{year}{2019}\natexlab{}.
\newblock \showarticletitle{Prism: Deconstructing the blockchain to approach
  physical limits}. In \bibinfo{booktitle}{\emph{Proceedings of the 2019 ACM
  SIGSAC Conference on Computer and Communications Security}}.
  \bibinfo{pages}{585--602}.
\newblock


\bibitem[Bai et~al\mbox{.}(2023)]%
        {bai2023blockchain}
\bibfield{author}{\bibinfo{person}{Qianlan Bai}, \bibinfo{person}{Yuedong Xu},
  \bibinfo{person}{Nianyi Liu}, {and} \bibinfo{person}{Xin Wang}.}
  \bibinfo{year}{2023}\natexlab{}.
\newblock \showarticletitle{Blockchain mining with multiple selfish miners}.
\newblock \bibinfo{journal}{\emph{IEEE Transactions on Information Forensics
  and Security}}  \bibinfo{volume}{18} (\bibinfo{year}{2023}),
  \bibinfo{pages}{3116--3131}.
\newblock


\bibitem[Bar-Zur et~al\mbox{.}(2022)]%
        {bar2022werlman}
\bibfield{author}{\bibinfo{person}{Roi Bar-Zur}, \bibinfo{person}{Ameer
  Abu-Hanna}, \bibinfo{person}{Ittay Eyal}, {and} \bibinfo{person}{Aviv
  Tamar}.} \bibinfo{year}{2022}\natexlab{}.
\newblock \showarticletitle{WeRLman: to tackle whale (transactions), go deep
  (RL)}. In \bibinfo{booktitle}{\emph{Proceedings of the 15th ACM International
  Conference on Systems and Storage}}. \bibinfo{pages}{148--148}.
\newblock


\bibitem[Canetti(2020)]%
        {canetti2020universally}
\bibfield{author}{\bibinfo{person}{Ran Canetti}.}
  \bibinfo{year}{2020}\natexlab{}.
\newblock \showarticletitle{Universally composable security}.
\newblock \bibinfo{journal}{\emph{J. ACM}} \bibinfo{volume}{67},
  \bibinfo{number}{5} (\bibinfo{year}{2020}), \bibinfo{pages}{1--94}.
\newblock


\bibitem[Cao and Guo(2025)]%
        {cao2025security}
\bibfield{author}{\bibinfo{person}{Shu-Jie Cao} {and} \bibinfo{person}{Dongning
  Guo}.} \bibinfo{year}{2025}\natexlab{}.
\newblock \showarticletitle{Security, Latency, and Throughput of Proof-of-Work
  {N}akamoto Consensus}.
\newblock \bibinfo{journal}{\emph{IEEE Transactions on Information Theory}}
  \bibinfo{volume}{71}, \bibinfo{number}{6} (\bibinfo{year}{2025}),
  \bibinfo{pages}{4708--4731}.
\newblock
\href{https://doi.org/10.1109/TIT.2025.3557761}{doi:\nolinkurl{10.1109/TIT.2025.3557761}}


\bibitem[Dembo et~al\mbox{.}(2020)]%
        {dembo2020everything}
\bibfield{author}{\bibinfo{person}{Amir Dembo}, \bibinfo{person}{Sreeram
  Kannan}, \bibinfo{person}{Ertem~Nusret Tas}, \bibinfo{person}{David Tse},
  \bibinfo{person}{Pramod Viswanath}, \bibinfo{person}{Xuechao Wang}, {and}
  \bibinfo{person}{Ofer Zeitouni}.} \bibinfo{year}{2020}\natexlab{}.
\newblock \showarticletitle{Everything is a Race and {Nakamoto} Always Wins}.
  In \bibinfo{booktitle}{\emph{Proceedings of the 2020 ACM SIGSAC Conference on
  Computer and Communications Security}}.
\newblock


\bibitem[Doger and Ulukus(2025a)]%
        {doger2024refined}
\bibfield{author}{\bibinfo{person}{Mustafa Doger} {and} \bibinfo{person}{Sennur
  Ulukus}.} \bibinfo{year}{2025}\natexlab{a}.
\newblock \showarticletitle{Refined Bitcoin Security-Latency Under Network
  Delay}.
\newblock \bibinfo{journal}{\emph{IEEE Transactions on Information Theory}}
  \bibinfo{volume}{71}, \bibinfo{number}{4} (\bibinfo{year}{2025}),
  \bibinfo{pages}{3038--3047}.
\newblock
\href{https://doi.org/10.1109/TIT.2024.3447576}{doi:\nolinkurl{10.1109/TIT.2024.3447576}}


\bibitem[Doger and Ulukus(2025b)]%
        {doger2025when}
\bibfield{author}{\bibinfo{person}{Mustafa Doger} {and} \bibinfo{person}{Sennur
  Ulukus}.} \bibinfo{year}{2025}\natexlab{b}.
\newblock \showarticletitle{When Should Selfish Miners Double-Spend?}
\newblock \bibinfo{journal}{\emph{arXiv preprint arXiv:2501.03227}}
  (\bibinfo{year}{2025}).
\newblock


\bibitem[Durrett(2010)]%
        {durrett2010probability}
\bibfield{author}{\bibinfo{person}{Rick Durrett}.}
  \bibinfo{year}{2010}\natexlab{}.
\newblock \bibinfo{booktitle}{\emph{Probability: Theory and Examples}
  (\bibinfo{edition}{4th} ed.)}.
\newblock \bibinfo{publisher}{Cambridge University Press},
  \bibinfo{address}{USA}.
\newblock
\showISBNx{0521765390}


\bibitem[Feng and Niu(2019)]%
        {feng2019selfish}
\bibfield{author}{\bibinfo{person}{Chen Feng} {and} \bibinfo{person}{Jianyu
  Niu}.} \bibinfo{year}{2019}\natexlab{}.
\newblock \showarticletitle{Selfish mining in ethereum}. In
  \bibinfo{booktitle}{\emph{Proceedings of the 39th IEEE International
  Conference on Distributed Computing Systems}}. \bibinfo{pages}{1306--1316}.
\newblock


\bibitem[Garay et~al\mbox{.}(2015)]%
        {garay2015bitcoin}
\bibfield{author}{\bibinfo{person}{Juan Garay}, \bibinfo{person}{Aggelos
  Kiayias}, {and} \bibinfo{person}{Nikos Leonardos}.}
  \bibinfo{year}{2015}\natexlab{}.
\newblock \showarticletitle{The bitcoin backbone protocol: Analysis and
  applications}. In \bibinfo{booktitle}{\emph{Annual International Conference
  on the Theory and Applications of Cryptographic Techniques}}. Springer,
  \bibinfo{pages}{281--310}.
\newblock


\bibitem[Garay et~al\mbox{.}(2020)]%
        {garay2020full}
\bibfield{author}{\bibinfo{person}{Juan~A Garay}, \bibinfo{person}{Aggelos
  Kiayias}, {and} \bibinfo{person}{Nikos Leonardos}.}
  \bibinfo{year}{2020}\natexlab{}.
\newblock \showarticletitle{Full Analysis of {N}akamoto Consensus in
  Bounded-Delay Networks.}
\newblock \bibinfo{journal}{\emph{IACR Cryptol. ePrint Arch.}}
  \bibinfo{volume}{2020} (\bibinfo{year}{2020}), \bibinfo{pages}{277}.
\newblock


\bibitem[Ga\v{z}i et~al\mbox{.}(2020)]%
        {gazi2020tight}
\bibfield{author}{\bibinfo{person}{Peter Ga\v{z}i}, \bibinfo{person}{Aggelos
  Kiayias}, {and} \bibinfo{person}{Alexander Russell}.}
  \bibinfo{year}{2020}\natexlab{}.
\newblock \showarticletitle{Tight Consistency Bounds for {B}itcoin}. In
  \bibinfo{booktitle}{\emph{Proceedings of the 2020 ACM SIGSAC Conference on
  Computer and Communications Security}} (Virtual Event, USA).
  \bibinfo{address}{New York, NY, USA}, \bibinfo{pages}{819–838}.
\newblock
\showISBNx{9781450370899}
\href{https://doi.org/10.1145/3372297.3423365}{doi:\nolinkurl{10.1145/3372297.3423365}}


\bibitem[Ga{\v{z}}i et~al\mbox{.}(2022)]%
        {gazi2022practical}
\bibfield{author}{\bibinfo{person}{Peter Ga{\v{z}}i}, \bibinfo{person}{Ling
  Ren}, {and} \bibinfo{person}{Alexander Russell}.}
  \bibinfo{year}{2022}\natexlab{}.
\newblock \showarticletitle{Practical Settlement Bounds for Proof-of-Work
  Blockchains}. In \bibinfo{booktitle}{\emph{Proceedings of the 2022 ACM SIGSAC
  Conference on Computer and Communications Security}} (Los Angeles, CA, USA).
  \bibinfo{address}{New York, NY, USA}, \bibinfo{pages}{1217–1230}.
\newblock
\showISBNx{9781450394505}
\href{https://doi.org/10.1145/3548606.3559368}{doi:\nolinkurl{10.1145/3548606.3559368}}


\bibitem[Ga{\v{z}}i et~al\mbox{.}(2023)]%
        {gazi2023practical}
\bibfield{author}{\bibinfo{person}{Peter Ga{\v{z}}i}, \bibinfo{person}{Ling
  Ren}, {and} \bibinfo{person}{Alexander Russell}.}
  \bibinfo{year}{2023}\natexlab{}.
\newblock \showarticletitle{Practical Settlement Bounds for Longest-Chain
  Consensus}. In \bibinfo{booktitle}{\emph{Advances in Cryptology -- CRYPTO
  2023}}, \bibfield{editor}{\bibinfo{person}{Helena Handschuh} {and}
  \bibinfo{person}{Anna Lysyanskaya}} (Eds.). \bibinfo{publisher}{Springer
  Nature Switzerland}, \bibinfo{address}{Cham}, \bibinfo{pages}{107--138}.
\newblock
\showISBNx{978-3-031-38557-5}


\bibitem[Guo and Ren(2022)]%
        {guo2022bitcoin}
\bibfield{author}{\bibinfo{person}{Dongning Guo} {and} \bibinfo{person}{Ling
  Ren}.} \bibinfo{year}{2022}\natexlab{}.
\newblock \showarticletitle{Bitcoin's Latency--Security Analysis Made Simple}.
  In \bibinfo{booktitle}{\emph{Proceedings of the 4th ACM Conference on
  Advances in Financial Technologies}}. \bibinfo{address}{Cambridge, MA},
  \bibinfo{pages}{244--253}.
\newblock
\href{https://doi.org/10.48550/ARXIV.2203.06357}{doi:\nolinkurl{10.48550/ARXIV.2203.06357}}


\bibitem[Kiffer et~al\mbox{.}(2018)]%
        {kiffer2018better}
\bibfield{author}{\bibinfo{person}{Lucianna Kiffer}, \bibinfo{person}{Rajmohan
  Rajaraman}, {and} \bibinfo{person}{Abhi Shelat}.}
  \bibinfo{year}{2018}\natexlab{}.
\newblock \showarticletitle{A better method to analyze blockchain consistency}.
  In \bibinfo{booktitle}{\emph{Proceedings of the 2018 ACM SIGSAC Conference on
  Computer and Communications Security}}. \bibinfo{pages}{729--744}.
\newblock


\bibitem[Kroese(1992)]%
        {kroese1992difference}
\bibfield{author}{\bibinfo{person}{D.~P. Kroese}.}
  \bibinfo{year}{1992}\natexlab{}.
\newblock \showarticletitle{The difference of two renewal processes level
  crossing and the infimum}.
\newblock \bibinfo{journal}{\emph{Communications in Statistics. Stochastic
  Models}} \bibinfo{volume}{8}, \bibinfo{number}{2} (\bibinfo{year}{1992}),
  \bibinfo{pages}{221--243}.
\newblock
\href{https://doi.org/10.1080/15326349208807222}{doi:\nolinkurl{10.1080/15326349208807222}}
\showeprint{https://doi.org/10.1080/15326349208807222}


\bibitem[Li et~al\mbox{.}(2021)]%
        {li2021close}
\bibfield{author}{\bibinfo{person}{Jing Li}, \bibinfo{person}{Dongning Guo},
  {and} \bibinfo{person}{Ling Ren}.} \bibinfo{year}{2021}\natexlab{}.
\newblock \showarticletitle{Close Latency-Security Trade-off for the {N}akamoto
  Consensus}. In \bibinfo{booktitle}{\emph{Proceedings of the 3rd ACM
  Conference on Advances in Financial Technologies}} (Arlington, Virginia).
  \bibinfo{address}{New York, NY}, \bibinfo{pages}{100–113}.
\newblock
\showISBNx{9781450390828}
\href{https://doi.org/10.1145/3479722.3480992}{doi:\nolinkurl{10.1145/3479722.3480992}}


\bibitem[Nakamoto(2008)]%
        {nakamoto2008bitcoin}
\bibfield{author}{\bibinfo{person}{Satoshi Nakamoto}.}
  \bibinfo{year}{2008}\natexlab{}.
\newblock \bibinfo{title}{Bitcoin: A peer-to-peer electronic cash system}.
\newblock \bibinfo{howpublished}{Whitepaper}.
\newblock
\urldef\tempurl%
\url{https://bitcoin.org/bitcoin.pdf}
\showURL{%
\tempurl}


\bibitem[Pass et~al\mbox{.}(2017)]%
        {pass2017analysis}
\bibfield{author}{\bibinfo{person}{Rafael Pass}, \bibinfo{person}{Lior Seeman},
  {and} \bibinfo{person}{Abhi Shelat}.} \bibinfo{year}{2017}\natexlab{}.
\newblock \showarticletitle{Analysis of the Blockchain Protocol in Asynchronous
  Networks}. In \bibinfo{booktitle}{\emph{Advances in Cryptology -- EUROCRYPT
  2017}}, \bibfield{editor}{\bibinfo{person}{Jean-S{\'e}bastien Coron} {and}
  \bibinfo{person}{Jesper~Buus Nielsen}} (Eds.). \bibinfo{publisher}{Springer
  International Publishing}, \bibinfo{address}{Cham},
  \bibinfo{pages}{643--673}.
\newblock
\showISBNx{978-3-319-56614-6}


\bibitem[Pass and Shi(2017a)]%
        {pass2017rethinking}
\bibfield{author}{\bibinfo{person}{Rafael Pass} {and} \bibinfo{person}{Elaine
  Shi}.} \bibinfo{year}{2017}\natexlab{a}.
\newblock \showarticletitle{Rethinking Large-Scale Consensus}. In
  \bibinfo{booktitle}{\emph{Proceedings of the 30th IEEE Computer Security
  Foundations Symposium}}. \bibinfo{pages}{115--129}.
\newblock
\href{https://doi.org/10.1109/CSF.2017.37}{doi:\nolinkurl{10.1109/CSF.2017.37}}


\bibitem[Pass and Shi(2017b)]%
        {pass2017sleepy}
\bibfield{author}{\bibinfo{person}{Rafael Pass} {and} \bibinfo{person}{Elaine
  Shi}.} \bibinfo{year}{2017}\natexlab{b}.
\newblock \showarticletitle{The Sleepy Model of Consensus}. In
  \bibinfo{booktitle}{\emph{Advances in Cryptology -- ASIACRYPT 2017}},
  \bibfield{editor}{\bibinfo{person}{Tsuyoshi Takagi} {and}
  \bibinfo{person}{Thomas Peyrin}} (Eds.). \bibinfo{publisher}{Springer
  International Publishing}, \bibinfo{address}{Cham},
  \bibinfo{pages}{380--409}.
\newblock
\showISBNx{978-3-319-70697-9}


\bibitem[Puterman(1994)]%
        {puterman1994markov}
\bibfield{author}{\bibinfo{person}{Martin~L. Puterman}.}
  \bibinfo{year}{1994}\natexlab{}.
\newblock \bibinfo{booktitle}{\emph{Markov Decision Processes: Discrete
  Stochastic Dynamic Programming} (\bibinfo{edition}{1st} ed.)}.
\newblock \bibinfo{publisher}{John Wiley \& Sons, Inc.},
  \bibinfo{address}{USA}.
\newblock
\showISBNx{0471619779}


\bibitem[Sapirshtein et~al\mbox{.}(2017)]%
        {sapirshtein2017optimal}
\bibfield{author}{\bibinfo{person}{Ayelet Sapirshtein},
  \bibinfo{person}{Yonatan Sompolinsky}, {and} \bibinfo{person}{Aviv Zohar}.}
  \bibinfo{year}{2017}\natexlab{}.
\newblock \showarticletitle{Optimal selfish mining strategies in bitcoin}. In
  \bibinfo{booktitle}{\emph{The 20th International Conference Financial
  Cryptography and Data Security, Revised Selected Papers}}.
  \bibinfo{publisher}{Springer}, \bibinfo{pages}{515--532}.
\newblock


\bibitem[Zur et~al\mbox{.}(2020)]%
        {zur2020efficient}
\bibfield{author}{\bibinfo{person}{Roi~Bar Zur}, \bibinfo{person}{Ittay Eyal},
  {and} \bibinfo{person}{Aviv Tamar}.} \bibinfo{year}{2020}\natexlab{}.
\newblock \showarticletitle{Efficient MDP analysis for selfish-mining in
  blockchains}. In \bibinfo{booktitle}{\emph{Proceedings of the 2nd ACM
  Conference on Advances in Financial Technologies}}.
  \bibinfo{pages}{113--131}.
\newblock


\end{thebibliography}
\appendix
\section{Proof of lemma}
\label{a:pr_of_lemma}
\subsection{Proof of Lemma~\ref{lm:SV->absorb}}
\label{a:proof_lm1}

    Consider the $t$-credible and $t'$-credible chains as defined in Definition~\ref{def:v}, where $t\le t'$. Let the height-1 block of the $t$-credible chain be denoted as block $b$. The height of branch $b$ at time $t$ is at least $k$. Define $s$ as the first time during $[t,t']$ for which there exists an $s$-credible chain $c$ of height $k$ or higher, which does not contain block $b$. If $s=t$, then the lemma holds trivially. We assume $s>t$ for the remainder of this proof. In this case, the arrival time of block $c$ is exactly $s$; if it were earlier, then $s$ cannot be the first time such a chain exists, contradicting the definition of $s$. Moreover, as the parent of block $c$, block $f_c$ arrives strictly before $s$. Evidently, chain $c$ cannot be strictly higher than branch $b$ at $s$; if it were, then chain $f_c$ would be $s'$-credible for some $s'\in(t,s)$ with height at least $k$, contradicting the assumption that $s$ is the first such time. This implies that chain $c$ and some chain in branch $b$ are both $s$-credible with height at least $k$, establishing the lemma.

\subsection{Proof of Lemma~\ref{lm:lim V(X_n)}}
\label{a:lmlimV}

Let us define
\begin{align} \label{eq:barG}
    \bar{G} = \sum_{i=1}^{\infty}r(X_i,Y_i)+1_{\{X_1 \in \mathscr{A}\}}
\end{align}
which is equal to 1 if the starting state $X_1$ is in $\mathscr{A}$ or the state will ever arrive in $\mathscr{A}$.
Let $\mathscr{F}_1,\mathscr{F}_2,\dots$ denote the filtration generated by the history of states and actions, i.e.,
$H_1,H_2,\dots$ defined in Sec.~\ref{sec:MDP-I}.

Let us also define
\begin{align} \label{eq:uHn}
    u(H_n) = E 
    ^{\pi^*}\left[\bar{G}|\mathscr{F}_n\right],
\end{align}
which denotes the expected future value conditioned on the history $H_n$. It is easy to verify that $u(H_1),u(H_2),\dots$
is a martingale with respect to the filtration. 
Note that $u(H_\infty) = E^{\pi^*}\left[\bar{G}|\mathscr{F}_\infty\right]$ is also well-defined. By Doob's optional stopping theorem~\cite{durrett2010probability}, for every stopping time $T$ and arbitrary policy $\pi$:
 \begin{align} \label{eq:doob2}
    E_s^{\pi}[u(H_1)]
    &= E_s^{\pi}[u(H_T)].
\end{align}
Moreover,
\begin{align}
    E_s^{\pi^*}[ u(H_1) ]
    &= E_s^{\pi^*}[ E^{\pi^*} [\bar{G}|\mathscr{F}_1] ] \\
    &= E_s^{\pi^*}[\bar{G}] \\
    &= V(s) \label{eq:EuH1}
\end{align}
where~\eqref{eq:EuH1} is by definition of $V$ 
in~\eqref{eq:Vs}.
Since $\mathscr{A}$ is absorbing, once the state arrives in $\mathscr{A}$, it remains in that state, which implies
    \begin{align} \label{eq:1XnX1}
        \sum_{i=1}^{n-1}r(X_i, Y_i) + 1_{\{X_1 \in \mathscr{A}\}} = 1_{\{X_n \in \mathscr{A}\}} . 
    \end{align}
    Using~\eqref{eq:barG} and~\eqref{eq:uHn}, we can write
    \begin{align}
       u(H_n)
       &=
       E
       ^{\pi^*}\left[\sum_{i=1}^{\infty}r(X_i,Y_i)+1_{\{X_1 \in \mathscr{A}\}} \big| \mathscr{F}_n\right] \label{eq:uHnXnX1}
       \\
       &=
        E
        ^{\pi^*}\left[\sum_{i=n}^{\infty}r(X_i,Y_i)+1_{\{X_n \in \mathscr{A}\}}\right] \label{eq:P,i}
        \\
        &= V(X_n) \label{eq:uHVX}
    \end{align} 
    where~\eqref{eq:P,i} is due to~\eqref{eq:1XnX1} and the fact that, under the stationary conserving policy $\pi^*$, conditioned on $X_n$, future states and actions are independent of $\mathscr{F}_n$, and~\eqref{eq:uHVX} is by definition of $V$.

Recall that $Z_n = V(X_n)$ for every $n\in\mathbb{N}$, where $\mathbb{N}$ represents the natural number set. By~\eqref{eq:uHn} and~\eqref{eq:uHVX}, we can write
\begin{align} \label{eq:VEG}
    Z_n 
    = E^{\pi^*}\left[\bar{G}|\mathscr{F}_n\right].
\end{align}
Using~\cite[Theorem 4.6.8]{durrett2010probability}, we have
\begin{align}
    \lim_{n\to\infty} Z_n 
    &= E^{\pi^*}\left[\bar{G}|\mathscr{F}_{\infty}\right]
    \;\; \text{almost surely} \\
    &= \bar{G}        \label{eq:VXT_Tinf}
\end{align}
where~\eqref{eq:VXT_Tinf} is by definition of conditional expectation. 
Hence the limit of $Z_n=V(X_n)$ exists and
$Z_\infty=\lim_{n\to \infty}V(X_n) = \bar{G}$ almost surely.

\subsection{Proof of Lemma~\ref{lm:Vbar=EVbar}}
\label{a:lemma7}

Using~\eqref{eq:uHVX}, we can write
    \begin{align}
    E_{s}^{\pi^*}\left[Z_T\right]
   &=
   E_{s}^{\pi^*}\left[
   u(H_T)\right] \\
   &=
   E_{s}^{\pi^*}\left[
   u(H_1)\right] \label{eq:doob3} \\
   &=
   V(s), \label{eq:EV=V}
   \end{align}
where~\eqref{eq:doob3} is due to~\eqref{eq:doob2}, and~\eqref{eq:EV=V} is due to~\eqref{eq:EuH1}. Hence the proof of~\eqref{eq:V=EVXT}.
   
We prove~\eqref{eq:Vbar*>EVbar} by contradiction. Suppose there exists a policy $\pi$ such that $E_s^{\pi}[Z_T] > V(s)$. Let $\pi'$ denote a new policy that follows $\pi$ until $T$ before switching to $\pi^*$ permanently. Then
\begin{align}
    \bar{v}^{\pi'}(s)
    &= E_s^{\pi'} [\bar{G}] \label{eq:vpi'G} \\
    &=E_s^{\pi}[Z_T] \label{eq:vpi'Z} \\
    &> V(s)
\end{align}
where~\eqref{eq:vpi'G} is by definition of $\bar{v}$ in~\eqref{eq:def_vbar},~\eqref{eq:vpi'Z} is due to~\eqref{eq:doob2}, contradicting the optimality of policy $\pi^*$. This completes the proof of Lemma~\ref{lm:Vbar=EVbar}.

\subsection{Proof of Lemma~\ref{lm:compare}}
\label{a:lemmacompare}

Using~\eqref{eq:V=EVXT}, we can write
\begin{align}
    V(s')
    &= E_{s'}^{\pi^*}[Z_T] \\ 
    &= E_{s'}^{\pi^*}\left[ Z_T 
    \cdot 1_{\{T<\infty\}}\right]
    +
    E_{s'}^{\pi^*}\left[ Z_T 
    \cdot 1_{\{T=\infty\}}\right]\\
    &=E_{s'}^{\pi^*}\left[ Z_T 
    \cdot 1_{\{T<\infty\}}\right]
    +
    E_{s'}^{\pi^*}\left[\bar{G}\cdot 1_{\{T=\infty\}}\right]\label{eq:V=EV+EG}
\end{align}
where \eqref{eq:V=EV+EG} is due to \eqref{eq:VXT_Tinf}.
For any $q$, if $T=\infty$, then $\overline{T} = \infty$ since $T\leq \overline{T}$ ($\overline{T}$ is the first time $C'$ arrives in $\mathscr{A}$). By \eqref{eq:reward}, $E_{s'}^{\pi^*}\left[\bar{G}\cdot 1_{\{\overline{T}=\infty\}}\right] = 0$. 
Thus,
\begin{align} \label{eq:VEY}
  V(s') =  E_{s'}^{\pi^*}\left[ Z_T 
  \cdot 1_{\{T<\infty\}}\right].
\end{align}

We denote the distribution of the arrival processes as $p(q)$, then using~\eqref{eq:Vbar*>EVbar},~\eqref{eq:VEY}, and $Z_T=V(X_T)$ for finite $T$, we write
\begin{align}
     V(s')
     &=\int_{T<\infty} V(x_{T}^{\pi^*}(s',q))d p(q) \\
     &\leq \int_{T<\infty} V(x_{T}^{\pi}(s,q))d p(q) \label{eq:intVs'<intVs} \\
     &\leq E_{s}^{\pi}[Z_T] \\ 
     &\leq V(s) \label{eq:EsVX_T<Vs}
\end{align}
where~\eqref{eq:intVs'<intVs} is by the assumption that $x_{T}^{\pi^*}(s',q) \preceq x_{T}^{\pi}(s,q)$ for $T<\infty$, and~\eqref{eq:EsVX_T<Vs} is due to \eqref{eq:Vbar*>EVbar}.

As a special case, let use define $T$ to be the smaller of $\overline{T}$ and the first time system $C$ arrives in $\mathscr{A}$, the preceding arguments show that $s\succeq s'$. Hence the proof is complete.

\subsection{Proof of Lemma~\ref{lm:h/l rule'}}
\label{a:lm_h/l}
Consider two systems, $C$ and $C'$, starting from states $s$ and $s'$, respectively. Let $C'$ adopt an optimal policy $\pi^*$. Suppose both systems receive the same arrival sequence. 
Let stopping time $T$ be defined as the first time $C'$ is in violation, i.e., $V(X_T)=1$ in $C'$ (for a sample path where $C'$ never arrives in violation, $T=\infty$).

We define a policy $\pi$ for $C$ to be such that $C$ 
replicates every action $C'$ takes. We note that $C$ can simulate $C'$ on the side to keep track of the current state and action of $C'$. Moreover, as long as every timer in $C$ is no less than the corresponding timer in $C'$, the block placement action $C'$ takes is also feasible in $C$, upon which
all timers in $C$ 
remain no less than their counterparts
in $C'$.
Therefore, $\pi$ is well-defined. 

If $T<\infty$, then at 
$T$, the two highest branches in $C$ have identical heights to the two highest branches in $C'$. Furthermore, the public height in $C$ is no greater than the public height in $C'$. Therefore, $C$ must also be in $\mathscr{A}$.
It follows from Lemma~\ref{lm:compare} that $s \succeq s'$.

\subsection{Proof of Lemma~\ref{lm:two_branches}}
\label{a:two_branches}

Consider two systems, $C$ and $C'$, starting at states $s'$ and $s$, respectively. Both systems have the same arrival processes. Suppose $C'$ follows an optimal policy $\pi^*$. We let $C$ adopt a policy $\pi$ defined as follows:
\begin{itemize}
    \item If $C'$ places a block on one of its two highest branches, $C$ replicates this action.
    \item Otherwise, $C'$ places the block on the same height as $C$, but on the second-highest branch of $C'$.
\end{itemize}
The policy $\pi$ is valid because, by following $\pi$ up to $T$, the public height of $C$ is no lower than that of $C'$.

It is not difficult to see that the highest branch of $C'$ is never lower than the highest branch of $C$; the second-highest branch of $C'$ is never lower than any branch of $C$ except its highest one. 
Consequently, 
when $C$ is in violation, so is $C'$. By Lemma~\ref{lm:compare}, $s' \succeq s$.

Conversely, it is easy to see that if $C'$ adopts an optimal policy, $C$ can always mirror $C'$'s actions. Hence $s \succeq s'$ is trivial.

\subsection{Proof of Lemma~\ref{lm:l_{mland d}-1 = 0}}
 \label{a:lemma l_{mland d}-1 = 0}

Let $s=[m,n,(l_1,\dots,l_d),I]$ and $s'=[m,n,(0,\dots,0,l_{m\land d},\dots,l_d),I]$. The goal here is to prove their equivalence.
By Lemma~\ref{lm:h/l rule'}, $s \succeq s'$. It suffices to show $s' \succeq s$. 
Consider two systems $C$ and $C'$ which start at $s$ and $s'$, respectively. Suppose both systems have the same arrival process. 
We define stopping time $T$ 
as the first time when one of the following occurs:
\begin{itemize}
    \item system $C$ reaches a state in $\mathscr{A}$;
    \item the timer on height $(m \land d)-1$ in system $C$ expires.
\end{itemize}
Let $C$ adopt the optimal policy $\pi^*$. We define policy $\pi$ for $C'$ as follows:
\begin{itemize}
    \item 
    When $\pi ^*$ places an H-block on a height lower than $m \land d$, $\pi$ places it on a highest branch on height $m \land d$.
    \item Under all other circumstances, $\pi$ replicates $\pi^*$.
\end{itemize}

It is not difficult to verify that policy $\pi$ is well-defined. Specifically, when $C$ places an H-block on a height lower than $m \land d$, all timers in both systems remain unchanged. Both systems have identical timers from height $m \land d$ onward until the stopping time $T$. Moreover, the two highest branches in both systems always have identical heights.

If the stopping time $T$ is reached when $C$ enters a state in $\mathscr{A}$, then the two highest branches in $C$ are higher than the public height and no lower than $k$. Since the two highest branches in $C'$ are at the same heights as in $C$, they are also no lower than $k$. The public height in $C$ is lower than $(m \land d) - 1$, then the public height in $C'$ is $(m \land d) - 1$, which is lower than the two highest branches in $C'$. This implies that $C'$ is also in $\mathscr{A}$ at $T$.

If the stopping time $T$ is reached when the timer of height $(m \land d) - 1$ expires in $C$, then $C$ and $C'$ have the same state representation with identical timers.

By Lemma~\ref{lm:compare}, it follows that $s \preceq s'$. The proof is complete.

\section{Proof of Propositions~ \ref{prop:posdelay0H} and~\ref{prop:delay0A}}
\label{a:proof_proposition12}
\subsection{Proof of Proposition~\ref{prop:posdelay0H}}
\label{a:posdelay0H}

By Lemma~\ref{lm:posHaction_0}, it suffices to identify an optimal action between the choices of $\underline{m+1}$, $\overline{n+1}$,  $\overline{d+1}$. We divide the proof into Case 1, $m<d$, and Case 2, $d\le m$.

We begin with Case 1.
Because the public height is higher than the lower branch, the new H-block can only be added to the higher branch with $\overline{d+1}$ or $\overline{n+1}$.
We further divide the case
into two subcases:

1) $m<d=n$. 
In this subcase, $\overline{d+1}=\overline{n+1}$, which causes
a transition to $[m,d+1,d+1]$. 
The proposition holds trivially.

2) $m<d<n$.
Action $\overline{d+1}$ causes a transition to $[m,d+1,n]$, while action $\overline{n+1}$ causes a transition to $[m,n+1,n+1]$. The former state can transition to a state as good as the latter one upon an H-block arrival by taking action $\overline{n+1}$. By Lemma~\ref{lm:A>=Hss'}, $[m,d+1,n] \succeq [m,n+1,n+1]$, so $\overline{d+1}$ is optimal in this subcase.

We have thus established Case 1 of the proposition. 

We next 
divide 
Case 2,
$d\le m \le n$,
into four subcases:

1) $d=m=n$. 
In this subcase, the two branches are of identical height and are as high as the highest public block, so $\overline{d+1}$, $\underline{m+1}$, and $\overline{n+1}$ are equivalent actions in the sense that they all transition to $[d+1,d+1,d+1]$.  Hence the proposition holds trivially in this case.

2) $d<m=n$.
In this subcase, the two branches are of identical height and higher than the highest public block, so $\underline{m+1}$ is equivalent to $\overline{n+1}$.
It suffices to show that action $\overline{d+1}$ is as good as action $\overline{n+1}$.  If $\overline{d+1}$ is taken, the system transitions to state $[n,d+1,n]$; if $\overline{n+1}$ is taken, the system transitions to state $[n,n+1,n+1]$.  The former state is superior because, upon an H-block arrival state $[n,d+1,n,H]$ can take action $\overline{n+1}$, 
transitioning to 
$[n,n+1,n+1]$.
Hence $\overline{d+1}$ is optimal in this subcase due to Lemma~\ref{lm:A>=Hss'}.

3) $d=m<n$.
The lower branch is as high as the highest public block in this subcase.  Action $\overline{d+1}$ causes a transition to state $[d,d+1,n]$, while action $\underline{m+1}$ causes a transition to a superior state $[d+1,d+1,n]$ according to Lemma~\ref{lm:h/l0}.  It remains to show that $\underline{m+1}$ is at least as good as $\overline{n+1}$, which causes a transition to $[d,n+1,n+1]$.  By Lemma~\ref{lm:A>=Hss'}, it suffices to show that state $[d+1,d+1,n]$ can transition to a state superior to $[d,n+1,n+1]$ upon a subsequent H-block arrival: Action $\overline{n+1}$ on $[d+1,d+1,n]$ transitions it to $[d+1,n+1,n+1]$, where $[d+1,n+1,n+1] \succeq [d,n+1,n+1]$ .

Hence $\underline{m+1}$ is optimal in this subcase.

4) $d<m<n$.
In this subcase, the two branches have different heights, and both are higher than the highest public block.  Action $\overline{d+1}$ cause transition to the same state $[m,d+1,n]$, while $\underline{m+1}$ causes a transition to $[m+1,m+1,n]$, and $\overline{n+1}$ causes a transition to $[m,n+1,n+1]$.  On $[m,d+1,n]$, with an H-block arrival, we can add it to either the lower branch to transition to $[m+1,m+1,n]$ or the higher branch to transition to $[m,n+1,n+1]$.  Hence $\overline{d+1}$ is optimal in this subcase too.

We have thus also established Case 2 of the proposition. 

\subsection{
Proof of Proposition~\ref{prop:delay0A}}
\label{a:delay0A}
By Lemma~\ref{lm:Aaction-0}, either $\underline{m+1}$ or $\overline{n+1}$ is optimal in $[m,d,n,A]$.

In case (1), where
$d\le m\le n$, 
if 
$m=n$, then the proposition holds trivially as 
$\underline{m+1}$ and $\overline{n+1}$ transition to the same state. 
Hence it suffices to establish case (1) under the assumption of $d\leq m<n$, where the choice is between $S = [m,d,n+1]$ upon action $\overline{n+1}$ and $S' = [m+1,d,n]$ upon $\underline{m+1}$.

    We next show that $S \succeq S'$.
    First, if $n+1 \geq k$, where $k$ stands for confirmation depth, then
    $V(S) = 1$, 
    thus $S \succeq S'$.
    Hence it suffices to consider the case of $n+1 < k$. 
    Consider two systems $C$ and $C'$, starting at state $S$ and $S'$, respectively, which have identical arrival sequences.
    We examine the evolution of their states, denoted as $X=[M,D,N,I]$ and $X'=[M',D',N',I]$, respectively, where $I$ can be omitted if $I=\emptyset$.
    At the beginning, $X=S$ and $X'=S'$, where the initial values of the height variables satisfy $M=m=M'-1$, $D=D'=d<M'$, and $N-1=n=N'$. With time, their values evolve according to the arrivals and their respective policies.
    
    Given an optimal policy $\pi^*$ for $C'$, which always takes action $\overline{D'+1}$ for an H-arrival (cf.~Proposition~\ref{prop:posdelay0H}) and always takes $\underline{M'+1}$ or $\overline{N'+1}$ for an A-arrival, we define 
    a corresponding policy $\pi$ for $C$ 
    as follows: 
    \begin{enumerate}[i)]
        \item If $I=H$, $D'=M'-1$, 
        then $\pi$ takes  $\underline{M+1}$.
        
        \item If $I=H$, $D'\neq M'-1$, 
        then $\pi$ takes  $\overline{D+1}$.

        \item If $I=A$, $M'<N'$, and $\pi^*$ takes  $
        \underline{M'+1}$,
        then 
        $\pi$ takes  $\underline{M+1}$.

        \item If $I=A$ and $\pi^*$ takes   
        $\overline{N'+1}$, then $\pi$ takes  
        $\underline{M+1}$.
    \end{enumerate}

    We define a stopping time $T$ as the first time when (i) or (iv) is executed.
    To verify $S\succeq S'$, it remains to check that $X_T\succeq X'_T$ according to Lemma~\ref{lm:compare}. 
    Before $T$, only (ii) and (iii) are executed, and it is not difficult to verify that $(M,D,N)$ rise in the same manner as $(M',D',N')$, which we formally put as the following:
    
    \begin{lemma}\label{lm:ind_claim}
    From the beginning of $C$ and $C'$, we always have $N'+1=N$ and $M'-1=M>D=D'$ before stopping time $T$.
    \end{lemma}

\begin{proof}
    The assertion of Lemma~\ref{lm:ind_claim} holds by definition from the beginning of $C$ and $C'$. Before $T$, upon an H-arrival, $\pi$ must take (ii) $\overline{D+1}$ ($\pi^*$ takes $\overline{D'+1}$); upon an A-arrival, $\pi$ must take (iii) $\underline{M+1}$ ($\pi^*$ takes $\underline{M'+1}$).
    Suppose the assertion also hold
    after the $i$-th arrival, and the $(i+1)$-st arrival $I$ is before $T$. 
    Note that the lower branch is higher than the public height in both systems $C$ and $C'$.
    If
    $I=H$, 
    system $C'$ makes transition 
    $X' \rightarrow[M',D'+1,N']$ and system $C$ makes transition $X \rightarrow [M,D+1,N]$.
    For $I=A$, since 
    $M'<N'$, 
    system $C'$ makes transition $X'\rightarrow [M'+1,D',N']$, and $C$ makes transition $X\rightarrow [M+1,D,N]$.
    It is easy to see that the public height and the lower and higher branches rise in the same manner in both systems. Hence the assertion in the lemma holds at all times before $T$.
    \end{proof}

    We next compare the states at $T$.
    If (iv) is executed, then $C'$ makes a transition $X' \rightarrow [M',D',N'+1]$ and $C$ makes a transition $X \rightarrow [M+1,D,N]$.
    If (i) is executed when $D'=M'-1=M$, then $X' \rightarrow [M',D'+1,N']$ and $X \rightarrow [M+1,D+1,N]$.
    In both cases, $X\succeq X'$ at time $T$ by Lemma~\ref{lm:ind_claim}, which implies the desired result
    $S\succeq S'$, i.e., $\overline{n+1}$ is superior in state $[m,d,n,A]$. 
    We have thus established case (1) 
    of Proposition~\ref{prop:delay0A}.

    As for case (2), the state is $[m,d,n,A]$ with $m<d\le n$. By Lemma~\ref{lm:Aaction-0}, the two potential optimal actions are $\underline{m+1}$, causing a transition to $S=[m+1,d,n]$, and $\overline{n+1}$, causing a transition to $S'=[m,d,n+1]$. Given an optimal policy $\pi^*$ for $S'$, we define the following policy $\pi$ for $S$:
    \begin{enumerate}[i)]
        \item If $I=H$, 
        ${\pi}$ takes  $\overline{D+1}$.

        \item If $I=A$ and 
         ${\pi^*}$ takes 
         $\overline{N'+1}$, then 
         ${\pi}$ takes 
         $\overline{N+1}$.
        
        \item If $I=A$, $M'<N'$, and 
        ${\pi^*}$ takes 
        $\underline{M'+1}$, then 
        ${\pi}$ takes 
        $\overline{N+1}$.
    \end{enumerate}

    Let $T$ be the first time when (iii) is executed. Since $M+1<D+1$ from the beginning, $\pi$'s action never places any block on the lower branch until $T$. Before $T$, (i) and (ii) guarantee that $(M,D,N)$ always increase the same way as $(M',D',N')$. At $T$, $M'$ and $N$ rise to yield
    $X \asymp X'$. By Lemma~\ref{lm:compare}, $S\succeq S'$, which implies that $\underline{m+1}$ is superior in state $[m,d,n,A]$. 
    We have thus established case (2) of Proposition~\ref{prop:delay0A}.

\section{The Case of Arbitrary Delays}
\label{a:pow_delta}

\subsection{Proof of Proposition~\ref{cl:ahead}}
\label{a:A-ahead}

\subsubsection{H-arrivals}

We first consider the case of an H-arrival. By Lemma~\ref{lm:must take the form of}, we must have $d\le m\le n$ and $l_d = l\in[0,\Delta]$ in the ahead state, which can be divided into cases (i)--(iii) in the proposition. Due to Lemma~\ref{lm:potential_action_H>0}, in the ahead state,
    at least one of the following four actions (along with applicable conditions and state transitions) is optimal:
    \begin{enumerate}[I)]
    \item $\overline{n+1}$, causing a transition to 
    $B'=[m,n+1,n+1,(\Delta,\dots,\Delta)]$ if $d<m$, or
    $B''=[m,n+1,n+1,(l,\Delta,\dots,\Delta)]$ if $d=m$.
    Let $B=[m,n+1,n+1,(0,\Delta,\dots,\Delta)]$. By Lemma~\ref{lm:lm=0}, $B'=B''=B$;

    \item $\underline{m+1}$ (applicable only if $m<n$), causing a transition to $C=[m+1,m+1,n,(\Delta)]$;

    \item $\overline{d}$ (applicable only if $l>0$), causing the system to transition to state $E=[m,d,n,(l)]$;
        
    \item $\overline{d+1}$ (applicable only if $l=0$ and $d<n$),  
    causing a transition to $F=[m,d+1,n,(\Delta)]$ if $d<m$,
    or $F'=[m,d 
    +1,n,(0,\Delta)]$ if $d=m<n$,
    or $F''=[n,n+1,n+1,(0,\Delta)]$ if $d=m=n$.
\end{enumerate}

We divide the proof into the following four cases:
\begin{enumerate}[a)]
    \item $d=m=n$.
    We further divide this case into two sub-cases:
    
    (a1) $l=0$. 
    Since $d=n$, $\overline{d+1}$ is optimal.
    
    (a2) $l>0$. It suffices to show that $\overline{d}$ is optimal. Only actions $\overline{n+1}$ and $\overline{d}$ are applicable. We prove that $\overline{d}$ is superior to $\overline{n+1}$ by showing that
    the subsequent states due to those two options satisfy
    \begin{align}
        E=[d,d,d,(l)] \succeq B=[d,d+1,d+1,(0,\Delta)] .
    \end{align} 
    To this end, by Lemma~\ref{lm:A>=Hss'}, it suffices to describe an action upon an H-block arrival as a subsequent block after state $E$ that causes a transition to a state as good as $B$.
    Consider action $\overline{d+1}$, causing transition to $[d,d+1,d+1,(l',\Delta)] \succeq B$ where $l'\in [0,l)$. 
    Hence $\overline{d}$ is optimal in this case.

    \item $d<m=n$. It suffices to show that $\overline{d}$ is optimal if $l > 0$ and $\overline{d+1}$ is optimal if $l=0$ in this case.
    Note that both $\overline{n+1}$ and either one of $\overline{d}$ and $\overline{d+1}$ are applicable depending on $l$. To determine which one is optimal, we compare their corresponding resulting states $B$, $E$, and $F$.
    By Lemma~\ref{lm:h/l rule}, $E \succeq F$. We want to show that $F \succeq B$ to prove $\overline{d}$ and $\overline{d+1}$ are superior to $\overline{n+1}$.  To this end, by Lemma~\ref{lm:A>=Hss'}, it suffices to describe an action upon an H-block arrival as a subsequent block after state $F$ that causes a transition to a state as good as $B$.
    Consider action to place the new block at height $n+1$, causing a transition to $[n,n+1,n+1,(l',\Delta,\dots,\Delta)]\asymp B$ by Lemma~\ref{lm:lm=0}, where $l' \in [0,l)$. 
    Thus, under this circumstance, $\overline{d}$ is superior if $l>0$, otherwise $\overline{d+1}$ is superior. We have thus proved the proposition in this case.

    \item  $d=m<n$. 
    It suffices to show that $\overline{d}$ is optimal if $l > 0$ and $\underline{m+1}$ is optimal if $l=0$ in this case.
    Note that $\overline{n+1}$ and $\underline{m+1}$ are always applicable, and either one of $\overline{d}$, $\overline{d+1}$ is applicable depending on $l$. We next compare their resulting states $B$, $C$, $E$, and $F'$.
    By Lemma~\ref{lm:h/l rule}, $E \succeq F'$. In the following, we want to show that $C \succeq F'$ and $C \succeq B$ to say $\underline{m+1}$ is superior if $l =0$ ($E$ is not applicable); 
    and $E \succeq C$ to say that $\overline{d}$ is superior if $l >0$ ($E$ is applicable).
    First, by Lemma~\ref{lm:h/l rule},
    $
        C
        \succeq [m,m+1,n,(0,\Delta)] 
        = F' .
    $

    To show $C \succeq B$, by Lemma~\ref{lm:A>=Hss'}, it suffices to describe an action upon an H-block arrival as a subsequent block after state $C$, causing a transition to a state no worse than $B$.
    Consider the action $\overline{n+1}$, causing a transition to $[m+1,n+1,n+1,(l',\Delta,\dots,\Delta)]$ for some $l' \in [0,\Delta)$. We have
    \begin{align}
        C
        &\succeq [m+1,n+1,n+1,(0,\Delta,\dots,\Delta)]\label{eq:G>} \\
        &\succeq B\label{eq:>B}
    \end{align}
    where~\eqref{eq:G>} is due to Lemma~\ref{lm:lm=0} and~\eqref{eq:>B} is due to Lemma~\ref{lm:h/l rule}.

    To show $E \succeq C$ when $l >0$, by Lemma~\ref{lm:A>=Hss'}, it suffices to describe an action upon an H-block arrival as a subsequent block after state $E$, causing a transition to a state no worse than $C$.
    Consider the action $\underline{m+1}$, causing a transition to $[m+1,m+1,n,(\Delta)]\asymp C$.
    
    We have thus proved the proposition in Case 3.

    \item $d<m<n$. 
    It suffices to show that $\overline{d}$ is optimal if $l > 0$ and $\overline{d+1}$ is optimal if $l=0$ in this case.
    $\overline{n+1}$ and $\underline{m+1}$ are always applicable, and one of $\overline{d}$, $\overline{d+1}$ is applicable depending on $l$. We compare their resulting states $B$, $C$, $E$, and $F$.
    By Lemma~\ref{lm:h/l rule}, $E \succeq F$. It then suffices to show that $F \succeq C$ and $F \succeq B$ to prove that $\overline{n+1}$ and $\underline{m+1}$ are inferior.

    To show $F \succeq C$, by Lemma~\ref{lm:A>=Hss'}, it suffices to describe an action upon an H-block arrival as a subsequent block after state $F$, causing a transition to a state as good as than $C$. 
     Consider the action $\underline{m+1}$, causing a transition to $[m+1,m+1,n,(\Delta)]=C $.
     
     To show $F \succeq B$, by Lemma~\ref{lm:A>=Hss'}, it suffices to describe an action upon an H-block arrival as a subsequent block after state $E$, causing a transition to a state as good as than $B$. 
     Consider the action $\overline{n+1}$, causing a transition to $[m,n+1,n+1,(l,\Delta,\dots,\Delta)]\asymp B $ due to Lemma~\ref{lm:lm=0} if $m=d+1$, or into $[m,n+1,n+1,(\Delta,\dots,\Delta)]= B $ if $m > d+1$.
\end{enumerate}
    
Thus the proof of the proposition for H-arrivals is completed.

\subsubsection{A-arrivals}

We now consider the case of an A-arrival. By Lemmas~\ref{lm:must take the form of} and~\ref{lm:potential_action_A>0}, in the ahead state $[m,d,n,(l),A ]$, we have $d \leq m \leq n$ and at least one of the following actions is optimal:
\begin{itemize}
    \item $\overline{n+1}$, causing a transition to $s = [m,d,n+1,(l)]$;
    \item $\underline{m+1}$, applicable if $m<n$, causing a transition to $s' = [m+1,d,n,(l)]$. 
\end{itemize}
If $m=n$, both $\underline{m+1}$ and $\overline{n+1}$ cause a transition to the same state. Hence it suffices to assume $m<n$ and
show that $s\succeq s'$, so that $\overline{n+1}$ is superior to $\underline{m+1}$. 

Consider two systems, $\mathcal{S}$ and $\mathcal{S}'$, 
which begin in states $s$ and $s'$, respectively, at time 0. Suppose that the two systems have identical arrival processes. Let $X = [M,D,N,(L_{M \land D},\dots,L_{D}),I]$ denote the state of system $\mathcal{S}$. Let $X' = [M',D',N',(L'_{M' \land D'},\dots,L'_{D'}),I]$ denote the state of $\mathcal{S}'$. While not explicit in notation, these states are random processes in time.

Suppose system $\mathcal{S'}$ adopts a specific optimal policy $\pi^*$ which satisfies the following: Upon each H-arrival in an ahead state, $\pi^*$ takes the optimal action prescribed by Proposition~\ref{cl:ahead}; upon each A-arrival, $\pi^*$ takes either $\underline{M'+1}$ or $\overline{N'+1}$, as one of them must be optimal by Lemma~\ref{lm:potential_action_A>0}.

We define the following policy $\pi$ for system $\mathcal{S}$, depending on the instantaneous state $X'$ and the action $\pi^*$ takes:
\begin{enumerate}[i)]
    \item If $I = H$, $D' = M' -1$, and $\pi^*$ takes action $\overline{D'+1}$, then $\pi$ takes action $\underline{M +1}$;
    \item If $I = H$, $D' < M' -1$, and $\pi^*$ takes action $\overline{D'}$ (or\ $\overline{D'+1}$), then $\pi$ takes the same action; 
    \item If $I = A$ and 
    $\pi^*$ takes action $\underline{M'+1}$, then $\pi$ takes action $\underline{M+1}$;
    \item If $I = A$ and $\pi^*$ takes action $\overline{N'+1}$, then $\pi$ takes action $\underline{M+1}$.
\end{enumerate}
We define a stopping time $T$ to be the first time when (i) or (iv) is executed or $\mathcal{S}'$ arrives in $\mathscr{A}$, whichever occurs first.

At time 0, $D'=d\le m=M'-1$ by assumption of the lemma, so $\mathcal{S}'$ begins in an ahead state. According to the optimal actions prescribed by Proposition~\ref{cl:ahead}, $D'\le M'-1$ continues to hold until (i) is executed. Thus until $T$, it suffices to consider $\overline{D'}$ and $\overline{D'+1}$ as potential optimal actions for $\pi^*$. Thus $\pi$ for $\mathcal{S}$ is well-defined.

Before $T$, only (ii) and (iii) are executed, so that during the entire interval $[0,T)$, we have $D=D'$, $M=M'-1$, $N=N'+1$, and all timers 
in $\mathcal{S}$ are equal to their 
counterparts in $\mathcal{S}'$. 
In particular, (ii) is only executed if $D'<M'-1$, yielding $D=D'\le M'-1=M$, both systems remain in ahead states.
At time $T$, one of the following events has to occur:
\begin{itemize}
    \item 
(i) is executed. In this case, the public height in $\mathcal{S'}$ rises to $D'=M'$, and the H-block in $\mathcal{S}$ extends the lower branch, yielding $D=M=M'=D'$. Since
$N=N'+1$, 
we have $X_T \succeq X'_T$.

    \item (iv) is executed. In this case, the A-block in $\mathcal{S}$ extends the lower branch, whereas the A-block in $\mathcal{S'}$ extends the higher branch, yielding $M=M'$, $N=N'$, \and $D=D'$ with no timer changes, so that $X_T \asymp X'_T$.
    
    \item $\mathcal{S}'$ arrives in $\mathscr{A}$. In this case, $M' \geq k$. 
    Since $M=M'-1\ge k-1$ and $M\ge D$, $\mathcal{S}$ is guaranteed to arrive in $\mathscr{A}$ by taking action $\underline{M+1}$ upon the first block arrival after $T$, yielding $M\ge k$, so that $\mathcal{S}$ also arrives in $\mathscr{A}$.
    Thus with probability $1$, $V(X'_T)=V(X_T)=1$.
\end{itemize}
We have thus proved $X_T \succeq X'_T$ in all cases. By Lemma~\ref{lm:compare}, $s \succeq s'$, hence Proposition~\ref{cl:ahead} is established.

\subsection{Proof of Propositions~\ref{cl:on-time} and~\ref{cl:behind}} 
\label{a:A-behind}

We note that Proposition~\ref{cl:ahead} is proved independently of Propositions~\ref{cl:on-time} and~\ref{cl:behind} because the state remains ahead within the proof. In this subsection, we prove Propositions~\ref{cl:on-time} and~\ref{cl:behind} together as we need to address state transitions between on-time and behind states within their proof.

\subsubsection{\bf{Proof of Proposition~\ref{cl:behind} with an A-Arrival}}
\label{s:behindA}

By Lemma~\ref{lm:potential_action_A>0}, in the behind state $[m,d,n,(0,0,l_{m+2},\dots,l_d),A]$, at least one of the following actions is optimal:
\begin{itemize}
    \item $\underline{m+1}$, causing a transition to $s = [m+1,d,n,(0,l_{m+2},\dots,l_d)]$;
    \item $\overline{n+1}$, causing a transition to $s' = [m,d,n+1,(0,0,l_{m+2},\dots,l_d)]$.
\end{itemize}
In order to show that $\underline{m+1}$ is superior to $\overline{n+1}$, it suffices to show that $s\succeq s'$. The proof uses the technique developed in Appendix~\ref{a:A-ahead}.

Consider two systems, $\mathcal{S}$ and $\mathcal{S}'$, which begin in states $s$ and $s'$, respectively, at time 0. Suppose the two systems have identical arrival processes. Suppose system $\mathcal{S}'$ adopts a specific optimal policy $\pi^*$ defined the same way as in Appendix~\ref{a:A-ahead} (based on Proposition~\ref{cl:ahead} and Lemma~\ref{lm:potential_action_A>0}).
Let state processes $X$ and $X'$ be defined the same way as in Sec.~\ref{a:A-ahead}. We define the following policy $\pi$ for system $\mathcal{S}$:
\begin{enumerate}[i)]
    \item If $\pi^*$ takes action $\overline{N'+1}$, then $\pi$ takes action $\overline{N+1}$;
    \item If $\pi^*$ takes action $\underline{m+1}$, then $\pi$ takes action $\overline{N+1}$;
    \item Otherwise, let $\pi$ takes the same action as $\pi'$.
\end{enumerate}
We define a stopping time $T$ to be the first time when (ii) is executed.

At time 0, $M=m+1$, $M'=m$, $N=n$, and $N'=n+1$. It is easy to verify that, before $T$, both systems extend their respective higher branches simultaneously, and neither system extends its lower branch. Thus during $[0,T)$, we always have $N=N'-1$, $M=m+1=M'+1$, and all timers in system $\mathcal{S}$ are at least as large as their counterparts in system $\mathcal{S}'$, so $\pi$ is well-defined.

Since $l_{m+1}=0$, to execute (ii), $I$ has to be A-block arrival. Thus, it will not affect height timers in system $\mathcal{S}$. Once (ii) is executed at stopping time $T$, $M' = M = m+1$, $N' = N$, $D' \geq D$, and height timers in system $\mathcal{S}$ remains at least as large as in system $\mathcal{S}'$. 
Hence, by Lemma~\ref{lm:h/l rule}, $X_T \succeq X'_T$. 
Note that system $\mathcal{S}'$ will not reach $\mathscr{A}$ before stopping time $T$.
So $s \succeq s'$ by Lemma~\ref{lm:compare}, and action $\underline{m+1}$ is optimal for behind state with A-arrival.



\subsubsection{\bf{Proof by Induction}}
For any given state
\begin{align}
z=[m,n,d,(l_{m\land d},\dots,l_{d}),I],    
\end{align}
we define
\begin{align}
    K(z) =
    \begin{cases}
        2k - m - n, \quad & \text{if } n<k \\
        0, & \text{if } n\ge k.
    \end{cases}
\end{align}
If $n<k$, $K(z)$ denotes the minimum number of blocks needed for both branches to reach the confirmation depth $k$.
\begin{proposition}\label{pr:ABCD}
    The following statements are true for every $J\in\{0,1,2,\dots\}$:
\begin{enumerate}
    \item [$(A_J)$:] Let $s=[m,n,n,(l_{m},\dots, l_{n})]$ and $s'=[m,n',n',(l'_{m},\dots, l'_{n'})]$, where $m \leq n \leq n'$ and $l_i \ge l'_i$ for every $i \in \{m,\dots, n\}$. If $K(s)\le J$, then $s\succeq s'$.

    \item [$(B_J)$:] For behind state $s=[m,d,n,(0,0,l_{m+2},\dots,l_d),H]$ with $K(s)\le J$, action $\overline{d}$ is optimal if $l_d>0$; action $\overline{d+1}$ is optimal if $l_d=0$.
    
    \item [$(C_J)$:] For on-time state $s=[m,d,n,(l_m, 
    l_{m+1},\dots,l_d),A]$ 
    with $K(s)\le J$, action $\underline{m+1}$ is optimal.

    \item [$(D_J)$:] For on-time state $s=[m,d,n,(l_m,l_{m+1},\dots,l_d),H]$ with $K(s)\le J$, action $\underline{m+1}$ is optimal.
\end{enumerate}
\end{proposition}

Before proving Proposition~\ref{pr:ABCD}, we show that it implies Propositions~\ref{cl:on-time} and~\ref{cl:behind}: If $(C_J)$ and $(D_J)$ hold for $J=0,1,\dots$, then extending the lower branch is optimal in an on-time state, establishing Proposition~\ref{cl:on-time}. If $(B_J)$ holds for every $J\in\{0,1,\dots\}$, then the case of H-arrival in Proposition~\ref{cl:behind} is established. Together with the A-arrival case proved in Appendix~\ref{s:behindA}, Proposition~\ref{cl:behind} is established.

In Secs.~\ref{s:proofi}--\ref{s:proofv}, we prove
Proposition~\ref{pr:ABCD} by establishing the following five claims, respectively, for every $J\in\mathbb{N}$:
\begin{enumerate}[I)]
    \item \label{item:I} $(A_0)$, $(B_0)$, $(C_0)$, and $(D_0)$ are true. 
    \item \label{item:II} Together, $(A_{J-1})$ 
    and $(C_{J-1})$ 
    imply $(A_J)$.
    \item \label{item:III} Together, $(B_{J-1})$, $(C_{J-1})$, and $(D_{J-1})$ imply $(D_J)$.
    \item \label{item:IV} $(A_J)$ implies $(B_J)$.
    \item \label{item:V} Together, $(A_J)$, $(B_J)$, $(C_{J-1})$, and $(D_{J-1})$ imply $(C_J)$.
\end{enumerate}
In particular, it is easy to see that these claims imply that, if $(A_{J-1})$, $(B_{J-1})$, $(C_{J-1})$, and $(D_{J-1})$ hold, then $(A_J)$, $(B_J)$, $(C_J)$, and $(D_J)$ hold. 
Hence Proposition~\ref{pr:ABCD} is established by induction on $J$.

\subsubsection{\bf{Proof of \eqref{item:I}}}
\label{s:proofi}

With $J=0$, we have $K(S)=0$, which implies that the higher branch is no lower than $k$ ($n\geq k$).  The higher branch already confirms a block on height 1.  

\begin{lemma}\label{lm:n>=k}
    Let
    \begin{align}
        s &= [m,d,n,(l_m,\dots,l_d),I] \\
        s' &= [m,d',n',(l'_{m},\dots,l'_{d'}),I] .
    \end{align}
    We have $s\succeq s'$ as long as the following conditions hold:
    \begin{itemize} 
        \item $n\geq m$ and $n'\geq m$;
        \item $n \geq k$ and $n' \geq k$;
        \item $d \leq d'$ and $l_i \geq l'_i$, for every $i\in\{m, \dots, d \}$.
    \end{itemize}
\end{lemma}

\begin{proof}
If $n\geq n'$, then
\begin{align}
    s
    &\succeq [m,d,n',(l_m,\dots,l_d),I] \label{eq:s<m,d,n'} \\
    &\succeq s' \label{eq:mdns}
\end{align}
where~\eqref{eq:s<m,d,n'} is due to~\eqref{eq:m'n'>mn} and~\eqref{eq:mdns} is due to~\eqref{eq:l'>l}. Moreover,
if $m = n \ge k$, then $V(s) = 1$ as the state $s$ is in violation. Thus, in the following proof, we only consider cases where $m<n<n'$.

Consider two systems $\mathcal{S}$ and $\mathcal{S'}$, which begin in states $s$ and $s'$,
respectively, at time 0. Suppose their arrival processes are identical. Let $X = [M,D,N,(L_{M \land D},\dots,L_{D}),I]$ denote the state of system $\mathcal{S}$. Let $X' = [M',D',N',(L'_{M' \land D'},\dots,L'_{D'}),I]$ denote the state of system $\mathcal{S'}$.
Let system $\mathcal{S'}$ take an optimal policy $\pi^*$.
Define the policy $\pi$ system $\mathcal{S}$ takes as follows:
\begin{enumerate}[i)]
    \item If $\pi^*$ takes action $\overline{r}$ with $r> N+1$, $\pi$ takes action $\overline{N+1}$;
    \item otherwise, $\pi$ takes the same action as $\pi^*$ takes.
\end{enumerate}
Define stopping time $T$ to be the first time one of the following occurs: (a) $L'_{N'} = 0$, (b) $N=N'$, or  (c) system $\mathcal{S}'$ arrives in $\mathscr{A}$, whichever occurs first.

At time 0, $M' = M$, and each timer in $\mathcal{S}$ is no smaller than its counterpart in $\mathcal{S}'$. According to the policies, each new block is placed no lower in $\mathcal{S'}$ than in $\mathcal{S}$, and every time the lower branch is extended in $\mathcal{S}'$, the lower branch in $\mathcal{S}$ is also extended. Thus, during $[0,T)$, $M' = M<N$, $L_i \ge L'_i$ for $i = M \dots, D$, which implies that the public height of $X'$ is never lower than that of $X$. At time $T$, one of the following events has to occur:
\begin{itemize}
    \item $L'_{N'} = 0$. In this case, 
    all heights of $X'_T$ are public. The higher branch of $X'_T$ is $(N'-N)$ higher than the higher branch of $X_T$. We invoke Lemma~\ref{lm:A>=Hss'} $(N'-N)$ times to obtain a state $X''$ that satisfies $X''\preceq X_T$: Specifically, all $(N'-N)$ H-blocks extend the higher branch to yield $X''$, whose higher branch is of height $N'$. 
    Since $X''$ and $X'_T$ share the same branch heights and each the timers of $X''$ are no smaller than their counterparts of $X'_T$, we have $X'_T \preceq X''$. So $X_T \succeq X'_T$ in this case.
    
    \item $N=N'$. In this case $X_T$ catches up with $X'_T$ in branch heights.
    By Lemma~\ref{lm:h/l rule}, $X_T \succeq X'_T$. 
    
    \item $\mathcal{S}'$ arrives in $\mathscr{A}$. In this case, $M=M'$ is no lower than the public heights of both $X_T$ and $X'_T$, which implies that $\mathcal{S}$ has also arrived in $\mathscr{A}$, i.e.,
    $V(X'_T) = V(X_T)=1$.
\end{itemize}

As a result, $X_T \geq X'_T$. By Lemma~\ref{lm:compare}, $s \succeq s'$.
\end{proof}

We next use Lemma~\ref{lm:n>=k} to prove $(A_0)$, $(B_0)$, $(C_0)$ and $(D_0)$ of Proposition~\ref{pr:ABCD}.

Evidently, $(A_0)$ holds as a special case of Lemma~\ref{lm:n>=k} with $d=n$, $d'=n'$. 

By Lemma~\ref{lm:potential_action_H>0}, for a behind state $[m,d,n,(0,0,l_{m+2},\dots,l_{d}),H]$ with $n\geq k$, there are three potential optimal actions: $\overline{d}$, $\overline{d+1}$ (applicable only if $d<n$), or $\overline{n+1}$. (Note that $\underline{m+1}$ is not applicable since $l_{m+1}=0$.) Action $\overline{d}$ transitions the state to $s=[m,d,n,(0,0,l_{m+2},\dots,l_{d})]$. Action $\overline{d+1}$ (under condition $d<n$) transitions the state to $s' = [m,d+1,n,(0,0,l_{m+2},\dots,l_{d+1})]$. Action $\overline{n+1}$ transitions the state to $s'' = [m,n+1,n+1,(0,0,l_{m+2},\dots,l_{n+1})]$. By Lemma~\ref{lm:n>=k}, we have $s\succeq s'\succeq s''$, which proves $(B_0)$.

By Lemma~\ref{lm:potential_action_A>0}, for an on-time state $[m,d,n,(l_m,l_{m+1},\dots,l_d),A]$, there are two potential optimal actions $\overline{n+1}$ and $\underline{m+1}$, which lead to states $s = [m,d,n+1,(l_m,l_{m+1},\dots,l_d)]$ and $s' = [m+1,d,n,(l_{m+1},\dots,l_d)]$, respectively.  We have
\begin{align}
    s'
    &\succeq [m,d,n,(0,l_{m+1},\dots,l_d)] \label{eq:s's1} \\
    &\asymp [m,d,n,(l_m,l_{m+1},\dots,l_d)] \label{eq:s's2} \\
    &\succeq s \label{eq:s's3}
\end{align}
where~\eqref{eq:s's1} is due to Lemma~\ref{lm:h/l rule}, \eqref{eq:s's2} is due to Lemma~\ref{lm:lm=0}, and \eqref{eq:s's3} is due to Lemma~\ref{lm:n>=k}. Hence, ($C_0$) is established.

By Lemma~\ref{lm:potential_action_H>0}, for an on-time state $[m,d,n,(l_m,l_{m+1},\dots,l_d),H]$, there are three potential optimal actions: $\overline{d}$, $\underline{m+1}$ and $\overline{n+1}$. (Since $l_d\ge l_{m+1}>0$, $\overline{d+1}$ is not applicable.) It is obvious that $\underline{m+1}$ is superior to $\overline{d}$ by Lemma~\ref{lm:h/l rule}.
We note that $\underline{m+1}$ causes a transition to $s = [m+1,d,n,(l_{m+1},\dots,l_d)]$, and $\overline{n+1}$ causes a transition to $s' = [m,n+1,n+1,(l_m,l_{m+1},\dots,l_d,\Delta,\dots,\Delta)]$. The two states satisfy
\begin{align}
    s
    &\succeq [m,d,n,(l_m,l_{m+1},\dots,l_d)] \label{eq:s's11}\\
    &\succeq s', \label{eq:s's12}
\end{align}
where~\eqref{eq:s's11} is due to Lemma~\ref{lm:h/l rule} and~\eqref{eq:s's12} is due to Lemma~\ref{lm:n>=k}. Hence, $\underline{m+1}$ is superior to $\overline{n+1}$, so $(D_0)$ is established.

Thus, the proof of (I) is completed.

\subsubsection{\bf{Proof of (II)}}
\label{s:proofii}

Suppose $(A_{J-1})$,$(B_{J-1})$, $(C_{J-1})$, and $(D_{J-1})$ hold. Let $s = [m,n,n,(l_m,\dots,l_n)]$ and $s' = [m,n',n',(l'_m,\dots,l'_{n'})]$, where $m \leq n\leq n'$, $l'_i \leq l_i$ for every $i \in \{m,\dots,n\}$, and $K(s)=J$. We prove $(A_J)$ holds by showing that $s \succeq s'$.

In the special case of $n'=n$, $(A_J)$ stands trivially by Lemma~\ref{lm:h/l rule}.

In the special case of $m=n<n'$, we have $s = [m,m,m,(l_m)] \succeq [m,m+1,m+1,(l_m,\Delta)]$ due to Lemma~\ref{lm:A>=H}, because the latter state is the result of adding an H-block on height $m+1$ of state $s$. We can further write $s\succeq s'$ due to assumption $(A_{J-1})$.

In the remainder of this proof, we assume $m<n<n'$.

Consider two systems, $\mathcal{S}$ and $\mathcal{S}'$, which begin in states $s$ and $s'$, respectively, at time 0. Supppose their arrival processes are identical. At any given time, let $X = [M,D,N,(L_{M \land D},\dots,L_{D}),I]$ denote the state of system $\mathcal{S}$ and let $X' = [M',D',N',(L'_{M' \land D'},\dots,L'_{D'}),I]$ denote the state of system $\mathcal{S}'$. Let system $\mathcal{S}'$ take a specific optimal policy $\pi^*$ that follows the action prescribed by $(C_{J-1})$ when the state satisfies $K(X')\le J-1$ as well as Proposition~\ref{cl:behind} upon an A-arrival (proved in Appendix~\ref{s:behindA}).
We define the policy $\pi$ system $\mathcal{S}$ takes as follows:
\begin{enumerate}[i)]
    \item If $\pi^*$ takes action $\overline{r}$ where $r >N$, $\pi$ takes $\overline{N+1}$;
    \item Otherwise, $\pi$ takes the same action as $\pi^*$.
\end{enumerate}
Define a stopping time $T$ as the first time action $\underline{m+1}$ is executed.

Before $T$, both systems stay on-time or behind. Since $n<n'$ and $K(s) \leq J$, we have $K(s') \leq J-1$.
By Proposition~\ref{cl:behind} with A-arrival and $(C_{J-1})$, if there were an A-arrival during $[0,T)$, the A-block would extend the lower branch. In other words, there can be no A-arrivals during $[0,T)$.
Thus, during $[0,T)$, $M'=M=m$, $N \leq N'$, $N=D$, $N'=D'$, and $L'_{i'} \leq L_{i}$ for $i = m,\dots, N$. In particular, when (i) is executed, i.e., $\pi^*$ takes an action $\overline{r}$ with $r>N$, the higher branch is extended in $\mathcal{S}$, where the resulting $N$ is no higher than $N'$.

At time $T$, when $\underline{m+1}$ is executed (in both systems), where $m+1\le D'$, no timer decreases in either system.
So we have $X_T = [m+1, N, N,(L_{m+1},\dots, L_N)]$ and $X'_T = [m+1,N',N',(L'_{m+1},\dots, L'_{N'})]$. Evidently, $K(X_T)\leq J-1$.
Due to $(A_{J-1})$, we have $X_T \succeq X'_T$. Since $m<k$, system $\mathcal{S}'$ can not arrive in $\mathscr{A}$ before $T$, Lemma~\ref{lm:compare} can be invoked to establish that $s \succeq s'$. Hence, $(A_J)$ is established. We have thus proved (II).

\subsubsection{\bf{Proof of (III)}}
\label{s:proofiii}

By Lemma~\ref{lm:potential_action_H>0}, in an on-time state with an H-arrival, there are three potential optimal actions $\overline{d}$, $\overline{n+1}$ and $\underline{m+1}$, where $\overline{d+1}$ is not applicable since $l_d>0$. Neither $\underline{m+1}$ and $\overline{d}$ decreases any timer. Since $\underline{m+1}$ also extend the lower branch, we have $\underline{m+1}$ is superior to $\overline{d}$ due to Lemma~\ref{lm:h/l rule}. It suffices to prove $\underline{m+1}$ is superior to $\overline{n+1}$, which is equivalent to that their resulting states 
$s=[m+1,d,n,(l_{m+1},\dots,l_d)]$ 
is superior to
$s'=[m,n+1,n+1,(l_m,l_{m+1},\dots,l_d,\Delta,\dots,\Delta)]$.

Consider system $\mathcal{S}$ and $\mathcal{S}'$, starting at state $s$ and $s'$ respectively, at time 0, with the same arrival process. Let $\mathcal{S}'$ take a specific optimal policy $\pi^*$ that follows $(B_{J-1}), (C_{J-1}), (D_{J-1})$ that satisfies $K(X') \leq J-1$, as well as Proposition~\ref{cl:behind} upon an A-arrival (proved in Appendix~\ref{s:behindA}).
Let $\mathcal{S}$ take policy $\pi$ defined as follows:
\begin{enumerate}[i)]
    \item If $\pi^*$ takes either $\underline{m+1}$ or $\overline{n+1}$, then $\pi$ takes action $\overline{N+1}$;
    \item Otherwise, $\pi$ takes the same action as $\pi^*$.
\end{enumerate}
Define stopping time $T$ to be the first time one of the following occurs: (a) the timer on height $n+1$ expires in system $\mathcal{S}'$, i.e., $L'_{n+1} = 0$; (b) the preceding option (i) is executed, whichever occurs first.
 

Before (i) is executed, $X'$ is either on-time or behind. Due to assumption $(C_{J-1})$, and Proposition~\ref{cl:behind} for A-arrival, $\pi^*$ would take action $\underline{M'+1}$ upon any A-arrival. This implies that no A-blocks arrive before $T$.

Due to $(D_{J-1})$ and $(B_{J-1})$, no H-block will be placed on height $n+2$ in system $\mathcal{S}'$ before $L'_{n+1}=0$. Thus, during $[0,T)$, $M'=M-1=m$, $N=n$, $D'=N'=n+1$, and $L'_i\leq L_i$ for $i = M,\dots,D$.

Consequently, before $T$, neither $\mathcal{S}$ nor $\mathcal{S}'$ can extend any of their branches. At time $T$, one of the following events must occur:
\begin{itemize}
    \item $L'_{n+1}=0$. 
    In this case, $X_T=[m+1,D,n,(0,\dots,0,L_{d+1},\dots,L_{D})]$ and $X'_T=[m,n+1,n+1,(0,\dots,0)]$. Since adding an H-block to $X_T$ on height $n+1$ will yield a state that is superior to $X'_T$, we can invoke Lemma~\ref{lm:A>=Hss'} to show that $X_T\succeq X'_T$.
    
    \item (i) is executed. In this case, at time $T$, we have $N=N'=n+1$, $M'\leq M=m+1$, and $L'_i\leq L_i$ for $i = M,\dots,n+1$. Thus $X_T \succeq X_{T'}$.
\end{itemize}

Before $T$, neither branch of $\mathcal{S}'$ grows, so $\mathcal{S}'$ cannot arrive $\mathscr{A}$ before $T$. Invoking Lemma~\ref{lm:compare}, we have $s \succeq s'$, so that $\underline{m+1}$ is optimal. 

Hence the proof of (III).

\subsubsection{\bf{Proof of (IV)}}
\label{s:proofiv}

If $d=n$, then the behind state of interest is expressed as $s=[m,n,n,(0,0,l_{m+2},\dots,l_n),H]$ with $K(s)=J$. By Lemma~\ref{lm:potential_action_H>0}, either $\overline{n}$ or $\overline{n+1}$ is optimal. In this case, $(B_J)$ is directly implied by $(A_J)$.

In the remainder of this proof, we assume $d<n$, and the behind state of interest is denoted as $s=[m,d,n,(0,0,l_{m+2},\dots,l_d),H]$.

There are three potentially optimal actions according to Lemma~\ref{lm:potential_action_H>0}, leading to three states:
\begin{enumerate}[i)]
    \item $\overline{d}$ (applicable when $l_d >0$), causing a transition to\\ $E = [m,d,n,(0,0,l_{m+2},\dots, l_{d})]$
    \item $\overline{d+1}$ (applicable when $l_d =0$), causing a transition to $F = [m,d+1,n,(0,\dots, 0,\Delta)]$
    \item $\overline{n+1}$, causing a transition to\\  $B=[m,n+1,n+1,(0,0,l_{m+2},\dots,l_d,\Delta,\dots,\Delta)]$.
\end{enumerate}
To compare these states,  we have
\begin{align}
    F
    &\succeq [m,n,n,(0,0,l_{m+2},\dots,l_d,\Delta,\dots,\Delta)] \label{eq:Fd<nB} \\
    &\succeq B \label{eq:O_1>B},
\end{align}
where~\eqref{eq:Fd<nB} is due to Lemma~\ref{lm:h/l rule} and~\eqref{eq:O_1>B} is due to $(A_J)$.
If $l_d >0$, clearly $E \succeq F$.
Hence the proof of (IV).

\subsubsection{\bf{Preliminary: Nakamoto Distribution}}
\label{s:nakamotodistribution}

In preparation for the proof of (V), we introduce a distribution, which we shall refer to as the Nakamoto distribution, and develop useful preliminary results. The  Nakamoto distribution is parameterized by $(a,h,\Delta,l)$, assuming that $(a,h,\Delta)$ satisfy~\eqref{eq:a>}.
Let $ N_1(t) $ denote a Poisson process with rate $ a $. Let $ N_2(t) $ denote a delayed renewal process where the inter-arrival times are given by $ \Delta + X $, where $ X $ follows an exponential distribution with rate $h$. However, the interval preceding the first arrival has the distribution of $ l + X $.
Let $ \tau $ denote the first time $ N_2 $ exceeds $ N_1 $ by 1, namely,
\begin{align}
\label{eq:tau}
    \tau = \inf \left\{ t \geq 0 : N_2(t) = N_1(t) + 1 \right\}.
\end{align}
We refer to the distribution of $N_2(\tau)$ as the Nakamoto distribution with parameter $(a,h,\Delta,l)$. It is easy to see that under condition~\eqref{eq:a>}, $\tau$ is finite with probability 1, so this distribution is well defined.


For every $l\in[0,\Delta]$, let $Z_l$ denote a Nakamoto random variable with parameters $(a,h,\Delta,l)$. Evidently, $Z_l\ge1$ always holds. It is easy to see that $Z_l$ is stochastically larger than $Z_t$ if $l\ge t$. Let $L$ denote a random variable in $[0,\Delta]$ independent of the collection of variables $(Z_l, \, l\in[0,\Delta])$. Then $Z_L$ is a random variable following a mixture of Nakamoto distributions. Evidently, $Z_\Delta$ is stochastically larger than $Z_L$.

Let $( Z_{\Delta,1}, Z_{\Delta,2}, \dots)$ denote a sequence of are independent and identically distributed (i.i.d.) random variables with the same distribution as \( Z_{\Delta} \). We define partial sums
\begin{align}
    Z^i = \sum_{j = 1}^i Z_{\Delta,j}
\end{align}
for $i=1,2,\dots$. Evidently, $Z_i\ge i$ always holds. We shall invoke the following lemma repeatedly in this section.

\begin{lemma} \label{lm:mmm}
    Let $L$ be a random variable in $[0,\Delta]$. Then
    \begin{align}
        [m,m,m,(L)] &\asymp [Z_L+m-1,Z_L+m,Z_L+m,(0,\Delta)], \label{eq:mmm}
    \end{align}
\end{lemma}

\begin{proof}
    Consider a system $\mathcal{S}$ which begins at time 0 in state $s=[m,m,m,(L)]$. Suppose it adopts an optimal policy $\pi^*$ that is consistent with Proposition~\ref{cl:ahead}. We define a stopping time $T$ to be the first time $\mathcal{S}$ arrives in an on-time state or $\mathscr{A}$, whichever occurs first. 

    Before $T$, $\mathcal{S}$ must be in an ahead state. By Proposition~\ref{cl:ahead}, an A-arrival extends the higher branch, while an H-arrival is placed one height above the public height (thereby extending the lower branch if it is at the public height). This process can be interpreted as a race between A-arrivals and jumpers (by Definition~\ref{def:jumper} and due to policy $\pi^*$, a jumper is the first H-block that arrives after the previous jumper expires). Starting from time \( 0 \), the first jumper arrives after time \( L \), with an interval \( L + X \), where \( X \) is exponentially distributed with rate \( \frac{1}{h} \). Subsequent intervals between jumpers are \( \Delta + X \).

    System \( \mathcal{S} \) arrives in an on-time state when a jumper becomes the highest block and the lower branch is exactly one height below, i.e., when there are exactly one more jumper than A-arrivals since time 0. At this time, the number of jumpers has the same distribution as $Z_L$, and, with slight abuse of notation, we express the state as $[Z_L+m-1,Z_L+m,Z_L+m,(0,\Delta)]$.
    
    We caution that $\mathcal{S}$ may arrive at $X_T \in \mathscr{A}$ before the arrival of $Z_L$ jumpers, which occurs in the event that $Z_L+m-1>k$. Nonetheless, conditioned on this event, the state $s'=[Z_L+m-1,Z_L+m,Z_L+m,(0,\Delta)]$ is also in $\mathscr{A}$, so that $V(s')=1=V(X_T)$.
    
    By Lemma~\ref{lm:Vbar=EVbar}, we have
    \begin{align}
    \label{eq:vs=v(Z_L...)}
        V(s)
        &= E_{s}^{\pi^*}\left[V(X_T)\right] \\
        &= E[ V([Z_L + m - 1, Z_L + m, Z_L + m, (0, \Delta)]) ] .
    \end{align}
    We have thus established~\eqref{eq:mmm}.
\end{proof}

Lemma~\ref{lm:mmm} plays the crucial role of equating the value of an ahead state to that of an on-time state. The insight here is that, as long as the system parameters are within the ultimate fault tolerance, the system cannot be permanently in ahead states. Indeed, with probability 1, H-arrivals will catch up to yield an on-time state. The lemma generalizes to the following:

\begin{lemma}
\label{lm:k*adelta1}
    Let $L$ be a random variable in $[0,\Delta]$. Then
    \begin{align}
        [m,m,m+i,(L)] \asymp [m+Z_L,m+Z_L,m+Z_L+i-1,(\Delta)]\label{eq:k*adelta1}
    \end{align}
\end{lemma}

\begin{proof}
    Consider a system begins in state $[m,m,m+i,(L)]$ at time 0. We define a stopping time to be the first time that the higher branch is $i-1$ higher than the lower branch, or that the system arrives in $\mathscr{A}$, whichever occurs first. It is straightforward to verify that it takes exactly one more jumper than A-arrivals for the higher and lower branches to have identical heights, where the number of jumpers has the same distribution as $Z_L$. With slight abuse of notation, the resulting state is expressed as $[Z_L+m,Z_L+m,Z_L+m+i-1,(\Delta)]$. Using essentially the same arguments in the proof of~\eqref{eq:mmm}, we have
    \begin{align}
        V([m,m,m+i,(L)])
        =
        E[ m+Z_L,m+Z_L,m+Z_L+i-1,(\Delta)] .
    \end{align}
    We have thus established~\eqref{eq:k*adelta1}.
\end{proof}
For special case $i=1$, we have
\begin{align}
     [m,m,m+1,(L)] &\asymp [Z_L+m,Z_L+m,Z_L+m,(\Delta)] \label{eq:mmm+1_1} .
\end{align}
The following is a corollary of the preceding lemmas:
\begin{align}
    & [m,m,m+i,(L)] \notag \\
    &\quad \asymp [m+Z^{i-1}+Z_L,
    m+Z^{i-1}+Z_L,
    m+Z^{i-1}+Z_L,
    (\Delta)] \label{eq:i-Z^{i-1}}\\
    &\quad \asymp  [m+Z^{i}+Z_L-1,
    m+Z^{i}+Z_L,
    m+Z^{i}+Z_L,
    (0,\Delta)]    \label{eq:k*adelta2}
\end{align}
where~\eqref{eq:i-Z^{i-1}} is obtained by applying~\eqref{eq:k*adelta1} $i$ times and~\eqref{eq:k*adelta2} is due to~\eqref{eq:mmm}.

\begin{lemma}
    Starting from state $s=[m,m+2,m+2+i,(0,0,\Delta)]$, where $m, i\in\mathbb{N}$ and $K(s) \leq J-1$, if the system arrives in an ahead state no later than arriving in $\mathscr{A}$, then the ahead state takes the form of:
    \begin{align}
    \label{eq:stateform}
        [m+Y,m+Y,m+Y+j,(R)]
    \end{align}
    for some random variable $R$, $Y \in \{2,3,\dots\}$, and $j = \max\{0,2+i-Y\}$.
\end{lemma}

\begin{proof}
    Consider a system $\mathcal{S}$ which begins in state $s$
at time 0. Suppose it adopts an optimal policy $\pi^*$ that is consistent with $(B_J)$ and Proposition~\ref{a:A-behind} with A-arrivals, $(C_{J-1})$ and $(D_{J-1})$. We define a stopping time $T$ to be the first time $\mathcal{S}$ arrives in an ahead state or $\mathscr{A}$, whichever occurs first.
By $\pi^*$, the higher branch with nonpublic will only be extended when the system arrives in an ahead state. Thus, during $[0,T]$, the higher branch's height is no lower than $D$, when $D$ catches up with $N$, $N$ increases together with $D$.
By Lemma~\ref{lm:must take the form of}, if the system $\mathcal{S}$ is in an ahead state at $T$, it has the state form of \eqref{eq:stateform}, where $Y$ stands for the height increase of the lower branch, which is at least $2$ to catch up with $D$. And $R$ stands for the maximum of 0 and the timer on height $m+Y$ when the lower branch reaches height $m+Y$, which is in $[0,\Delta]$.
\end{proof}

\subsubsection{\bf{Proof of (V)}}
\label{s:proofv}

With new tools developed in Appendix~\ref{s:nakamotodistribution}, we next assume that $(A_0),\dots,(A_J)$, $(B_0),\dots,(B_J)$, $(C_0),\dots,(C_{J-1})$, and $(D_0),\dots,(D_{J-1})$ hold for some $J\in\mathbb{N}$, and finish proving Proposition~\ref{pr:ABCD} by showing that $(C_J)$ must also hold.

Consider the on-time state $s=[m,d,n,(0,l_{m+1},\dots,l_d),A]$ with $K(s)=J$. By Lemma~\ref{lm:potential_action_A>0}, either $\underline{m+1}$ or $\overline{n+1}$ or both are optimal. The proof boils down to proving that $\underline{m+1}$ is at least as good as $\overline{n+1}$.
Since $m<d\le n$, the on-time state 
must be one of the following four cases: 
\begin{itemize}
    \item $S_1 = [m,m+1,m+1,(0,\Delta),A]$;
    \item $S_2 = [m,m+1,m+1,(0,l_{m+1}),A]$ with $l_{m+1}<\Delta$;
    \item $S_3 = [m,m+1,m+i,(0,\Delta),A]$ for some $i\in\{2,3,\dots\}$;
    \item $S_4 = [m, d, m+i, (0,l_{m+1},\dots,l_d), A]$ for some $i\in\{2,3,\dots\}$ and $d\in\{m+1,m+2,\dots\}$.
\end{itemize}
We prove $(C_J)$ holds in each case separately.

{\bf Case S1:}

In $S_1 = [m,m+1,m+1,(0,\Delta),A]$, at least one of the following actions is optimal:
\begin{enumerate}[i)]
    \item $\overline{m+2}$, causing a transition to state $E = [m,m+1,m+2,(0,\Delta)]$;
    \item $\underline{m+1}$, causing a transition to state $F = [m+1,m+1,m+1,(\Delta)]$.
\end{enumerate}

Consider a system $\mathcal{S}$, which begins in state $E$ at time 0. Suppose $\mathcal{S}$ adopts a specific optimal policy $\pi^*$ that is consistent with $(B_{J-1})$, $(C_{J-1})$ and $(D_{J-1})$. Consider the state upon the first arrival. There are two possibilities:
\begin{itemize}
    \item 
    The first arrival is an H-block and arrives no earlier than $\Delta$. This case occurs with probability $q=(h/\lambda) e^{-\lambda\Delta}$. Evidently, the timer on height $m+1$ expires at $\Delta$ by this H-arrival, when the system is in a behind state. 
    Due to $(B_{J-1})$, 
    $\overline{m+2}$ is optimal, causing a 
    transition to $E_2=[m,m+2,m+2,(0,0,\Delta)]$.
    
    \item Either the first arrival is during $[0,\Delta)$, when the system remains in an on-time state or it is an A-block and arrives no earlier than $\Delta$, when the system is in a behind state. This case occurs with probability $1-q$. Due to assumptions $(C_{J-1})$ and $(D_{J-1})$ as well as Proposition~\ref{cl:behind} with an A-arrival (proved in Appendix~\ref{s:behindA}), $\underline{m+1}$ is optimal in this case, where the system transitions to state $E_1=[m+1,m+1,m+2,(L)]$ with some random $L\in[0,\Delta)$.
\end{itemize}
By Lemma~\ref{lm:Vbar=EVbar},
\begin{align} \label{eq:VE12}
    V(E) = q V(E_2) + (1-q) V(E_1) .
\end{align}

Let $F' = [Z_\Delta+m,Z_\Delta+m+1,Z_\Delta+m+1,(0,\Delta)]$. By~\eqref{eq:mmm}, we have $F \asymp F'$. 

Consider a system $\mathcal{S}'$, which begins in state $F'$ and adopts the same optimal policy $ \pi^* $ as $\mathcal{S}$.
Consider the state upon the first arrival under the same two possibilities:
\begin{itemize}
    \item 
    With probability $q$, the timer on height $Z_\Delta+m+1$ expires by the first arrival, which is an H-arrival. Due to $(B_{J-1})$, it is optimal to place the block on height $Z_\Delta+m+2$ in this behind state, causing a transition to $F_2=[Z_\Delta+m,Z_\Delta+m+2,Z_\Delta+m+2,(0,0,\Delta)]$. 
    
    \item With probability $1-q$, either the first arrival is an A-block, or an H-block arrives before the timer on height $Z_\Delta+m+1$ expires. 
    In either case, $\underline{Z_\Delta+m+1}$ is optimal, where the system transitions to state $F_1=[Z_\Delta+m+1,Z_\Delta+m+1,Z_\Delta+m+1,(L')]$ with some random $L'\in[0,\Delta]$. It is important to note that $L'$ in $\mathcal{S'}$ is identically distributed as $L$ in $\mathcal{S}$.
\end{itemize}
By Lemma~\ref{lm:Vbar=EVbar},
\begin{align} \label{eq:VF12}
    V(F) = q V(F_2) + (1-q) V(F_1) .
\end{align}

Using~\eqref{eq:VE12} and~\eqref{eq:VF12}, in order to show $E\preceq F$, it suffices to show that $E_1\asymp F_1$ and $E_2\preceq F_2$. First, we have
\begin{align}
    E_1
    &\asymp
    [Z_L+Z^1+m, Z_L+Z^1+m+1, Z_L+Z^1+m+1, (0,\Delta)] \label{eq:e1l}\\
    &\asymp F_1 \label{eq:ef1},
\end{align}
where \eqref{eq:e1l} is due to \eqref{eq:k*adelta2} 
and \eqref{eq:ef1} is due to \eqref{eq:mmm}.

Next, we consider two systems $\mathcal{S}''$ and $\mathcal{S}'''$, which begin in states $E_2$ and $F_2$, respectively, at time 0. Suppose the two systems have identical arrival processes. Suppose system $\mathcal{S}''$ employ an optimal policy. Whenever $\mathcal{S}''$ takes action $\underline{r}$ (resp.~$\overline{r}$), let $\mathcal{S}'''$ take  $\underline{r+Z_\Delta}$ (resp.~$\overline{r+Z_\Delta}$). It is easy to see that all states and actions in $\mathcal{S}''$ are shifted $Z_\Delta$ heights higher in $\mathcal{S}'''$. Evidently, $\mathcal{S}'''$ arrives in $\mathscr{A}$ no later than $\mathcal{S}''$. Hence $E_2\preceq F_2$ by Lemma~\ref{lm:compare}.

We have thus proved 
\begin{align}
\label{eq:E<F}
    E \preceq F
\end{align}
and thus the optimality of $\underline{m+1}$ in on-time state $S_1$. Hence $(C_J)$ is established in Case S1.

{\bf Case S2:}

In $S_2 = [m,m+1,m+1,(0,l_{m+1}),A]$ with $l_{m+1}<\Delta$, at least one of the following actions is optimal:
\begin{enumerate}[i)]
    \item $\overline{m+2}$, causing a transition to $U= [m,m+1,m+2,(0,l_{m+1})]$;
    \item $\underline{m+1}$, causing a transition to $W = [m+1,m+1,m+1,(l_{m+1})]$.
\end{enumerate}

States $U$ and $W$ here are the same as $E$ and $F$ in Case S1 except that $U$ and $W$'s timers on height $m+1$ is smaller. 
Consider two systems $\mathcal{S}$ and $\mathcal{S}'$, which begin in states $E$ and $F$ at time 0, respectively. Suppose the two systems have identical arrival processes. Consider their states upon the first arrival or time $\Delta-l_{m+1}$, whichever occurs first. There are two possibilities:
\begin{itemize}
     \item 
     In the absence of any arrival up to time $\Delta-l_{m+1}$, the timers on height $m+1$ reduces to $l_{m+1}$. In this case, system $\mathcal{S}$ arrives in state $U$, while system $\mathcal{S}'$ arrives in state $W$. This occurs with probability $ r = e^{-\lambda (\Delta - l_{m+1})}$.
     
     \item A first block arrives at $\Delta-\bar{l}_{m+1}$ with $\bar{l}_{m+1}>l_{m+1}$. 
     In system $\mathcal{S}$, due to assumptions $(C_{J-1})$ and $(D_{J-1})$, action $\underline{m+1}$ is optimal, causing a transition to $U' = [m+1,m+1,m+2,(\bar{l}_{m+1})]$.
     In system $\mathcal{S}'$, if the arrival is an H-block, $\overline{m+1}$ is optimal by Proposition~\ref{cl:ahead}, causing a transition to $W' = [m+1,m+1,m+1,(\bar{l}_{m+1})]$.
    On the other hand, if the arrival is an A-block, 
      action $\overline{m+2}$ is optimal due to Proposition~\ref{cl:ahead}, causing a transition to $U'$.
      By Lemma~\ref{lm:A>=H}, $W' \preceq U'$.
 \end{itemize}
 
By Lemma~\ref{lm:Vbar=EVbar}, we have
\begin{align}
    V(E) &= r V(U) + (1-r) V(U') \label{eq:u'u=e}
\end{align}
and
\begin{align}
    V(F) 
    &= r V(W) + (1-r) \left( \frac{h}{\lambda}V(W') + \frac{a}{\lambda} V(U') \right) \label{eq:f=w'u'w}\\
    &\le r V(W) + (1-r) \, V(U') \label{eq:VF<},
\end{align}
where~\eqref{eq:VF<} is due to $W'\preceq U'$. By~\eqref{eq:E<F}, we have $V(E)\le V(F)$. Comparing~\eqref{eq:u'u=e} and~\eqref{eq:VF<} yields $U \preceq W$. This completes the proof of $(C_J)$ in Case S2.

{\bf Case S3:}\\
In $S_3 = [m,m+1,m+i,(0,\Delta),A]$ for some $i\in\{2,3,\dots\}$, as least one of the following actions is optimal:
\begin{enumerate}[i)]
    \item $\overline{m+\eyetwo}$, causing a transition to state $E\Sthree = [m,m+1,m+\eyetwo,(0,\Delta)]$;
    \item $\underline{m+1}$, causing a transition to state $F\Sthree = [m+1,m+1,m+\eyeone,(\Delta)]$.
\end{enumerate}

Consider a system $\mathcal{S}$, which begins in state $E\Sthree$ at time 0. Suppose $\mathcal{S}$ adopts a specific optimal $\pi^*$ that is consistent with $(B_{J})$, $(C_{J-1})$, $(D_{J-1})$, and Proposition~\ref{cl:ahead}. Consider the state upon the first arrival. There are two possibilities:
\begin{itemize}
    \item The first arrival is an H-block and arrives no earlier than $\Delta$. This case occurs with probability $q = (h/\lambda)e^{-\lambda\Delta}$. The timer on height $m+1$ expires at $\Delta$. So this H-arrival arrives into a behind state. Due to $(B_{J})$, $\overline{m+2}$ is optimal, causing a transition to $E\Sthree_1=[m,m+2,m+\eyetwo,(0,0,\Delta)]$.
    \item Either the first arrival is during $[0,\Delta)$, when the system remains in an on-time state, or it is an A-block and arrives no earlier than $\Delta$, when the system is in a behind state. This case occurs with probability $1-q$. Due to assumptions $(C_{J-1})$ and $(D_{J-1})$ 
    as well as Proposition~\ref{cl:behind} with an A-arrival (proved in Appendix~\ref{s:behindA}), $\underline{m+1}$  is optimal in either case, causing a transition to $E\Sthree_2 =[m+1,m+1,m+\eyetwo,(L)]$ for some random $L\in[0,\Delta)$.
\end{itemize}
By Lemma~\ref{lm:Vbar=EVbar}, we have
\begin{align} \label{eq:E=E1E2}
    V(E\Sthree) = q V(E\Sthree_1)+(1-q) V(E\Sthree_2) .
\end{align}

Let $F'\Sthree = [m+Z^\eyeone,m+1+Z^\eyeone,m+1+Z^\eyeone,(0,\Delta)]$. By~\eqref{eq:k*adelta2}, we have $F\Sthree\asymp F'\Sthree$. Consider a system $\mathcal{S}'$, which begins in state $F'\Sthree$ at time 0. Suppose $\mathcal{S}'$ adopts the same optimal $\pi^*$ as system $\mathcal{S}$. Consider the state upon the first arrival under the same two possibilities:
\begin{itemize}
    \item With probability $q$, the timer on height $m+Z^\eyeone+1$ expires by the first arrival, which is an H-arrival. Due to $(B_{J-1})$, it is optimal to place the block on height $m+Z^\eyeone+2$ in this behind state, causing a transition to $F\Sthree_1=[m+Z^\eyeone,m+Z^\eyeone+2,m+Z^\eyeone+2,(0,0,\Delta)]$.
    
    \item With probability $1-q$, either the first arrival is an A-block, or an H-block arrives before the timer on height $m+Z^\eyeone+1$ expires. In either case, $\underline{m+Z^\eyeone+1}$ is optimal, causing a transition to $F\Sthree_2 =[m+1+Z^\eyeone,m+1+Z^\eyeone,m+1+Z^\eyeone,(L')]$. Note that $L'$ in $F\Sthree_2$ is identically distributed as $L$ in $E\Sthree_2$.
\end{itemize}
A careful reader might notice that $m+Z^\eyeone$ could be greater or equal to $k$ such that  $[m+Z^\eyeone,m+1+Z^\eyeone,m+1+Z^\eyeone,(0,\Delta)] \in \mathscr{A}$, which means $F'\Sthree$ transitions to a state of value 1, which is no lower than the values of $F\Sthree_1$ and $F\Sthree_2$. By Lemma~\ref{lm:Vbar=EVbar}, we have
\begin{align} \label{eq:F>F1F2}
    V(F\Sthree) \geq qV(F\Sthree_1)+(1-q)V(F\Sthree_2).
\end{align}

We first show that $E\Sthree_2$ and $F\Sthree_2$ are of identical values:
\begin{align} 
  E\Sthree_2
  &\asymp [Z_L+m+Z^\eyeone,Z_L+m+1+Z^\eyeone,Z_L+m+1+Z^\eyeone,(0,\Delta)] \label{eq:E2S3} \\
    &\asymp [m+1+Z^\eyeone,m+1+Z^\eyeone,m+1+Z^\eyeone,(L)] \label{eq:F2S3} \\
    &\asymp F\Sthree_2 \label{eq:E2=F2}
\end{align}
where~\eqref{eq:E2S3} is due to~\eqref{eq:k*adelta2},~\eqref{eq:F2S3} is due to~\eqref{eq:mmm}, and~\eqref{eq:E2=F2} is because $L$ and $L'$ are identically distributed.

Comparing~\eqref{eq:E=E1E2} and~\eqref{eq:F>F1F2}, one easily sees that $E\preceq F$ as long as the following lemma holds:

\begin{lemma} \label{lm:m+z}
    For every $m\in\mathbb{N}$ and $i\in\{2,3,\dots\}$, 
    \begin{align}
    E\Sthree_1 \preceq F\Sthree_1
    \end{align}
\end{lemma}

\begin{proof}
    Consider systems $\mathcal{S}$ and $\mathcal{S}'$, which begin in $F\Sthree_1 = [m+Z^\eyeone,m+Z^\eyeone+2,m+Z^\eyeone+2,(0,0,\Delta)]$ and $E\Sthree_1 =[m,m+2,m+\eyetwo,(0,0,\Delta)]$, respectively, at time 0.
    Suppose the two systems have identical arrival processes. At any given time, let $X=[M,D,N,(L_{M \land D}, \dots, L_D),I]$ denote the state of $\mathcal{S}$ and let $X'=[M',D',N',(L'_{M '\land D'}, \dots, L'_{D'}),I]$ denote the state of $\mathcal{S}'$.
    Let system $\mathcal{S}'$ adopt a specific optimal policy $\pi^*$ that is consistent with assumptions $(B_{J-1})$, $(C_{J-1})$, $(D_{J-1})$, as well as Proposition~\ref{cl:ahead} and Proposition~\ref{cl:behind} for A-arrivals. We define a policy $\pi$ for $\mathcal{S}$ as follows:
    \begin{itemize}
    \item If $\pi^*$ takes action $\overline{D'}$, then $\pi$ takes action $\overline{D}$; 
    \item If $\pi^*$ takes action $\overline{D'+1}$, then $\pi$ takes action $\overline{D+1}$; 
    \item If $\pi^*$ takes action $\underline{M'+1}$, then $\pi$ takes action $\underline{M+1}$. 
    \end{itemize}

    We define a stopping time $T$ to be the first time system $\mathcal{S}'$ arrives in $\mathscr{A}$ or an ahead state, whichever occurs first. Since system $\mathcal{S}'$ is never in an ahead state during $[0,T)$, we must have $M'<D'\le N'$ before $T$ and it suffices to consider actions $\overline{D'}$, $\overline{D'+1}$, and $\underline{M'+1}$ for $\pi^*$. Hence the policy $\pi$ is well-defined. It is also easy to verify that $M-M' = P-P' = D-D'$ during $[0,T)$.

There are two possibilities at $T$:
\begin{itemize}
    \item    $\mathcal{S}'$ arrives in $\mathscr{A}$. In this case, $M' \geq \max\{ P',k\}$, 
    which implies that $M \geq \max\{ P,k\}$, so $\mathcal{S}$ is also in $\mathscr{A}$.

    \item $\mathcal{S}'$ arrives in an ahead state. In this case, $M'$ catches up to $D'$, so $M\ge D$. Thus system $\mathcal{S}$ is either in $\mathscr{A}$ or arrives in an ahead state $F_T = [m+Z^\eyeone+Y,m+Z^\eyeone+Y,m+Z^\eyeone+Y,(R)]$ for some random variables $Y\in\{2,3,\dots\}$ and $R\in[0,\Delta]$. Depending on $i$ and the realization of $Y$, we have two cases:
    \begin{itemize}
        \item If the height of lower branch is still lower than the higher branch, the system is at $E_T = [m+Y,m+Y,\eyetwo + m ,(R)]$
    where $Y\in\{2,\dots,\eyeone\}$. 
    \item If the height of lower branch is on the same height as the higher branch, the system is at 
    $E'_T = [m+Y,m+Y,m+Y,(R)]$ where $Y \in \{\eyetwo,i+2,\dots\}$.
    \end{itemize}
    \end{itemize}

    To show that $E_1\preceq F_1$, it suffices to show that $E_T\preceq F_T$ and $E'_T\preceq F_T$. It is easy to see that $E'_T\preceq F_T$.
By \eqref{eq:k*adelta2},
\begin{align}
\begin{split}
   E_T 
   &\asymp [m+Y+Z_R+Z^{\eyetwo-Y}-1,m+Y+Z_R+Z^{\eyetwo-Y},\\
   &\qquad m+Y+Z_R+Z^{\eyetwo-Y},(0,\Delta)].    
\end{split}
\end{align}
By Lemma~\ref{lm:mmm}, 
\begin{align}
    F_T 
    &\asymp [m+Z^\eyeone+Y+Z_R-1,m+Z^\eyeone+Y+Z_R,m+Z^\eyeone+Y+Z_R,(0,\Delta)].
\end{align}
Since $\eyetwo-Y \leq \eyeone$, we have $E_T \preceq F_T$. 
Thus, by Lemma~\ref{lm:compare}, $E\Sthree_1 \preceq F\Sthree_1$.
\end{proof}

With $E\Sthree \preceq F\Sthree$ proven, $(C_J)$ is established in Case S3.

{\bf Case S4:}

Consider state $S_4 = [m,d, m+\eyeone,(0,l_{m+1},\dots,l_{d}), A]$ for some $i\in\{2,3,\dots\}$ and $d\in\{m+1,m+2,\dots\}$.

If $d = m+1$, then $S_4 = [m,m+1,m+\eyeone,(0,l_{m+1}),A]$. Moreover, if $l_{m+1} = \Delta$, $S_4$ degenerates to $S_3$ in Case S3. If $l_{m+1} < \Delta$, we let two systems, $\mathcal{S}$ and $\mathcal{S'}$, start in states $E\Sthree$ and $F\Sthree$ given in Case S3, respectively, at time 0. We then derive an optimal action in Case S4 based on the optimal action in Case S3 using the same technique applied earlier to derive the optimal action in Case S2 based on the optimal action in Case S1. As a result, $\underline{m+1}$ is shown to be superior in the special case of $d=m+1$ and $l_{m+1}<\Delta$. We omit the details here.

In the remainder of Case S4, we assume $d \geq m+2$.
At least one of the following actions is optimal:
\begin{enumerate}[i)]
    \item $\overline{m+\eyetwo}$, causing a transition to
    \begin{align} \label{eq:ES4}
        E\Sfour = [m,d,m+\eyetwo,(0,l_{m+1},\dots,l_d)];
    \end{align}
    \item  $\underline{m+1}$, causing a transition to
    \begin{align} \label{eq:FS4}
        F\Sfour = [m+1,d,m+\eyeone,(l_{m+1},\dots,l_d)] .
    \end{align} 
\end{enumerate}

\begin{lemma} \label{lm:E<=F}
    If $d>m$ and $i\in\{1,2,\dots\}$, then
    \begin{align} \label{eq:E<=F}
        [m,d,m+\eyetwo,(0,l_{m+1},\dots,l_d)]
        \preceq
        [m+1,d,m+\eyeone,(l_{m+1},\dots,l_d)] .
    \end{align}
\end{lemma}

\begin{proof}
The special case of~\eqref{eq:E<=F} with $i=1$ (which implies $d=m+1$) has already been proved as $U\preceq W$ in Case S2.

We assume $i\in\{2,3,\dots\}$ in the remainder of this proof. Let $E$ and $F$ be defined as in~\eqref{eq:ES4} and~\eqref{eq:FS4}, respectively. Let
\begin{align}
    E'\Sfour = [m,d,m+\eyetwo,(0,l_{m+2},l_{m+2},\dots,l_d)] .
\end{align}
Since $l_{m+1}\leq l_{m+2}$, we have $E\Sfour \preceq E'\Sfour$ by Lemma~\ref{lm:h/l rule}.

Consider a system $\mathcal{S}$, which begins in state $E'\Sfour$ at time 0. Suppose $\mathcal{S}$ adopts a specific optimal policy $\pi^*$ which is consistent with $(C_{J-1})$ and $(D_{J-1})$. Consider the state at time $l_{m+2}$ or upon the first arrival, whichever occurs first. There are two possibilities:
\begin{itemize}
    \item There is at least one arrival during $[0,l_{m+2})$. This occurs with probability $q = 1-e^{\lambda l_{m+2}}$. In this case, let the first arrival be at time $l_{m+2}-L$, where $0\leq L < l_{m+2}$. 
    Since the system is in an on-time state at the time of arrival, action $\underline{m+1}$ is optimal due to assumptions $(C_{J-1})$ and $(D_{J-1})$, causing a transition to $E_1\Sfour = [m+1,d,m+\eyetwo,(L,L,l'_{m+3},\dots,l'_d)]$, where $l'_i = l_i -l_{m+2}+L$ for $i=m+3,\dots,d$.

    \item No block arrives during $[0,l_{m+2})$. This occurs with probability $1-q$. In this case, at $l_{m+2}$, the system arrives in a behind state $E_2\Sfour = [m,d,m+\eyetwo,(0,0,0,l''_{m+3},\dots,l''_d)$, where $l''_i = l_i -l_{m+2}$ for $i=m+3,\dots,d$.
\end{itemize}

Consider a system $\mathcal{S}'$, which begins in state $F\Sfour$ at time 0. Suppose $\mathcal{S}'$ adopts the same optimal $\pi^*$ as system $\mathcal{S}$. Consider the state at time $l_{m+2}$ or upon the first arrival, whichever occurs first. Consider the same two possibilities:
\begin{itemize}
    \item With probability $q$, there is at least one arrival during $[0,l_{m+2})$. Let the first arrival be at time $l_{m+2}-L'$, where $0\leq L' < l_{m+2}$. At this time, the system is in an on-time state. Due to $(C_{J-1})$ and $(D_{J-1})$, action $\underline{m+2}$ is optimal, causing a transition to $F\Sfour_1 = [m+2,d,m+\eyeone,(L',l'''_{m+3},\dots,l'''_d)],$, where $l'''_i = l_i-l_{m+2}+L'$ for $i = m+3, \dots, d$. Note that $L'$ and $L$ in $E_1\Sfour$ are i.i.d., $l_i$ and $l'''_d$ are i.i.d., for $i = m+3,\dots,d$.
    
    \item With probability $1-q$, no block arrives during $[0,l_{m+2})$. At $l_{m+2}$, the system arrives in a behind state $F_2\Sfour = [m+1,d,m+\eyeone,(0,0,l''_{m+3},\dots,l''_{d})]$, where $l''_i$ is the same as in $E_2\Sfour$.
\end{itemize}

First, we compare states $E\Sfour_1$ and $F\Sfour_1$ by considering an auxiliary
on-time state: $s_1=[m+1,d,m+\eyeone,(0,L',l'''_{m+3},\dots,l'''_d),A]$. In state $s_1$, by assumption $(C_{J-1})$, action $\underline{m+2}$, which would cause a transition to $F\Sfour_1$, is superior to action $\overline{m+\eyetwo}$, which would cause a transition to $[m+1,d,m+\eyetwo,(0,L',l'''_{m+3},\dots,l'''_d)]$. By Lemma~\ref{lm:lm=0}, we have $E\Sfour_1 \asymp [m+1,d,m+\eyetwo,(0,L,l'_{m+3},\dots,l'_d)]$.
Hence, 
\begin{align}
    E_1\Sfour \preceq F_1\Sfour . \label{eq:E1<F1four}
\end{align}

We next compare states $E\Sfour_1$ and $F\Sfour_1$ by considering an auxiliary state $s_2=[m,d,m+\eyeone,(0,0,0,l''_{m+3},\dots,l''_d),A]$. In state $s_2$, action $\underline{m+1}$ causes a transition to $ F\Sfour_2$, and action $\overline{m+\eyetwo}$ causes a transition to $E\Sfour_2$. By Lemma~\ref{cl:behind} for $I = A$, action $\underline{m+1}$ is optimal in state $s_2$, so
\begin{align}
     E\Sfour_2 \preceq  F\Sfour_2. \label{eq:E2<F2four}
\end{align}

We can then write:
\begin{align}
   V(E\Sfour) &\leq V(E'\Sfour)\\
   &=  q \, V(E\Sfour_1) + (1-q)\, V(E\Sfour_2) 
   \label{eq:E=E1E2four} \\
    & \leq q \, V(F\Sfour_1) + (1-q)\, V(F\Sfour_2)
    \label{eq:E<=Ffour} \\
    & = V(F\Sfour)\label{eq:qv+1-qv < v},
\end{align}
where~\eqref{eq:E=E1E2four} and~\eqref{eq:qv+1-qv < v} are due to Lemma~\ref{lm:Vbar=EVbar},~\eqref{eq:E<=Ffour} is due to~\eqref{eq:E1<F1four}.    
\end{proof}

Lemma~\ref{lm:E<=F} implies $E\Sfour\preceq F\Sfour$, which completes the proof of proposition $(C_J)$ in Case 4. We have thus proved $(C_J)$ in all four cases, which completes the proof of Proposition~\ref{pr:ABCD}, which in turn proves Propositions~\ref{cl:on-time} and~\ref{cl:behind}. Together with Proposition~\ref{cl:ahead}, these results collectively prove that bait-and-switch is an optimal strategy.

\section{The Exact Probability of Violation}
\label{a:trade-off}

As demonstrated in Appendix~\ref{a:pow_delta}, the attack defined by Propositions~\ref{cl:ahead}--\ref{cl:behind} is a specific form of bait-and-switch in PoW. We shall study the probability of violation due to this attack, hereby referred to as {\em the optimal attack}, 
and henceforth prove Theorem~\ref{th:pow}. We first make the following observation:

\begin{lemma} \label{lm:d=m or d=n}
    In PoW, under the optimal attack, if the system is in a behind or an on-time state denoted as $[m,d,n,(l_m,\dots,l_d),I]$, it must satisfy $d=n$; also, $\Delta$ time after the arrival of each jumper, the state must take the form of $[m,d,n,(0,\dots,0),I]$ with either $d = m$ or $d = n$. 
\end{lemma}

\begin{proof}
    According to the state transition rule (Table~\ref{tb:I=H}), only the arrival of a jumper will decrease the timer of a height from infinity to a finite number. Hence 
    if the system is in state $[m,d,n,(l_m,\dots,l_d),I]$, there must have been a jumper that arrived on height $d$.
    
    Suppose block $j$ is a jumper.
    In general, the state upon its arrival at $t_j+\Delta$ can be denoted as $[m,d,n,(l_{m\land d},\dots,l_d),I]$. Since the optimal attack places no H-blocks higher than the jumper before $t_j+\Delta$, when jumper block $j$ becomes public at $t_j+\Delta$, its height $h_j$ becomes the public height and the timer on this height is the highest finite timer. Hence $d=h_j$ and $l_{m\land d}=\dots=l_d=0$.
    
    We next prove that either $m=d$ or $n=d$ by showing that both $m>d$ and $m < d < n$ contradict the optimal attack.

    Suppose $m>d$. In this case, each of the two highest branches has an A-block on height $d+1$ at time $t_j+\Delta$. Between those two A-blocks, the 
    later one arrives in a 
    state with 
    a lower branch on height $d$, so that it is placed on height $d+1$, which implies that it arrives in an ahead state according to Definition~\ref{def:State}. However, this contradicts Proposition~\ref{cl:ahead} for the optimal attack in ahead states, 
    which uses the second A-block to 
    extend the higher branch instead of extending the lower branch. Hence $m\le d$. 

    Suppose $m<d<n$ instead, which requires that jumper block $j$ on height $d$ to be placed in the higher branch. In this case, the higher branch's height-$n$ block, referred to as block $b$, is an A-block. Since block $b$ extends the higher branch and that $l_d>0$ at $t_b<t+\Delta$, block $b$ must arrive in an ahead state according to Propositions~\ref{cl:ahead}--\ref{cl:behind}. Now suppose that at time $t_b$, the lower branch is at height $m' \leq m < n$, which implies $l_{m'+1} = \infty$ at $t_b$. From time $t_b$ onward, once the lower branch reaches the public height, it becomes the public fork-choice according to Proposition~\ref{cl:ahead}. Consequently, the next jumper will extend the lower branch and will also arrive in an ahead state. This process continues: as long as the lower branch has not yet reached height $n$, the system remains in the ahead state, and all new jumpers extend the lower branch. Therefore, the jumper block $j$ must also extend the lower branch, contradicting the assumption that it is placed in the higher branch.

    Since both $m>d$ and $m<d<n$ are not possible, we must have either $d=m$ or $d=n$.

    Next, consider a system that arrives in either a behind or an on-time state, represented as $[m,d,n,(l_m,\dots,l_d),I]$. Consider the jumper which has arrived on height $d$. Since $d<n$ contradicts the optimal attack as we have argued, we must have $d=n$.
\end{proof}

\subsection{Chain Increment under the Optimal Policy}
\label{a:tradeoff_chainincre}

Let $T_0=0$. For $j\in\mathbb{N}$, let $T_j$ denote $\Delta$ units of time after the arrival of the $j$-th jumper. We refer to $(T_{j-1},T_j]$ as the $j$-th epoch. By Lemma~\ref{lm:d=m or d=n}, at $T_j$, either the higher or the lower branch or both are as high as the public height. We refer to one such branch as {\em the jumper branch} and the other one as {\em the adversarial branch}. 

Upon any arrival during $(T_j,T_{j+1}-\Delta]$, every A-block extends the adversarial branch; an H-block (aka the $(j+1)$-st jumper) arrives at $T_{j+1}-\Delta$, extending the jumper branch.

Upon any arrival during $(T_{j+1}-\Delta,T_{j+1})$, we have two cases:
\begin{itemize}
    \item If the adversarial branch is as high as the public height, since the jumper branch is one height higher, the adversarial branch is made the public fork choice, and every A-block or H-block extend the adversarial branch;
    \item If the adversarial branch is either higher or lower than the public height, every A-block extends the adversarial branch; all H-blocks are placed on the jumper branch on the jumper's height, so they are inconsequential in the sense that their placements yield no state change.    
\end{itemize}

For every $j\in\mathbb{N}$, we use $X_j$ and $Y_j$ to denote the public height and the height of the adversarial branch at time $T_j$.
From $T_j$ to $T_{j+1}$, the public height rises by 1, so $X_{j+1}= j+1$.

We fix $j$ and let $U$ and $W$ denote the height increment of the adversarial branch during $(T_j,T_{j+1}-\Delta]$ and $(T_{j+1}-\Delta,T_{j+1}]$, respectively. Hence at $T_{j+1}$, the state is represented as $(X+1,Y+U+W)$. As $U$ denotes the number of A-arrivals after $T_j$ before the first H-arrival, it is a geometric random variable with the following pmf:
\begin{align}
    P(U = x)
    = \left(\frac{a}{\lambda}\right)^x \frac{h}{\lambda},
    \quad x=0,1,\cdots.
    \label{eq:x1}
\end{align}

Let $L$ denote the public height minus the height of the adversarial branch at time $T_{j+1}-\Delta$, i.e.,
\begin{align}
    L = X_j - Y_j - U .
\end{align}
We derive the pmf of $W$ conditioned on $L=l$ as follows:
\begin{enumerate}[I)]
    \item If $l<0$, the adversarial branch is higher than the public height, only A-arrivals are placed extending the adversarial branch. The probability of having $i$ A-arrivals during $(T_{j+1}-\Delta,T_{j+1})$ is:
    \begin{align}
       f_1(i,a\Delta) \label{eq:x2_1}
    \end{align}
    \item If $l>0$, there are three possibilities:
\begin{enumerate}
    \item [II-1] The number of A-arrivals $i<l$. In this case, the adversarial branch rises by $i$ heights.     The probability is the same as \eqref{eq:x2_1}.
    \item [II-2] Exactly $l$ A-arrivals during $(T_{j+1}-\Delta,T_{j+1})$, where the last of them is followed by no H-arrivals by $T_{j+1}$. In this case, while the adversarial branch is made the public fork choice, no H-block arrives to be exploited, so the adversarial branch rises by exactly $l$ heights. Let $T_{j+1}-\Delta+T$ denote the time of the $l$-th A-arrival. The random variable $T$ has the Erlang pdf with parameter shape $l$ and rate $a$.  During $(T_{j+1}-\Delta+T, T_{j+1})$, there is no arrivals of any type. The probability of this event is
     \begin{align}
        \int_{0}^{\Delta}f_2(t;l,a)e^{-(\Delta -t)\lambda} dt . \label{eq:x2_2}
    \end{align}
    
    \item [II-3] More than $l$ A-arrivals during $(T_{j+1}-\Delta,T_{j+1})$, denoted as $x$. In this case, the adversarial branch may rise by either $x$ or $x+1$, depending on whether an H-block arrives to be exploited right after the $l$-th A-arrival. Suppose the $l$-th A-arrival is at time $t<T_{j+1}$, then the next arrival, regardless of A- or H-arrival, will add an additional height to the adversarial branch; afterwards, only A-arrivals will add heights to this branch. In this case, the probability for the adversarial branch to rise by $i$ heights where $i>l$ is given by:
    \begin{align}
        \int_{0}^{\Delta}f_2(t;l,a)\int_{0}^{\Delta - t}\lambda e^{-\lambda s}f_1(i-l-1,a(\Delta -t-s)) ds dt.\label{eq:x2_3}
    \end{align}
\end{enumerate}

    \item If $l=0$, the adversarial branch is made a public fork choice right away. There are two possibilities:
    \begin{enumerate}
        \item [III-1] A first A- or H-arrival will lead to an immediate rise of the adversarial branch; subsequent A-blocks will add additional heights. (This is equivalent to a degeneracy of Case (ii-3).) The probability is
        \begin{align}
         \int_{0}^{\Delta }\lambda e^{-\lambda s}f_1(i-l-1,a(\Delta-s)) ds 
        \end{align}
        
        \item [III-2] No blocks arrive during $(T_{j+1}-\Delta,T_{j+1})$.(This is equivalent to a degeneracy of Case (ii-2)) The probability is 
        \begin{align}
        \label{eq:x_2end}
           e^{-\lambda\Delta}. 
        \end{align}
    \end{enumerate}    
\end{enumerate}
Since $W$ denotes the height increment of the adversarial branch during $(T_{j+1}-\Delta, T_{j+1}]$, combining \eqref{eq:x2_1} to \eqref{eq:x_2end}, $  P(W = i|L=l)$ is given by~\eqref{eq:PW}.

\subsection{Markov Chain}

Let us consider a nonstationary Markov chain over epochs $1,\dots,k$. In epoch $j$, i.e, during $(T_{j-1},T_j]$, if the height of the adversarial branch rises from $y$ to $y'$, we say the Markov chain makes a transition from state $y$ to state $y'$. For our purposes, once the adversarial branch rises to height $k$ by epoch $k$, a safety violation is guaranteed to occur, so we curtail the state space to $0,1,\dots,k$, using state $k$ to represent any adversarial branch with height of at least $k$. As a result, we have a nonstationary Markov chain of $k+1$ states.

We first show that it suffices to consider the first $k$ epochs of this $(k+1)$-state Markov chain. First, at time $T_k$, right after the $k$ epochs, the jumper branch becomes public with height $k$. If the adversarial branch is of height at least $k$, i.e., the Markov chain is at state $k$, a safety violation is attained. Second, if the adversarial branch has height $y<k$ at $T_k$, i.e., it has a deficit of $k-y$ relative to the jumper branch, whether a violation occurs is determined by the outcome of a race between two renewal processes: One of them is a Poisson process $(A_t)$ with rate $a$, where each arrival represents an increase in the height of the lower branch. The other one, referred to as the \textit{jumper process}, $(J_t)$, has inter-arrival times following an exponential distribution with rate $1/h$ plus a constant $\Delta$, with each arrival corresponding to a new jumper becoming public. The jumper process is a renewal process, whose inter-arrival times have a shifted exponential distribution 
with the following pdf:
\begin{align}
\label{eq:fj}
    f_J(t) = he^{-h(t-\Delta)}1_{\{t>\Delta\}}.
\end{align}
Let us define 
\begin{align} \label{eq:M=sup}
    M=\sup_{t \geq 0}\{A_t-J_t\} .
\end{align}
If the Poisson process ever overtakes the jumper process by $k-y$ heights, i.e., $M\ge k-y$, a violation occurs; otherwise, a violation never occurs.

By~\cite[Theorem 3.4]{kroese1992difference}, the moment generating function of $M$ is given by~\eqref{eq:E(r)} and its pmf is given by~\eqref{eq:e(i)}.

A key distinction between the periods before and after $T_k$ is that before $T_k$, the adversary can \textit{bait} honest miners into switching the chain they are mining on, but after $T_k$, any remaining opportunity for baiting would result in a safety violation. Consequently, the \textit{bait-and-switch} phenomenon need not be considered beyond $T_k$. As a result, it suffices to run the Markov chain for $k$ epochs, while renewal theory accounts for the subsequent dynamics. In the following, we focus on the transition probability during $k$ epochs.

We now establish the transition probabilities for the Markov chain over epochs $1,\dots,k$. In the $j$-th epoch, we use
$P^{(j)}_{y,y'}$ to denote the probability of a transition from state $y$ to $y'$. Evidently, $P^{(j)}_{y,y'}=0$ if $y'<y$. Note that $y'=y+U+W$, we have that for all $y'\in\{y,\dots,k-1\}$,
\begin{align}
   P^{(j)}_{y,y'}
   &= P( Y_j = y' | Y_{j-1}=y )\\
   &= P(U+W=y'-y | Y_{j-1}=y) \\
   &= \sum_{i=0}^{y'-y}P(U = i)P(W = y'-y-i | Y_j=y, U=i)    \label{eq:UWYU}
\end{align}
where the distributions of $U$ 
is given by~\eqref{eq:x1}. It is easy to verify that~\eqref{eq:UWYU} can be written as~\eqref{eq:P^{j}_1}. Moreover, for every $y$, we define $P^{(j)}_{y,k}$ using~\eqref{eq:P^{j}_2}.
We have thus completed the description of transition matrices $P^{(j)}$, $j=1,\dots,k$, whose entry on the $(y+1)$-st row and $(y'+1)$-st column is equal to $P^{(j)}_{y,y'}$.

To calculate the probability of a safety violation on height 1, we investigate the state distribution at the end of the $k$-th epoch, where the Markov chain starts at state 0. 
It is easy to see that
\begin{align}
    P(Y_k=y)
    = \left[ P^{(1)} \times P^{(2)} \times \dots \times P^{(k)} \right]_{1,y
    +1}
\end{align}
where $[\cdot]_{1,y}$ takes the entry of a matrix on the first row and the $y$-th column.
A violation does not occur if and only if the state $Y_k$ plus $M$ (which is independent of $Y_k$) is strictly less than $k$, the probability of which can be calculated as
\begin{align}
    P( Y_k + M < k ) =
    \sum_{i=0}^{k-1} \, \sum_{y=0}^{k-1-i} P(M =i) P(Y_k=y) .
\end{align}
The probability of a violation is thus given
by~\eqref{eq:exact_bitcoin} in Theorem~\ref{th:pow}.


\section{Safety of the Target Block} 
\label{a:general}

\subsection{Generalized State Representation}

As discussed in Sec.~\ref{s:target}, a target transaction is created and immediately made public at time $s$, and is subsequently included in the first H-block mined after $s$, denoted block $b$. We use $t_b$ and $h_b$ to denote the arrival time and height of target block $b$. For the purpose of analyzing the safety of block $b$'s, the effective ``genesis'' height is $h_b-1$. If block $b$ is the first block placed at height $h_b$, then the height of the challenger branch at time $t_b$ is exactly $h_b-1$. Otherwise, at least one other block has already been placed on height $h_b$ by $t_b$, in which case the challenger branch contains a highest block at time $t_b$, i.e., its height is $\delta+L_{t_b}$, where $\delta$ is the height of the highest jumper and $L_{t_b}$ is the pre-mining lead at time $t_b$.

Starting from time $t_b$, the adversary seeks to grow both the target and challenger branches in order to maximize the probability that the following two conditions are met simultaneously: 1) both branches are credible; and 2) both reach at least height $h_b+k-1$. By Lemma~\ref{lm:eqtwobranches}, the probability of violating safety at height $h_b$ is equivalent to the probability of violating safety at height 1, starting from a transformed initial state in which the target branch contains the target block at height $1$ and the challenger branch has height either 0 or $\delta+L_{t_b}+1-h_b$, depending on whether the challenger branch is lower than the target branch.

Throughout this appendix, we use a generalized representation
\begin{align}
    [m,\delta,n,(l_{m\land \delta},\dots,l_\delta),I]_\eta
\end{align}
to denote the state when the target block is at height $\eta+1$. Here, $m$ and $n$ represent the lower and higher heights between the target and challenger branches, respectively; $\delta$ and the timers are defined as in the original state in Sec.~\ref{s:pru}. When $\eta=0$, the subscript $\eta$ can be omitted for consistency with the notation when the target block is at height 1. Thus, the subscript $\eta$ indicates the effective ``genesis'' height for a target block at height $\eta+1$. The following lemma allows us to equate a problem of attacking height $\eta+1$ to a problem of attacking height 1.

\begin{lemma}
\label{lm:eta=1}
    We have
    \begin{align}
    \begin{split}
    [m+\eta,\delta+\eta,\,& n+\eta,(l_{(m\land \delta)+\eta},\dots,l_{\delta+\eta})]_\eta        \\
    &\qquad \asymp
    [m,\delta,n,(l_{\eta+ m\land \delta},\dots,l_{\delta+\eta})]
    \end{split}
    \end{align}
    as long as both sides of the equality are valid states.
\end{lemma}

\begin{proof}
We claim that the probability of violation is invariant under a uniform height shift of $\eta$, where all branches---including the target block---are raised by $\eta$ heights, while keeping block timers unchanged. To establish this, we construct a one-to-one correspondence between trajectories in the original system (where the target block is at height 1) and those in the shifted system (where the target is at height $1+\eta$). Specifically, for every block placement in the original system, we consider the corresponding placement $\eta$ heights higher in the shifted system.

Since the dynamics, including block arrivals, actions, and timer updates, remain identical across the two systems---except for the uniform height offset---they transition through structurally equivalent states at the same times. In particular, any trajectory that leads to a safety violation in the original system leads to a corresponding violation in the shifted system at the same time, and vice versa.

Therefore, the probability of safety violation remains unchanged under this transformation.
\end{proof}

For any general target block, the state is equivalent to a state in a system where height~1 is under attack. Since the optimal attack on safety at height~1, bait-and-switch, is Markovian, we can see that from the moment the target block including the target transaction is on chain, bait-and-switch is optimal for violating a target block/height's safety. Hence the proof of Theorem~\ref{th:general}.

\subsection{Target Placement}

We first note that if the highest jumper is public at time $s$, then the target block must be strictly higher. We claim that even if the highest jumper is nonpublic, placing the target block strictly lower than that jumper yields a less favorable state. Intuitively, because that highest jumper becomes public soon after (by $s+\Delta$), placing the target block any lower will require the adversary to catch up more heights in order to create a violation. This argument is formalized as follows:

\begin{lemma}
\label{lm:eta<d}
    Let $d$ denote the height of the highest jumper at time $s$. If the target block can be placed at a lower height $e<d$, then doing so yields a probability of safety violation that is no greater than placing it at height $d$.
\end{lemma}

\begin{proof}
    Let $t$ denote the target block's arrival time. 
    The lead immediately before the arrival, $\Ltm$ does not depend on the target's placement (in contrast to $L_t$). With the highest jumper at height $d$, if the target block is also placed at the height, then a target branch at height $d$ is created, with the challenger branch at height $d+\Ltm$. 
    It is not difficult to see that the system state is represented as $[d,d,d+\Ltm, 
    (l_d)]_{d-1}$, which is equivalent to
    \begin{align} \label{eq:etaB}
        B=[1,1,1+\Ltm,(l_d)]
    \end{align}
    by Lemma~\ref{lm:eta=1}.

    Suppose the target block is placed at height $e<d$. We use $\eta=e-1$ to denote the effective ``genesis'' height. By Lemma~\ref{lm:eta=1}, the state $[e,d,d+\Ltm,(l_e,\dots,l_d)]_{\eta}$ is equivalent to
    \begin{align}
        C
        &=[1,d-\eta,d+\Ltm-\eta,(l_{\eta+1},\dots,l_d)] \\
        &=[1,d-\eta,d+\Ltm-\eta,(0,l_{\eta+2},\dots,l_d)] \label{eq:Clm=0}
    \end{align}
   where~\eqref{eq:Clm=0} is due to Lemma~\ref{lm:lm=0}.
    
    To establish this lemma, it suffices to prove that $B\succeq C$.

    We have the following two cases:
\begin{itemize}
    \item 
    Case $d-\eta>\Ltm$: In Sec.~\ref{s:proofv}, we have proved Lemma~\ref{lm:E<=F}, which states that $[m,d,m+i+1,(0,l_{m+1},\dots,l_d)] \preceq [m+1,d,m+i,(l_{m+1},\dots,l_d)]$ for all natural number $i$, i.e., moving one highest block from the challenger branch to the target branch improves the state for the adversary. Invoking Lemmas~\ref{lm:E<=F} and~\ref{lm:lm=0} repeatedly for $\Ltm$ times, we have
    \begin{align}
    \label{eq:1,d,d+1<2,d,d}
    C
    & \preceq
    [2,d-\eta,d+\Ltm-\eta-1,(l_{\eta+2},\dots,l_d)] \\   
    & \preceq
    [3,d-\eta,d+\Ltm-\eta-2,(l_{\eta+3},\dots,l_d)] \\
    & \preceq \dots \\
    & \preceq
    C_1
    \end{align}
    where
    \begin{align}
        C_1 = [1+\Ltm,d-\eta,d-\eta, (l_{\Ltm+\eta+1},\dots,l_d)].
    \end{align}
    We next show $C_1\preceq B$ in this case.

    If $d-\eta=\Ltm+1$, then
    \begin{align}
        C_1
        &= [1+\Ltm,1+\Ltm,1+\Ltm,(l_d)] \\
        &\preceq B \label{eq:C1<=B}
    \end{align}
    where~\eqref{eq:C1<=B} is due to Lemma~\ref{lm:A>=Hss'} (state $C_1$ can be obtained by adding $\Ltm$ H-blocks in the lower branch of state $B$).

    We next prove $C_1\preceq B$ assuming $d-\eta\ge\Ltm+2$:
    \begin{align}
        C_1
        &\preceq  [1+\Ltm,2+\Ltm,2+\Ltm, (l_{\Ltm+1},l_{\Ltm+2})] \label{eq:C1<=aaa} \\
        &\preceq  [1+\Ltm,2+\Ltm,2+\Ltm, (0,\Delta)] \label{eq:C1<=bbb} \\
        &\preceq  [1+\Ltm,1+\Ltm,1+\Ltm, (0)] \label{eq:C1<=ccc} \\
        &\preceq  B \label{eq:C1<=Btoo}
\end{align}
where~\eqref{eq:C1<=aaa} is due to $(A_{2k})$ in Proposition~\ref{pr:ABCD},~\eqref{eq:C1<=bbb} is due to Lemma~\ref{lm:h/l rule}, and~\eqref{eq:C1<=ccc} and~\eqref{eq:C1<=Btoo} are both due to Lemma~\ref{lm:A>=Hss'}.

    \item 
    Case $d-\eta\le \Ltm$: Invoking Lemmas~\ref{lm:E<=F} and~\ref{lm:lm=0} repeatedly for $d-\eta-1$ times yields $ C  \preceq  C_2$ where
    \begin{align}
        C_2
        &= [d-\eta,d-\eta,1+\Ltm, (l_d)] \\
        &\preceq [d-\eta,d-\eta,1+\Ltm, (\Delta)] \label{eq:C2<=aaa} \\
        &\preceq B \label{eq:C2<=B}
    \end{align}
    where~\eqref{eq:C2<=aaa} is due to Lemma~\ref{lm:h/l rule} and~\eqref{eq:C2<=B} is due to Lemma~\ref{lm:A>=Hss'}. 
\end{itemize}
    Since $B\succeq C_1$ and $B \succeq C_2$, it follows that $B\succeq C$ in all cases. In other words, placing the target block at height $d$ is at least as good as placing it below height $d$.
\end{proof}

We next claim that if the target block can be placed on the height of the highest jumper, it will not be placed any higher either. Intuitively, placing the target block higher reduces the pre-mining lead. This argument is formalized as follows:

\begin{lemma}
\label{lm:eta>d}
    Let $d$ denote the height of the highest jumper at time $s$. If the target block can be placed at height $d$, then doing so yields a probability of safety violation that is no less than placing it at any higher height $e>d$.
\end{lemma}

\begin{proof}
    As in the proof of Lemma~\ref{lm:eta<d}, if the target block is placed at height $d$, then the state at its arrival time $t$ is 
    equivalent to the state $B$ given by~\eqref{eq:etaB}. 

    Now suppose the target block is instead placed at a higher height $e>d$. In that case, the pre-mining lead reduces at time $t$ to $d+\Ltm-e<\Ltm$. Let $\eta=e-1$ denote the effective genesis height. Depending on whether the target block is the first to reach its height, the state at time $t$ falls into one of two categories:
    \begin{itemize}
        \item If the target block is the first to reach its height, then the challenger branch is one height lower, and the state is
        \begin{align}
            [\eta,e,e, (l_\eta,\Delta)]_\eta 
            &\asymp [0,1,1,(l_\eta,\Delta)] \label{eq:<=Bxxx} \\
            &\preceq B \label{eq:<=Byyy}
        \end{align}
        where~\eqref{eq:<=Bxxx} is due to Lemma~\ref{lm:eta=1}  and~\eqref{eq:<=Byyy} is due to Lemma~\ref{lm:h/l rule}.

        \item Otherwise, there has already been another block on height $e$, then the challenger branch is no lower than the target branch, and the state is
        \begin{align}
            [e,e,d+\Ltm, (\Delta)]_\eta
            &\asymp [1,1,d+\Ltm-\eta,(\Delta)] \label{eq:<=Buuu} \\
            &\preceq B \label{eq:<=Bvvv}
        \end{align}
        where~\eqref{eq:<=Buuu} is due to Lemma~\ref{lm:eta=1}  and~\eqref{eq:<=Bvvv} is due to Lemma~\ref{lm:h/l rule}.
    \end{itemize}
    In both cases, the resulting state is dominated by $B$. This concludes the proof.
\end{proof}

We have thus shown that if the target block can be placed at height $d$, the height of the highest jumper, before it becomes public, then doing so is optimal. However, if the target block arrives after the highest jumper becomes public, it must be placed higher (as a jumper). We now claim that, in this case, placing it at height $d+1$ is optimal. This is formalized as follows:

\begin{lemma}
\label{lm:d+1>eta+1}
    Let $d$ denote the height of the highest jumper at time $s$. If the target block arrives after the height-$d$ jumper has becomes public, then placing the target block at height $d+1$ yields a probability of safety violation that is no less than placing it at any higher height $e>d+1$.
\end{lemma}

\begin{proof}
    Suppose the target block is placed at height $d+1$ at time $t$, yielding the following possible state representations:
    \begin{itemize}
        \item If the target block is the first to reach height $d+1$, then at time $t$, the state is $[d,d+1,d+1,(0,\Delta)]_d$, which is equivalent to $[0,1,1,(0,\Delta)]$ by Lemma~\ref{lm:eta=1}.
        
        \item Otherwise, there must be at least one A-block at height $d+1$. In this case, the target branch is at height $d+1$, and the challenger branch is at height $d+\Ltm$. Thus, the state at time $t$ is $[d+1,d+1,d+\Ltm,(\Delta)]_d$, which is equivalent to $[1,1,\Ltm,(\Delta)]$ by Lemma~\ref{lm:eta=1}. 
    \end{itemize}
    
    Now suppose the target block is placed at height $e>d+1$ instead, with $\eta=e-1$ representing the effective genesis height. Since the previous highest jumper is at height $d$, there must be at least one A-block at height $d+1$. At time $t$, the state is:
    \begin{itemize}
        \item If the target block is the first to reach its height, the challenger branch is at height $\eta$, so the state is represented as $[\eta,e,e,(\Delta,\Delta)]_\eta$, which is equivalent to $[0,1,1,(0,\Delta)]$ by Lemma~\ref{lm:eta=1} and Lemma~\ref{lm:lm=0}.

        \item Otherwise, the challenger branch is at height $d+\Ltm$, and the state is represented as $[e,e,d+\Ltm,(\Delta)]_\eta$, which is equivalent to $[1,1,d+\Ltm-\eta,(\Delta)] \preceq [1,1,\Ltm,(\Delta)]$.
    \end{itemize}

    In either case, the adversary is no worse off---and potentially better off---by placing the target block at height $d+1$.
\end{proof}

Combining Lemmas~\ref{lm:eta<d}--\ref{lm:d+1>eta+1}, we arrive at the following conclusion:

\begin{lemma}
\label{lm:placetarget}
    The target block is placed at the height of the highest jumper if it is not yet public, and one height above the highest jumper if it is already public.
\end{lemma}

\subsection{Pre-mining Lead}

In the following, we first analyze the system state upon the first jumper arrives after time $s$, and then we analyze 
the distribution of pre-mining lead right before time $t$, $L_{t^-}$, where a jumper arrives at time $t$.

\begin{lemma}
\label{lm:L_t}
    For every $t>0$,
    \begin{align}
    \label{eq:L_t}
    L_t \leq \sup_{r \in [0,t]}\{A_{r,t}-J_{r,t}\}
    \end{align}
    where the equality holds when each A-block extends a highest chain and each jumper is placed at one height above the previous jumper.
\end{lemma}

\begin{proof}
    By Definition~\ref{def:pre}, the pre-mining lead increases by at most 1 for each A-block, which is achieved by extending a highest chain. Clearly, to maximize the pre-mining lead, every A-block should extend a highest chain.

    Suppose a jumper is placed at time $t$. If $\Ltm=0$, then no A-block is higher than the highest jumper just before $t$, hence the new jumper must be exactly 1 height higher. If $\Ltm>0$, then a new jumper reduces the pre-mining lead by at least 1. To maximize the lead $L_t$, the new jumper must be placed 1 height above the previous highest jumper, so that $L_t=\max\{0,\Ltm-1\}$.
    
    If 
    $L_t = 0$, 
    then~\eqref{eq:L_t} holds trivially by setting $r=t$ on the right hand side of the equation.

    If $L_t>0$, since $L_0=0$ and the lead can only increase one at a time,  there exists a time $v\ge0$ when $L_v = 0$, but $L_q >0$ for $q \in (v,t]$. In this case, every jumper arriving during $(v,t]$ decreases the lead by 1. Thus,
    \begin{align}
        L_t
        &\le L_v + A_{v,t} - J_{v,t} \label{eq:LtLv} \\
        &= A_{v,t}-J_{v,t} \label{eq:Lt:A-J}
    \end{align}
    where the equality in~\eqref{eq:LtLv} holds if every A-block extends a highest chain.

    Evidently, taking the supremum on the right hand side of~\eqref{eq:Lt:A-J} leads to inequality~\eqref{eq:L_t}. One can also show that the equality in~\eqref{eq:L_t} is achieved by placing jumpers at one height above the previous jumper and using every A-block to extend a highest chain.
\end{proof}

By Lemma~\ref{lm:L_t}, right before time $t$ we have
\begin{align}
\label{eq:Lt-}
    L_{t^-} =  \sup_{r \in [0,t)}\{A_{r,t^-}-J_{r,t^-}\},
\end{align}
where $J_{r,t^-}$ denotes the number of jumpers arrive during $(r,t)$.  

\begin{lemma}
\label{lm:state_start}
    Consider the state's equivalence upon the optimal placement of the first jumper after $s$. We have the following two case:
\begin{enumerate}
    \item [i)]
    If the target block is a jumper, where $t$ denote its arrival time. Then the state at time $t$ is equivalent to state $[0,1,1,(0,\Delta)]$ if $L_{t^-} = 0$ or $[1,1,L_{t^-},(\Delta)]$ if $L_{t^-} > 0$.
    
    \item [ii)]
    If the target block is not a jumper, let $r$ denote the arrival time of the first jumper after time $s$. Then the state at time $r$ is equivalent to  state $[1,2,2,(0,\Delta)]$ if $L_{r^-}=0$ or state $[2,2,1+L_{r^-},(\Delta)]$ if $L_{r^-}>0$.
\end{enumerate}
\end{lemma}

\begin{proof}
    Let $d$ denote the height of the highest jumper before the target block arrives at time $t$.
    
    Consider Case (i) where the target block is a jumper. By Lemma~\ref{lm:placetarget}, it must be at height $d+1$.

    If $\Ltm=0$, then the target block is the first to reach height $d+1$.
    In this case, the state at time $t$ can be represented as $[d,d+1,d+1,(0,\Delta)]_d$, which is equivalent to $[0,1,1,(0,\Delta)]$ by Lemma~\ref{lm:eta=1}.
    
    If $L_{t^-}>0$, then at least one other block has reached height $d+1$. The state at time $t$ can be represented as $[d+1,d+1,d+1+L_t,(\Delta)]_d$. Since $\Ltm=1+L_t$, the state is equivalent to
    \begin{align}
        [d+1,d+1,d+L_{t^-},(\Delta)]_d
        \asymp
        [1,1,L_{t^-},(\Delta)]
    \end{align}
    by Lemma~\ref{lm:eta=1}.

    Now consider Case (ii) where the target block is not a jumper. By Lemma~\ref{lm:placetarget} the target block is place at height $d$. Hence the state at time $t$ is $[d,d,d+L_t,(l)]_{d-1}$, which is equivalent to $[1,1,1+L_t,(l)]$ by Lemma~\ref{lm:eta=1}. 

Starting from an ahead state at $[1,1,1+L_t,(l)]$ at time $t$, denote the arrival time of the first H-block that arrives after $t+l$ as time $r$.
Before time $r$, the state stays ahead.
For blocks arriving during $(t,r)$, every A-block extends the higher branch, and every H-block is placed a height~1 of the lower branch according to Proposition~\ref{cl:ahead}.
At time $r$, height $1$ is already public, so this H-block arriving at time $r$ is a jumper at height~2.

Since the subsequent jumper after $t$ arrives at time $r$, and all A-blocks extend the highest chain, by Lemma~\ref{lm:L_t}, $ L_t+A_{t,r}=L_{r^-}$.

If $L_{r^-} = L_t+A_{t,r} = 0$, then right before $r$, the highest block is at target block's height, $1$, which implies that there is only the new jumper block at height~2 at time $r$. At time $r$, the state can be represented as $[1,2,2,(0,\Delta)]$.

If $L_{r^-} = L_t+A_{t,r} > 0$, then right before $r$, the highest block is higher than the target block's height, $1$, which implies that there is not only the new jumper block at height~2. The state can be represented as
$[2,2,1+L_t+A_{t,r},(\Delta)] = [2,2,1+L_{r^-},(\Delta)]$ at time $r$.
\end{proof}

Let $\mathscr{J}$ denote the event that the target block is a jumper; let $\mathscr{J}^C$ denote its complement.

Let
\begin{align}
\label{eq:fG}
  f_G(g) =
\begin{cases}
\displaystyle \frac{1}{\Delta + \frac{1}{h}}, & 0 \le g \le \Delta \\
\displaystyle \frac{e^{-h(g - \Delta)}}{\Delta + \frac{1}{h}}, & g > \Delta.
\end{cases}  
\end{align}

Let
\begin{align}
    s(0) &= e(1)+e(0)\label{eq:s0}\\
    s(i) &= e(i+1) \text{ for } i>0,\label{eq:s1}
\end{align}
where $e(i)$ is given in \eqref{eq:e(i)}.

\begin{lemma}
Given time $s$, let $t$ be the arrival time of the first jumper block arriving after $s$, we have
\begin{align}
\label{eq:L,J}
&P(L_{t^-} = n, \mathscr{J}) \notag\\
&= \sum_{i=0}^{n} s(i) 
    \sum_{j=0}^{n-i} \Bigg(  \int_{0}^{\Delta} 
    f_1(ga; j) \, e^{-h(\Delta - g)}\times \notag \\
&\quad
    \int_{0}^{\infty} 
    f_1((t + \Delta - g)a; n - i - j) \, 
    h e^{-h t} \, dt \, f_G(g) \, dg \notag \\
&\quad + \int_{\Delta}^{\infty} 
    f_1(ga; j) 
    \int_{0}^{\infty} 
    f_1(t a; n - i - j) \, 
    h e^{-h t} \, dt \, f_G(g) \, dg 
\Bigg).
\end{align}
    For convenience, we denote $P(L_{t^-}=n,\mathscr{J})$ as $f_3(n)$. Then
    \begin{align}
    \label{eq:L,Jc}
      &P(L_{t^-}=n,\mathscr{J}^C)\notag \\ &= \sum_{i=0}^{n}s(i)\sum_{j=0}^{n-i} \Bigg( \int_{0}^{\Delta}f_1(ga;j)(1-e^{-h(\Delta-g)})\notag \\
      &\times \int_{0}^{\infty}f_1((t+\Delta-g)a;n-i-j)he^{-ht}dtf_G(g)dg\Bigg)
    \end{align}
    where $f_1(\lambda;\cdot)$ denotes the pmf of a Poisson distribution with expectation $\lambda$.
    For convenience, we denote $P(L_{t^-}=n,\mathscr{J}^C)$ as $f_4(n)$.
\end{lemma}

\begin{proof}
We suppose at the given time $s$, the jumper process reaches the stationary. Let $r$ denote the last jumper arrival before $s$ and $t$ denote the next jumper arrival after $s$. 
Since $J_{r,s^-} = 0$, by Lemma~\ref{eq:Lt-}, we have
\begin{align}
    L_{t^-} = L_r +A_{r,s}+A_{s,t}.
\end{align}

Looking at the joint probability of $L_{t^-} = 0$ and $\mathscr{J}$:
\begin{align}
    P(L_{t^-} = n, \mathscr{J}) &= P(L_r +A_{r,s}+A_{s,t}= n,\mathscr{J})\\
    &=\sum_{i = 0}^{n}P(L_r = i, A_{r,s}+A_{s,t}= n-i,\mathscr{J})\\
    & = \sum_{i = 0}^{n}P(L_r = i| A_{r,s}+A_{s,t}= n-i,\mathscr{J})\notag\\
    &\quad \quad \times P(A_{r,s}+A_{s,t}= n-i,\mathscr{J})\\
    &=\sum_{i = 0}^{n}P(L_r = i)P(A_{r,s}+A_{s,t}= n-i,\mathscr{J}),\label{eq:L|=L}
\end{align}
where \eqref{eq:L|=L} is due to that $r$ is a renewal point of the jumper process, it is independent with events happen
after $r$.

Similarly, we have
\begin{align}
\label{eq:L|=Lc}
    L(L_{t^-}=n,\mathscr{J}^C) = \sum_{i = 0}^{n}P(L_r = i)P(A_{r,s}+A_{s,t}= n-i,\mathscr{J}^C).
\end{align}

By Lemma~\ref{lm:L_t}, $L_{r^-} = \sup_{v\in[0,r)}\{A_{v,r^-}-J_{v,r^-}\} $. Since $r$ is a renewal point of the jumper process, which is also stationary by $r$, we have
$J_{v,r^-} \overset{d}{=} J_{r,r+r-v} \overset{d}{=}J_{0,r-v}$. 
Since A, the adversarial arrival, is a Poisson process, we have $A_{v,r^-} \overset{d}{=} A_{0,r-v}$.
Thus,
\begin{align}
    L_{r^-} \overset{d}{=} \sup_{v\in[0,r)}\{A_{0,r-v}-J_{0,r-v}\} .
\end{align}
Since $A$ is stochastically dominated by $J$ due to \eqref{eq:a>}, for $r$ big enough we have
\begin{align}
    L_{r^-} \overset{d}{=} \sup_{v>0}\{A_{0,v}-J_{0,v}\}.
\end{align}
Thus, $L_{r^-} \overset{d}{=} M$ defined in \eqref{eq:M=sup}, so 
\begin{align}
    P(L_{r^-} = i) = e(i),
\end{align}
where $e(i)$ is given in \eqref{eq:e(i)}.
Since there is one jumper arrives at $r$, we have $L_r = L_{r^-} -1$ if $L_{r^-}>0$ and $L_r = 0$ if $L_{r^-}=0$. 
Thus, for $i >0$
\begin{align}
    P(L_r = i) = e(i+1),
\end{align}
and
\begin{align}
     P(L_r = 0) = e(1)+e(0).
\end{align}
Thus, the pmf of $L_r$ is given as $s(i)$ in \eqref{eq:s0} and \eqref{eq:s1}.

The age at time $s$ is $s-r$, let $G$ denote $s-r$.
Since the jumper process is stationary at $s$, $G$ has the distribution with pdf~\eqref{eq:fG}.
We have
\begin{align}
  &P(A_{r,s}+A_{s,t}= n-i,\mathscr{J})\notag 
  \\
  &= \sum_{j = 0}^{n-i}  P(A_{r,s}=j,A_{s,t}= n-i-j,\mathscr{J})\\
  & = \sum_{j = 0}^{n-i} \int_{0}^{\infty}P(A_{r,s}=j,A_{s,t}= n-i-j,\mathscr{J}|G=g)f_G(g)dg\\
   &= \sum_{j = 0}^{n-i} \int_{0}^{\infty}P(A_{s-G,s}=j|A_{s,t}= n-i-j,\mathscr{J},G=g) \notag \\
   &\quad \quad \quad \times P(A_{s,t}= n-i-j,\mathscr{J}|G=g)f_G(g)dg\\
   &=\sum_{j = 0}^{n-i} \int_{0}^{\infty}P(A_{s-g,s}=j) P(A_{s,t}= n-i-j,\mathscr{J}|G=g)f_G(g)dg\label{eq:P(A_{s-g,s}=j)}
\end{align}
where \eqref{eq:P(A_{s-g,s}=j)} is due to that Poisson process $A$ is memoryless.

Since $A$ is Poisson process, 
\begin{align}
\label{eq:As-g}
    P(A_{s-g,s}=j) = f_1(ga;j),
\end{align}
where $f_1$ is the pmf of Poisson point distribution.

If $g \geq \Delta$, then at time $s$, the highest jumper at time $s$ is already public. The next H-block arrives at time $t$ after time $s$ is the new jumper. Thus, $t-s$ follows the exponential distribution with rate $h$. And during $(s,t]$, the number of A-arrivals follows a Poisson point distribution.
For $g \geq \Delta$:
\begin{align}
\label{eq:PJG>}
   & P(A_{s,t}= n-i-j,J|G=g) \notag\\
   &= \int_{0}^{\infty}f_1((t-s)a;n-i-j)he^{-h(t-s)}d(t-s)\\
    &= \int_{0}^{\infty}f_1(ta;n-i-j)he^{-ht}dt.
\end{align}

If $g<\Delta$, at time $s$, the highest jumper still need $\Delta-g$ to become public. $\mathcal{J}$ implies that there is no H-block arriving during $(s,\Delta-g)$, which occurs with probability $e^{-h(\Delta-g)}$. After time $s+\Delta-g$, the first H-block is the new jumper. Thus, $t-(s+\Delta-g)$ follows the exponential distribution with rate $h$. And during $(s,t]$, the number of A-arrivals follows a Poisson point distribution.
For $g < \Delta$:
\begin{align}
\label{eq:PJG<}
    &P(A_{s,t}= n-i-j,J|G=g) \notag \\&= e^{-h(\Delta-g)}\int_{0}^{\infty}f_1((t-s)a;n-i-j)\notag \\
    & \quad \times he^{-h(t-(s+\Delta-g))}d(t-(s+\Delta-g)) \\
    &= e^{-h(\Delta-g)}\int_{0}^{\infty}f_1((t+\Delta-g)a;n-i-j)he^{-ht}dt .
\end{align}
Combining \eqref{eq:PJG<}, \eqref{eq:PJG>},  \eqref{eq:P(A_{s-g,s}=j)} and \eqref{eq:As-g}, we have
\begin{align}
    &P(A_{r,s}+A_{s,t} = n-i,\mathscr{J})\notag\\
    &=\sum_{j=0}^{n-i}\Bigg(\int_{0}^{\Delta}f_1(ga;j) e^{-h(\Delta-g)}\notag \\
    &\times\int_{0}^{\infty}f_1((t+\Delta-g)a;n-i-j)he^{-ht}dt f_G(g)dg\notag \\
    &+\int_{\Delta}^{\infty}f_1(ga;j) \int_{0}^{\infty}f_1(ta;n-i-j)he^{-ht}dtf_G(g)dg\Bigg) .
\end{align}
Plugging $P(A_{r,s}+A_{s,t} = n-i,\mathscr{J})$ into \eqref{eq:L|=L}, we have proved \eqref{eq:L,J}. 

Similar to the case of $\mathscr{J}$, for $\mathscr{J}^C$, we also have
\begin{align}
\label{eq:As-gc}
  &P(A_{r,s}+A_{s,t} = n-i,\mathscr{J}^C) \notag \\
  &=   \sum_{j=0}^{n-i}\int_{0}^{\infty}P(A_{s-g,s}=j)P(A_{s,t}=n-i-j,\mathscr{J}^C|G=g)f_G(g)dg.
\end{align}

If $g\geq \Delta$, at time $s$, the highest jumper at time $s$ is already public, so the next H-block which is the target block is jumper, contradict to $\mathscr{J}$.
Thus, for $g\geq \Delta$
\begin{align}
\label{eq:PJG>c}
    P(A_{s,t}=n-i-j,\mathcal{J}^C|G=g) =0.
\end{align}

If $g<\Delta$, at time $s$, the highest jumper still need $\Delta-g$ to become public. $\mathcal{J}$ implies that there exist H-block arrives during $(s,\Delta-g)$, which occurs with probability $1-e^{-h(\Delta-g)}$. After time $s+\Delta-g$, the first H-block is the new jumper. Thus, $t-(s+\Delta-g)$ follows the exponential distribution with rate $h$. During $(s,t]$, the number of A-arrivals follows a Poisson point distribution.
For $g < \Delta$:
\begin{align}
\label{eq:PJG<c}
    &P(A_{s,t}= n-i-j,J|G=g) \notag \\&= (1-e^{-h(\Delta-g)})\int_{0}^{\infty}f_1((t-s)a;n-i-j)\notag \\
    & \quad \times he^{-h(t-(s+\Delta-g))}d(t-(s+\Delta-g)) \\
    &= (1-e^{-h(\Delta-g)})\int_{0}^{\infty}f_1((t+\Delta-g)a;n-i-j)he^{-ht}dt .
\end{align}
Combining \eqref{eq:PJG<c}, \eqref{eq:PJG>c},\eqref{eq:P(A_{s-g,s}=j)} and \eqref{eq:As-gc}, we have
\begin{align}
    &P(A_{r,s}+A_{s,t} = n-i,\mathscr{J}^C) \notag \\
    &=\sum_{j=0}^{n-i} \int_{0}^{\Delta}f_1(ga;j)(1-e^{-h(\Delta-g)})\notag\\
    &\times\int_{0}^{\infty}f_1((t+\Delta-g)a;n-i-j)he^{-ht}dtf_G(g)dg.
\end{align}
Plug $P(A_{r,s}+A_{s,t} = n-i,\mathscr{J}^C)$ into \eqref{eq:L|=Lc}, we have proved \eqref{eq:L,Jc}.
\end{proof}

\subsection{Probability Calculation}
In this subsection, we analyze the probability of safety violation caused by bait-and-switch attacks targeting a specific block at an arbitrary height. 
Our analysis considers two distinct cases:
\begin{itemize}
    \item The target block is a \emph{jumper};
    \item The target block is \emph{not} a jumper.
\end{itemize}
For both scenarios, we extend the Markov chain analysis developed in Appendix~\ref{a:trade-off} to compute the exact violation probabilities.

We first look at the case where the target block is a jumper, where the target block arrives at time $t$.

At time~$t$, the system state is characterized by the challenger branch height $L_{t^-}$, where by Lemma~\ref{lm:state_start}:
\begin{enumerate}
    \item For $L_{t^-} = 0$, the challenger branch is at height~0 with the target jumper block at height~1 (timer $\Delta$)
    \item For $L_{t^-} > 0$, the challenger remains at height~$L_{t^-}$
\end{enumerate}

We extend Appendix~\ref{a:trade-off}'s Markov chain analysis, initializing at $t = T_1 - \Delta$. Unlike the height~1 safety analysis starting from $T_0$, we examine arbitrary target heights with jumper blocks, beginning at the jumper's arrival time $T_1 - \Delta \ (= t)$ from initial state $Y_{1-\Delta} = L_{t^-}$.

For both cases, the public height at $T_1-\Delta$ is~0, and we conclude that the height difference is $L = -L_{t^-}$ (between public height and challenger/adversary branch).

From $T_1-\Delta$ to $T_1$, with the height increment of the adversarial branch denoted as $W$, at time $T_1$, the height of the adversarial branch is $W+L_{t^-}$. And the distribution of $W$ conditioned on $L$ is the same as in Appendix~\ref{a:tradeoff_chainincre}.
To establish the transition probability over epoch 1, starting from $T_1-\Delta$ tp $T_1$, for all $y'\in\{y,\dots,k-1\}$, we have
\begin{align}
\label{eq:P1l}
    P^{(1)}_{L_{t^-},y'} &= (Y_1 = y'|Y_{1-\Delta} = L_{t^-})\\
    &=P(W = y'-L_{t^-}|L = -L_{t^-}),
\end{align}
where the distribution of $W$ given $L$ is given by \eqref{eq:PW}. 
For $i\neq L_{t^-}$, 
\begin{align}
\label{eq:P1i}
  P^{(1)}_{i,y'}=0 .
\end{align}
Starting from time $T_1$, the state transition is the same as in Appendix~\ref{a:trade-off}. Thus,
for $j>1$,
\begin{align}
\label{eq:Pjyy}
    P^{(j)}_{y,y'} = 
\sum_{i=0}^{y'-y}
   \frac{h}{\lambda}(\frac{a}{\lambda})^i
   P(W = y'-y-i | L=j-1-y-i) .
\end{align}
 For $j = 1,\dots,k$,
 \begin{align}
 \label{eq:Pjyk}
     P^{(j)}_{y,k} = 1 - \sum_{i=y}^{k-1} P^{(j)}_{y,i} .
 \end{align}
We have thus completed the description of transition matrices $P^{(j)}$, $j = 1,\dots,k$, whose $y+1$-st row and $(y'+1)$-st column is equal to $P^{(j)}_{y,y'}$.

Thus,
\begin{align}
    P(Y_k = y|L_{t^-} = l, \mathscr{J}) = \left[P^{(1)}\times \dots,P^{(k)} \right]_{l+1,y+1} .
\end{align}
For this case, the safety violation does not occur given $L_{t^-}=l$ and target transaction is a jumper can be calculated as
\begin{align}
    &P(Y_k+M<k|L_{t^-} = l, \mathscr{J}) \notag\\
    &= \sum_{i = 0}^{k-1}\sum_{y = l}^{k-1-i} P(M=i) P(Y_k = y|L_{t^-} = l, \mathscr{J}).
\end{align}
Finally,
\begin{align}
\label{eq:Y+M<k,J}
    &P(Y_k+M<k,L_{t^-} = l, \mathscr{J}) \notag\\
    &= P(Y_k+M<k|L_{t^-} = l, \mathscr{J})P(L_{t^-} = l, \mathscr{J}),
\end{align}
where $P(L_{t^-} = l, \mathscr{J})$ is given in \eqref{eq:L,J}.

In the following, we look at the case where the target block is not a jumper. Consider the first jumper arrives after time $s$ at time $t$ at height~2.
 
At time $t$, the system is characterized by the challenger branch height $1+L_{t^-}$, where by Lemma~\ref{lm:state_start}:
\begin{enumerate}
    \item For $L_t^{-} = 0$, the challenger branch is at height $1$
    \item For $L_t^- > 0$, the challenger branch is at $1+L_t^{-}$
\end{enumerate}
We again extend the Markov analysis in Appendix~\ref{a:general}, initializing at $t= T_2-\Delta$ (the jumper's arrival time, $t$) from initial state $Y_{2-\Delta} = 1+L_t^{-}$.

For both cases, the public height at $T_2-\Delta$ is height~$1$. We conclude that $L= -L_t^{-}$. 

Similar to the jumper target block case, we analyze the interval from $T_2 - \Delta$ to $T_2$ to derive transition probabilities for epoch~2. 
(Note that epoch~1 is omitted here since the state at $T_2 - \Delta$ is already known.) 
For all $y' \in \{y, \dots, k-1\}$, we define:
\begin{align}
\label{eq:P'2l}
    P'^{(2)}_{L_t^{-}+1,y'} 
        &= \mathbb{P}(Y_2 = y' \mid Y_{2-\Delta} = L_t^{-} + 1) \\
        &= \mathbb{P}(W = y' - L_t^{-} - 1 \mid L = -L_t^{-}),
\end{align}
where the distribution of $W$ given $L$ is given by \eqref{eq:PW}. 
For $i \neq L_t^{-}+1$, 
\begin{align}
\label{eq:P'2y}
    P'^{(2)}_{i,y'}=0.
\end{align}
Starting from time $T_2$, the state transition is the same as in Appendix~\ref{a:general}. Thus,
for $j>2$,
\begin{align}
\label{eq:P'yy}
    P'^{(j)}_{y,y'} = \sum_{i=0}^{y'-y}
   \frac{h}{\lambda}(\frac{a}{\lambda})^i
   P(W = y'-y-i | L=j-1-y-i),
\end{align}
and for $j>1$,
\begin{align}
\label{eq:P'yk}
   P'^{(j)}_{y,k} =  1 - \sum_{i=y}^{k-1} P'^{(j)}_{y,i}.
\end{align}
We have thus completed the description of transition matrices $P'^{(j)}$, $j = 1,\dots,k$, whose $y+1$-st row and $(y'+1)$-st column is equal to $P'^{(j)}_{y,y'}$.

Thus,
\begin{align}
  P(Y_k=y|L_t^{-}=l, \mathscr{J}^C)= \left[P'^{(2)}\times \dots,P'^{(k)} \right]_{l+2,y+1}.
\end{align}
For this case, the safety violation does not occur given $L_t^{-}=l$ and target block is not a jumper can be calculated as
\begin{align}
\begin{split}
    P&(Y_k+M<k \,|\, L_t^{-}=l, \mathscr{J}^C)\\
   &= \sum_{i=0}^{k-1}\sum_{y=l}^{k-1-i} P(M=i)P(Y_k = y \,|\, L_t^{-}=l, \mathscr{J}^C) .    
\end{split}
\end{align}
So we have
\begin{align}
\label{eq:Y+M<k,Jc}
\begin{split}
    P&(Y_k+M<k, L_t^{-}=l, \mathscr{J}^C) \\
    &= P(Y_k+M<k|L_t^{-}=l, \mathscr{J}^C) P(L_t^{-}=l, \mathscr{J}^C),
\end{split}
\end{align}
where $P(L_t^{-}=l, \mathscr{J}^C)$ is given in \eqref{eq:L,Jc}.

Finally, we can derive the probability of safety violation does not occur with probability as the following:
\begin{align}
\begin{split}
    P(Y_k+M<k) = \sum_{l = 0}^{k-1} & \Big(P(Y_k+M<k,L_{t^-}=l,\mathscr{J}) \\
    &+P(Y_k+M<k,L_{t^-}=l,\mathscr{J}^C)\Big)
\end{split}
\end{align}
where $P(Y_k+M<k,L_{t^-}=l,\mathscr{J})$ is given by \eqref{eq:Y+M<k,J} and $P(Y_k+M<k,L_{t^-}=l,\mathscr{J}^C)$ is given by \eqref{eq:Y+M<k,Jc}. Thus, the probability of safety violation of a target height including the target transaction is given in \eqref{eq:exact_bitcoin_l}, hence we finish the prove of Theorem~\ref{th:trade-offgeneral}.


\end{document}